\documentclass{aa}  
\usepackage{caption}  
\usepackage{graphicx}

\usepackage{color}
\usepackage{makecell}
\usepackage{txfonts,amsmath}
\usepackage{hyperref}
\usepackage{booktabs}
\usepackage{multirow}
\usepackage{threeparttable} 
\usepackage{etoolbox}
\makeatletter
\patchcmd{\enddocument}
{\makeatletter \input\jobname.aux}
{\NAT@swatrue\let\bibcite\NAT@testdef\makeatletter \input\jobname.aux}
{}{\PackageError{aa-natbib-compat}{Failed to patch auxiliary-file reread}{Check the aa class version.}}
\makeatother

\newenvironment{systemfigurepage}
{\begin{figure*}[p]\vbox to \textheight\bgroup\centering\vfil}
	{\par\vfil\egroup\end{figure*}}

\newcommand{\zs}[1]{$z_s = #1$}

\newcommand{\mr}[1]{$m_r = #1$}
\newcommand{\mg}[1]{$m_g = #1$}
\newcommand{\mz}[1]{$m_z = #1$}
\newcommand{\mi}[1]{$m_i = #1$}
\newcommand{\TE}[1]{$\theta_{\rm E} = #1^{\prime\prime}$}
\newcommand{\Sc}[1]{$S_{\rm color} = #1$}

\begin{document}

	\title{Confirming lensed-quasar candidates with DESI and P200 spectroscopy}
	
	\subtitle{I. 62 confirmed and likely lensed quasars}

	\author{Zizhao He
		\inst{1,2,3}
		\and
		Qihang Chen
		\inst{4,5}
		\and
		Xiaosheng Huang
		\inst{6}
		\and
		Xinzhong Er
		\inst{7}
		\and
		Christopher J. Storfer
		\inst{8}
		\and
		Nan Li
		\inst{9}
		\and
		Shen Li
		\inst{3,10}
		\and
		Qinwen Tang
		\inst{1,2}
		\and
		Fuwen Shu
		\inst{1,2}
	}
	
	\institute{Department of Physics, Nanchang University, Nanchang, 330031, China\\
		\email{zzhe@ncu.edu.cn}
		\and
		Center for Relativistic Astrophysics and High Energy Physics,
		Nanchang University, Nanchang, 330031, China
		\and
		Purple Mountain Observatory,
		Chinese Academy of Sciences, Nanjing, Jiangsu, 210023, China
		\and
		School of Physics and Astronomy, 
		Beijing Normal University, Beijing, 100875, China
		\and
		Institute for Frontier in Astronomy and Astrophysics, 
		Beijing Normal University, Beijing, 102206, China
		\and
		Department of Physics \& Astronomy, University of San Francisco, San Francisco, CA 94117, USA
		\and
		Tianjin Astrophysics Center, Tianjin Normal University, Tianjin 300387, China
		\and
		Institute for Astronomy, University of Hawaii, 2680 Woodlawn Drive, Honolulu, HI 96822, USA
		\and
		National Astronomical Observatories, Chinese Academy of Sciences, 20A Datun Road, Chaoyang District, Beijing 100012, China
		\and
		School of Astronomy and Space Sciences, University of Science and Technology of China, Hefei, Anhui, 230026, China 
	}
	\date{Received September 15, 1996; accepted March 16, 1997}
	
	
	\abstract
	{We present a sample of 15 confirmed lensed quasars based on a spectroscopic and imaging analysis of previously selected 1,724 lensed-quasar candidates from the KiDS, HSC, and DESI Legacy Imaging Surveys. The spectroscopic data come from the Dark Energy Spectroscopic Instrument Data Release 1 (DESI DR1) and targeted observations with the Double Spectrograph on the Palomar 200-inch telescope (P200/DBSP). Of the 15 systems, 13 belong to the DESI-selected sample and two to the separate P200 sample. The systems span source redshifts of $z_s\simeq1.13$--3.23, and preliminary singular isothermal ellipsoid mass models for all 15 systems yield Einstein radii of $\theta_{\rm E}\simeq0.41\arcsec$--$2.49\arcsec$. We also provide the first complete spectroscopic deflector--source redshift pairs for four lensed galaxies, with $z_d=0.409$--0.612 and $z_s=1.101$--1.567, and the first spectroscopic deflector redshifts for 16 additional systems, spanning $z_d=0.325$--0.925, within the literature and catalogues checked. In addition, we retain 47 likely lensed quasars, comprising 46 DESI-selected systems and one P200 target. These systems cannot be confirmed with the available data because of weak or undetected deflectors, insufficient or ambiguous spectroscopic evidence, discrepant image colours, possible stellar contamination, or atypical lens configurations. Our results demonstrate a systematic route to assessing compact lens candidates with general-purpose survey spectroscopy. Assuming a comparable lensing yield among the remaining candidates, further spectroscopic follow-up could confirm approximately 100--170 additional lensed quasars beyond the systems analysed in this work.}
	
	\maketitle
	%
	
	\section{Introduction}
	
	Strongly lensed quasars are invaluable astrophysical and cosmological tools that allow researchers to explore fundamental questions ranging from galaxy evolution and dark matter distribution \citep{Oguri2014, Suyu2014, Sonnenfeld2021, Vyvere2021, Hou2026} to precise measurements of cosmological parameters, including the Hubble constant \citep{Oguri2010,Suyu2014,Shajib2020,Xie2025, TDCOSMO2026Measurement}. These systems, characterized by multiple images of distant quasars produced by gravitational lensing due to an intervening massive galaxy, offer insights into the co-evolution of supermassive black holes and their host galaxies through microlensing effects \citep{Anguita2008, Sluse2011, Motta2012, Guerras2013, Braibant2014, Fian2021, Hutsemekers2021, Wu2026}.
	
	Despite their strong scientific potential, the number of confirmed lensed quasars remains modest. To increase statistical power and fully realize their scientific utility, extensive efforts have focused on identifying candidates in wide-field imaging surveys, including the Dark Energy Spectroscopic Instrument (DESI) Legacy Imaging Surveys \citep[DESI-LS;][]{Dey2019}, the Kilo-Degree Survey \citep[KiDS;][]{Kuijken2019}, and the Hyper Suprime-Cam (HSC) Subaru Strategic Program \citep[HSC-SSP;][]{Aihara2018}; see, e.g., \citep{Sheu2024,He2023,Chan2023}. These surveys have yielded thousands of candidate systems, but spectroscopic follow-up remains indispensable for secure confirmation \citep{Shu2025,He2025-AA,Dux2024,lemon2022}.
	
	Spectroscopic surveys such as the Sloan Digital Sky Survey \citep[SDSS,][]{Blanton2017} and DESI \citep{DESICollaboration2016} have significantly advanced the confirmation process \citep{Inada2003,Inada2008}. DESI, in particular, with its extensive sky coverage and deep spectroscopic capabilities, provides unprecedented opportunities for systematically verifying candidate lenses and dual-quasar (DQ) systems, thereby substantially increasing the pool of confirmed lensed quasars available for detailed astrophysical studies \citep{He2025-AA,Shu2025}.
	
\begin{figure*}[!t]
		\centering
		\includegraphics[width=\textwidth]{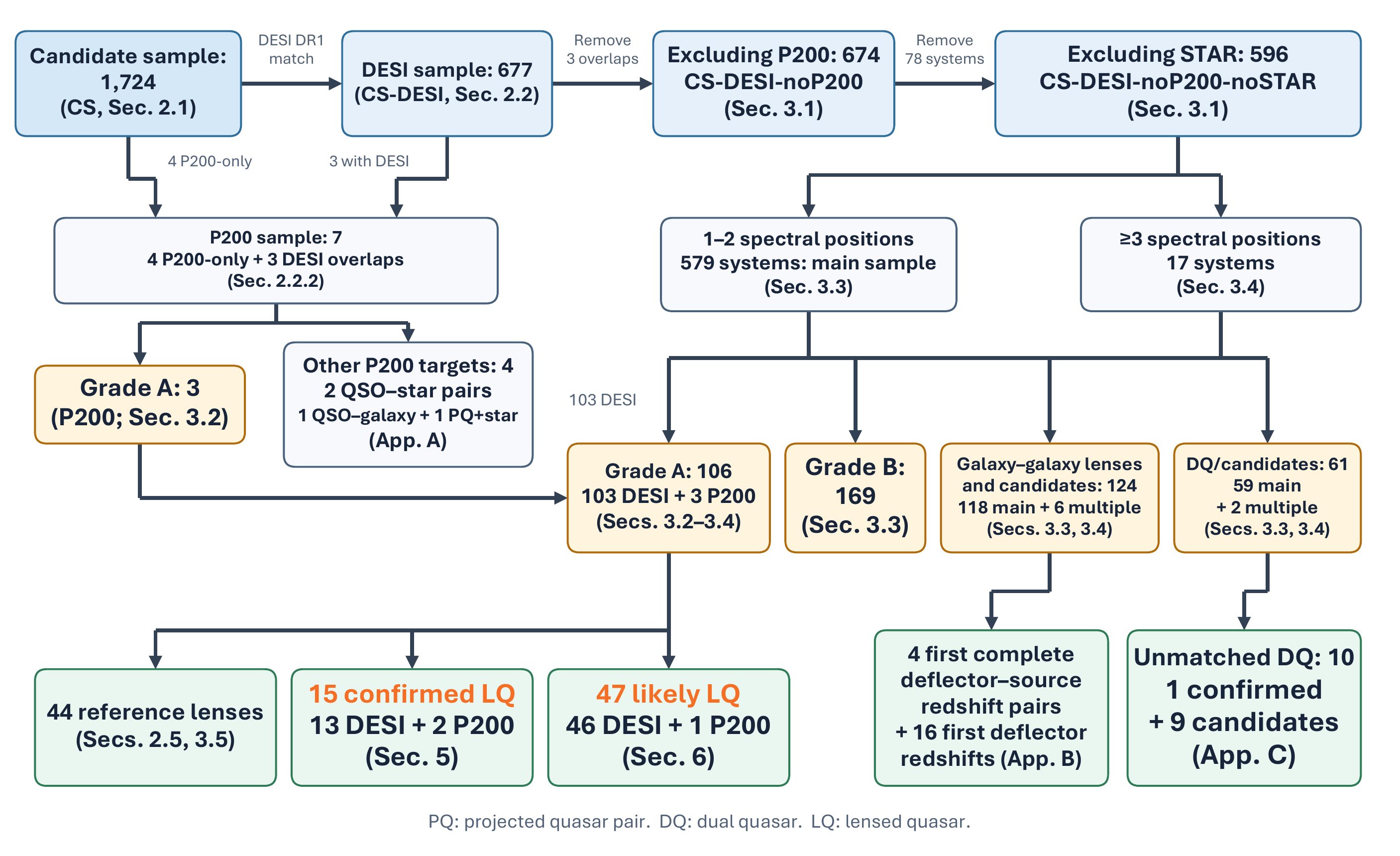}
		\caption{Sample selection and classification flow, with the DESI and P200 Grade~A systems combined before final classification. The three DESI--P200 overlaps are counted only in the P200 sample. The ten literature-unmatched DQs refer to the original cross-match described in Appendix~\ref{app:dq}.}
		\label{fig:sample_selection_overview}
	\end{figure*}

	We reassess 1724 unique lensed-quasar candidates from five imaging searches in KiDS, HSC, and DESI-LS \citep{He2025-Apj,He2023,Dawes2023,Chan2023,Andika2023}. Figure~\ref{fig:sample_selection_overview} summarises the selection. Screening the 677 DESI Data Release 1 (DR1) matches yields 59 systems for further assessment, excluding stellar contaminants, the independent reference subset, and systems analysed with the 200-inch Hale Telescope at Palomar Observatory (P200). Applying common spectroscopic, photometric, and astrometric criteria to these candidates and the three grade-A targets selected from seven observations with the P200 Double Spectrograph (DBSP) yields 15 confirmed and 47 likely lensed quasars (Sect.~\ref{sec:final_classification}). For all 15 confirmed lenses, we fit singular isothermal ellipsoid \citep[SIE;][]{Kormann1994} mass models using component positions and weakly weighted quasar flux ratios to guide follow-up observations. As by-products, we supplement the spectroscopic redshift information for previously identified galaxy--galaxy lens candidates. Within the literature and catalogues checked, we provide four newly completed deflector--source redshift pairs and the first spectroscopic deflector redshifts for 16 additional systems. We also identify 50 spectroscopically confirmed DQs, 11 DQ candidates, and 22 projected quasar pairs (PQs) from DESI. 
	
	The paper is organized as follows.
	Section~\ref{sec:data} describes the construction of the parent sample and the spectroscopic, imaging, and astrometric data used in our analysis.
	Sections~\ref{sec:pre_classification} and~\ref{sec:final_classification} describe the preliminary and final classifications, respectively.
	Sections~\ref{sec:confirmed_lq} and~\ref{sec:likely_lq} present the confirmed and likely lensed quasars, respectively.
	Section~\ref{sec:diss} presents the discussion and conclusions.
	Throughout this paper, we adopt the cosmological parameters of \citet{Planck18}.
	Unless otherwise stated, all magnitudes are in the AB system.
	\section{Data-sets and Observations}
	\label{sec:data}
	
	We combine spectroscopy and imaging to test the lens interpretation of previously selected candidates. 
	We first assemble a consolidated candidate catalogue from published imaging searches (Sect.~\ref{subsec:lensed_quasar_candidates}). 
	DESI DR1 and targeted P200/DBSP observations provide the spectral classifications and redshifts used in this assessment (Sect.~\ref{subsec:spectroscopic_data}): cross-matching the candidate catalogue with DESI identifies systems with survey spectra, while independently selected P200 targets provide long-slit coverage of multiple components. 
	Optical imaging (Sect.~\ref{subsec:optical_imaging}) allows us to assess whether the observed morphology and component configuration are consistent with gravitational lensing and provides inputs for subsequent modelling. 
	For putative quasar images without spectroscopic coverage, we assess whether they are active galactic nuclei (AGN) using Wide-field Infrared Survey Explorer (WISE) photometry and Gaia astrometry (Sect.~\ref{subsec:gaia_wise_data}). 
	To evaluate these diagnostics, we use known lenses to measure recovery by the preliminary classification and set the image-colour threshold, while reference AGN and stars define the comparison populations in optical--infrared colour space and reference galaxies constrain the deflector contribution to WISE photometry (Sect.~\ref{subsec:reference_samples}). 
	
	\subsection{Lensed quasar candidates}
	\label{subsec:lensed_quasar_candidates}
	
	We constructed a consolidated catalogue of lensed-quasar candidates by combining previously published candidate lists from wide-field imaging surveys, including KiDS \citep{He2025-Apj}, DESI-LS \citep{Dawes2023,He2023}, and HSC-SSP \citep{Chan2023,Andika2023}. 
	We homogenized the right ascension (RA), declination (Dec), and associated metadata, and resolved duplicate entries by clustering sources within $6\arcsec$ into a single candidate system. 
	This procedure resulted in a total of 1724 unique systems, hereafter referred to as the consolidated sample ({\tt CS}). 
	The consolidated catalogue is provided in \texttt{consolidated\_sources.csv}.\footnote{\label{fn:catalogue_repository}The machine-readable catalogues accompanying this paper are available at \mbox{\url{https://github.com/EigenHermit/Lqso_DESI_26}}.} 
	
	Within {\tt CS}, 671 systems are unique to \cite{He2023}, 253 to \cite{He2025-Apj}, 227 to \cite{Andika2023}, 140 to \cite{Dawes2023}, and 111 to \cite{Chan2023}. 
	In addition, 288 systems are common to both \cite{Dawes2023} and \cite{He2023}, while the remaining 34 systems appear in multiple catalogues. 
	
	\subsection{Spectroscopic data}
	\label{subsec:spectroscopic_data}
	
	\subsubsection{DESI DR1}
	\label{subsec:desi_dr1}
	
	DESI is a multi-object spectroscopic survey conducted with 5000 robotic fibres on the 4-m Mayall Telescope \citep{DESICollaboration2016}. DR1 includes observations from May 2021 to June 2022, together with reprocessed Survey Validation data \citep{DESI2025}. Its Main Survey component contains 1,553,713 unique quasars with reliable pipeline redshifts.\footnote{Official counts after primary-spectrum and redshift-quality selections: \url{https://data.desi.lbl.gov/doc/releases/dr1/}.} This is approximately $52\%$ of the three million quasars originally expected from the full five-year DESI survey \citep{Napolitano2023DESIQSO}.
	
	For this work, we used the \texttt{zall-pix-iron.fits} redshift catalogue,\footnote{\url{https://data.desi.lbl.gov/public/dr1/spectro/redux/iron/zcatalog/v1/}} which provides redshifts and galaxy (\texttt{GALAXY}), quasi-stellar object (\texttt{QSO}), or star (\texttt{STAR}) classifications from spectral fitting \citep{Anand2024}. We retrieved the corresponding DESI DR1 spectra through the SPectra Analysis and Retrievable Catalog Lab (SPARCL) service at the National Optical-Infrared Astronomy Research Laboratory (NOIRLab).\footnote{\url{https://astrosparcl.datalab.noirlab.edu/}} Matching the 1724 \texttt{CS} candidates to \texttt{TARGET\_RA} and \texttt{TARGET\_DEC} within $6\arcsec$, retaining all matches, yielded 937 spectra for 677 systems, forming \texttt{CS-DESI}. The pipeline classifies these as 511 quasars, 342 galaxies, and 84 stars. The full cross-match before these exclusions is provided in \texttt{crossmatched\_withDESI\_6''.csv}.\textsuperscript{\ref{fn:catalogue_repository}}
	
	\subsubsection{P200/DBSP observations}
	\label{subsec:p200_targets}
	
	\begin{table*}[t]
		\caption{P200/DBSP observation log of the seven targets analysed in this work. The exposure column gives the blue-arm integration time. For J0225+0513, the spectrum in Fig.~\ref{fig:p200_nonlenses} uses one 3000~s blue exposure and three 1000~s red exposures, giving 3000~s in each arm. The quoted redshift of J2225+0409 is tentative, as discussed in Sect.~\ref{sec:likely_lq}.}
		\tiny
		\centering
		\begin{tabular}{@{}lcccccclc@{}}
			\hline
			Name & RA & Dec & UT date & \makecell{Blue-arm\\exposure (s)} &
			Separation ($^{\prime\prime}$) & Redshift & Outcome & \makecell{Overlap with\\ DESI DR1} \\ \hline
			
			J0225+0513 & 36.402242 & +5.224956 & 2024-09-05 & 3000 &
			2.231 & 1.988 (A) &
			quasar + star & No \\
			
			J1652+0608 & 253.234753 & +6.143977 & 2024-09-04 & 1800 &
			2.065 & 2.427 (B) &
			quasar + star & No \\
			
			J1743+3744 & 265.776101 & +37.746769 & 2024-09-04 & 3000 &
			5.448 & 1.916 (A), 0.507 (B) &
			QSO (A) + gal (B) & Yes \\
			
			J1747+2212 & 266.771095 & +22.204578 & 2024-09-04 & 1500 &
			4.625 & 1.012 (B), 0.3575 (C) &
			projected QSO & Yes \\
			
			J1753+7333 & 268.484581 & +73.564684 & 2024-09-05 & 3000 &
			1.948 & 2.215 (A+B) &
			lensed quasar & No \\
			
			J1800+5305 & 270.014805 & +53.083527 & 2024-09-04 & 3000 &
			2.251 & 3.2293 (A), 3.2293 (B) &
			lensed quasar & Yes \\
			
			J2225+0409 & 336.484656 & +4.160525 & 2024-09-04 & 3600 &
			1.599 & 1.534 (A+B) &
			likely lensed quasar & No \\
			
			\hline
		\end{tabular}
		\label{tab:p200}
	\end{table*}
	P200 has a 5.1-m aperture and is equipped with DBSP \citep{Oke1982}, a moderate-resolution optical spectrograph whose two arms record blue and red spectra simultaneously using a selectable dichroic. We selected candidates for P200/DBSP follow-up before the public release of DESI DR1, primarily on the basis of their observability from Palomar and the presence of a visible central galaxy that could act as a lens. Of the ten candidates considered for this analysis, we retained seven. We excluded the other three because their spectra were of insufficient quality for a reliable classification. 
	
	The observations were obtained on 2024 September 4 and 5 UT (principal investigator Zizhao He). Of the seven targets in Table~\ref{tab:p200}, five were observed on September 4, and J0225+0513 and J1753+7333 were observed on September 5. After the release of DESI DR1, our cross-match identified two spectra for J1743+3744, classified as a quasar and a galaxy by the pipeline; two quasar spectra for J1747+2212; and one quasar spectrum for J1800+5305. The remaining four targets have no DESI DR1 counterparts in the adopted sample.
	
	For each target, we aligned the slit position angle to pass through both point sources (PSs) and centered the pair so that both spectra could be recorded simultaneously. For our observations, we adopted the D-55 dichroic to divide the light into the blue and red channels. The blue arm employed a 300 lines mm$^{-1}$ grating blazed at 3990\,\AA\ (2.108\,\AA\ pixel$^{-1}$), while the red arm used a 316 lines mm$^{-1}$ grating blazed at 7150\,\AA\ (1.535\,\AA\ pixel$^{-1}$). The typical seeing was $0.9^{\prime\prime}$--$1.5^{\prime\prime}$, and we used a slit width of $1.5^{\prime\prime}$. Note that the red-side detector contains bad pixels over the wavelength range $\sim$5800--6300\,\AA; consequently, the DBSP data are unusable in this interval.
	
	We report the preliminary and final classification results for these seven P200 systems in Sect.~\ref{subsec:p200_preliminary} and Sect.~\ref{sec:final_classification}, respectively.

	\subsection{Optical imaging data}
	\label{subsec:optical_imaging}
	
	We used optical images from DESI-LS \citep{Dey2019} and HSC-SSP \citep{Aihara2018} to examine the candidate configurations and measure component photometry. DESI-LS combines imaging from the Dark Energy Camera (DECam) Legacy Survey, the Beijing--Arizona Sky Survey, and the Mayall $z$-band Legacy Survey. We write $m_z$ for the apparent $z$-band magnitude, $z_s$ and $z_d$ for source and deflector redshifts, respectively, and $z_{\rm QSO}$ for the redshift of an individual quasar whose lensing status is unconfirmed. The principal DESI-LS optical bands are $grz$, with additional $i$-band imaging available in parts of the footprint. The DESI-LS images used here comprise Data Release 9 (DR9) cutouts distributed with \citet{He2023} and supplementary cutouts retrieved from Data Release 10 (DR10); the standard coadded images have a pixel scale of $0.262\arcsec$.
	
	HSC-SSP is a multiband imaging survey with HSC on the 8.2-m Subaru Telescope, with five broad-band filters ($grizy$) and a native pixel scale of $0.168\arcsec$ \citep{Aihara2018}. We used the available HSC cutouts for systems within its footprint, including the $griz$ images used in the component decompositions.
	
	We inspected the images for lensing features, such as arcs or multiple images, to guide the preliminary classification. For Grade~A systems, we performed multiband photometric decomposition using the calibrated Flexible Image Transport System (FITS) images, as described in Sect.~\ref{subsec:psf_subtraction}.
	
	\subsection{Gaia and WISE data}
	\label{subsec:gaia_wise_data}
	
	\textit{Gaia} Data Release 3 (DR3) provides source positions, parallaxes, proper motions, and astrometric-quality indicators \citep{GaiaDR3overview}. We used this astrometric information to help assess whether the compact components are consistent with AGN and to identify possible stellar contaminants.
	
	We also used the unWISE NEO6 $W1$ and $W2$ images \citep{Lang2014unWISE,Meisner2017unWISE}, together with HSC or DESI-LS $grz$ photometry, to assess the AGN nature of the components through their optical--infrared colours. The colour and astrometric criteria are described in Sect.~\ref{subsec:colour_astrometry}.

	\subsection{Reference samples}
	\label{subsec:reference_samples}
	
	We use reference lenses to assess the preliminary classification (Sect. \ref{sec:pre_classification}). Reference AGN, stars, and galaxies support colour–colour checks in the final classification (Sect. \ref{sec:final_classification}), providing an additional test for possible stellar contamination in systems with a detected foreground galaxy and similarly coloured putative images, but without spectra for all images. Additional galaxy samples constrain the deflector contribution to the blended WISE flux.
	
	\subsubsection{Reference lensed quasars}
	\label{subsubsec:reference_lensed_quasars}
	
	We constructed the reference lensed-quasar sample by cross-matching {\tt CS-DESI} with the confirmed lensed quasars listed in the \textit{Gaia} Gravitational Lenses (GraL) catalogue \citep{Ducourant2026}, using a matching radius of $6\arcsec$,; this yielded 50 matched systems. We add J0329-0208 and J0907+6224, which were confirmed by \citet{Lemon2023} but are absent from the adopted GraL table, and we remove the systems from \citet{Shu2025} from this reference subset. We intend to re-analyse these systems because \citet{Shu2025} reported only the cross-matching results and did not conduct detailed analyses such as re-photometry or lens modelling (see the notes on J1215-0058 and J1220+0112 in Sect.~\ref{subsec:confirmed_individual_notes}). This leaves 51 independent reference systems.
	
	All reference systems were established as gravitationally lensed quasars independently of the present work, as documented in the GraL catalogue and in \citet{Lemon2023}. We use this sample to assess recovery by the preliminary DESI classification (Sect.~\ref{subsec:desi_classification_results}) and to set the image-colour threshold for the final classification (Sect.~\ref{subsec:psf_subtraction}).
	
	\subsubsection{Reference AGN}
	\label{subsubsec:reference_agn}
	
	The reference AGN sample comprises 7,015 spectroscopically identified QSOs/AGN from DESI DR1 \citep{DESI2025}. We joined the primary spectra to the associated DESI-LS photometry through \texttt{TARGETID}, requiring \texttt{SPECTYPE=QSO}, \texttt{ZWARN=0}, \texttt{COADD\_FIBERSTATUS=0}, a positive redshift, and positive fluxes with signal-to-noise ratios (S/N) above five in each of the five bands $g$, $r$, $z$, $W1$, and $W2$. We used the catalogue random-identifier subset below 0.5 on its 0--100 scale and retained one entry per target. No additional colour cut was applied. This sample provides the QSO/AGN comparison population in the colour diagrams.
	
	For the WISE-flux predictions in Sect.~\ref{subsec:colour_astrometry}, we separately queried the online DESI DR1 spectroscopic and DESI-LS photometric tables through the NOIRLab Astro Data Lab. We selected primary, securely classified quasars with the same spectroscopic quality cuts, $grz$ S/N above five, and valid $W1/W2$ flux measurements. For each spectroscopic component, we required $|\Delta z_{\rm QSO}|<0.2$, $|\Delta m_r|<0.5$~mag, and $\sqrt{[\Delta(m_g-m_r)]^2+[\Delta(m_r-m_z)]^2}<0.3$~mag. We removed duplicate photometric sources and matches within $10\arcsec$ of the target. No WISE S/N cut was imposed on these matched samples. Their paired $W1/W2$ fluxes, scaled to the target's $r$-band flux, provide an empirical prediction and its object-to-object scatter. target attenuation is restored to the WISE predictions. The other matched samples use observed photometry.
	
	\subsubsection{Reference stars}
	\label{subsubsec:reference_stars}
	
	For a descriptive comparison with ordinary stars, we used a shared regional reference of 10,000 sources selected using significant Gaia astrometry and clean DESI-LS DR9 photometry, without a colour preselection. The reference requires $16<G<20.5$, a positive parallax or the absolute value of either proper-motion component exceeding five times its formal uncertainty, and S/N above five in $g$, $r$, $z$, $W1$, and $W2$. This sample provides the empirical stellar comparison used in Sect.~\ref{subsec:colour_astrometry}.
	
	\subsubsection{Reference galaxies}
	\label{subsubsec:reference_galaxies}
	
    We constructed a reference sample of elliptical galaxies from the combined South Galactic Cap $u$-band Sky Survey (SCUSS), SDSS, and WISE catalogue of \citet{Gao2018Photometry}. The catalogue provides Galactic-extinction-corrected model magnitudes on the AB system, with $W1$ and $W2$ magnitudes derived from unWISE forced photometry. We cross-matched this catalogue with Galaxy Zoo~1 \citep[GZ1,][]{Lintott2011GalaxyZoo} using a matching radius of $0\farcs5$ and selected galaxies with a clean elliptical classification (\texttt{GZ1\_fE=1}). We required finite magnitudes and uncertainties in all five bands, with uncertainties below 0.1~mag in each optical band ($g$, $r$, and $z$) and below 0.2~mag in each WISE band ($W1$ and $W2$). These criteria yielded 2,731 elliptical galaxies. For disc-like deflectors, we applied the same photometric cuts to galaxies with a clean spiral classification (\texttt{GZ1\_fS=1}), yielding a reference sample of 6,898 spiral galaxies.
	
	For the colour diagrams in Figs.~\ref{fig:confirmed_colours} and~\ref{fig:transferred_diagnostics}, we matched each deflector to galaxies in the corresponding reference sample using its deblended, Galactic-extinction-corrected $grz$ magnitudes. We fitted a single magnitude offset to each reference galaxy across the three bands, so the match depended on the two independent colours $m_g-m_r$ and $m_r-m_z$. We ranked the galaxies by the resulting $\chi^2$, adding the deflector and reference magnitude errors and a 0.05~mag term in quadrature in each band. In each Monte Carlo realisation we perturbed the deflector photometry, repeated the ranking, and drew uniformly from the 100 closest matches without a fixed colour or redshift cut. We applied the fitted offset to both WISE magnitudes of the selected galaxy and restored Galactic attenuation along the target sightline to predict the deflector's paired $W1$ and $W2$ fluxes.
	
	\section{Preliminary classification}
	\label{sec:pre_classification}

	We used the spectroscopy and imaging assembled in Sect.~\ref{sec:data} to select promising lensed-quasar candidates and to identify alternative interpretations. Firstly, we define some terms used in the classification (Sect. \ref{subsec:definition_used_in_pre_class}). We then grouped the DESI sample by the number of independent spectroscopic positions (Sect.~\ref{subsec:sample_selection}) after separating out the P200 targets and excluding systems with DESI stellar classifications. The P200 targets were examined separately (Sect.~\ref{subsec:p200_preliminary}). We applied a common decision tree to systems with one or two positions (the main sample; Sect.~\ref{subsec:one_two_positions}), assessed systems with more positions individually (the multiple-position sample; Sect.~\ref{subsec:multi_positions}), and used the reference lenses as a check on recovery by the DESI selection (Sect.~\ref{subsec:desi_classification_results}).
	
	After this entire process, we identified 103 grade-A candidates among the 596 DESI-selected systems: 101 from the main sample and two from the multiple-position sample. Excluding the 44 grade-A reference lenses leaves 59 DESI candidates for further assessment. Together with the three retained P200 targets, these form the 62-system sample used for the photometric analysis and final classification in Sect.~\ref{sec:final_classification}.
	
	\subsection{Some definitions}
	\label{subsec:definition_used_in_pre_class}
	\paragraph{Grade A/B.} For lens candidates, grade A required a plausible configuration of putative quasar images and similar colours among the multiple quasar images. For systems with relatively large separations, an evident foreground galaxy was also required; for small-separation systems, a visible deflector was not required prior to PSF subtraction. Grade B indicated less convincing configurations, for example, large color differences or a configuration inconsistent with commonly observed lensed-quasar configurations. This grading was qualitative and did not require detailed lens modelling. Grade-A candidates proceed to the final assessment in Sect.~\ref{sec:final_classification}, whereas grade-B candidates are left for future study.
	
	\paragraph{Dual quasars and projected quasar pairs.} 
	Dual quasar (DQ) systems contain two components with quasar spectra that differ in their emission-line profiles or continuum shapes. We require their radial-velocity difference to satisfy \(|\Delta v| \leq 2000\,\mathrm{km\,s^{-1}}\), calculated as
	\begin{equation}
		|\Delta v| = \frac{c|z_2-z_1|}{1+(z_1+z_2)/2},
	\end{equation}
	where \(z_1\) and \(z_2\) are the redshifts of the two quasars, and \(c\) is the speed of light in \(\mathrm{km\,s^{-1}}\). We classify two spectroscopically identified quasars with \(|\Delta v| > 2000\,\mathrm{km\,s^{-1}}\) as a projected quasar pair (PQ).
	
	\paragraph{Dual quasar candidates.}
	We consider pairs of point-like sources in which only one component is spectroscopically identified as a quasar. We retain pairs with no visible deflector despite a separation large enough that one would be detectable if the pair were lensed. In our sample, the components of all such pairs have similar optical colours, so we list them as DQ candidates and report no PQ candidates. Spectroscopy of the companion is needed to confirm its quasar nature and measure \(|\Delta v|\) to distinguish a DQ from a PQ.
	
	\paragraph{Galaxy--galaxy systems and galaxy--quasar systems.}
	We classify a system as galaxy--galaxy (G+G) when separate spectra identify both components as galaxies, and as galaxy--quasar (G+Q) when one component is a galaxy and the other is a quasar. No restriction is imposed on the radial-velocity difference between the components; these classifications therefore do not imply a physical association.
	
	\paragraph{Galaxy--galaxy candidates and galaxy--quasar candidates.}
	When only one component is spectroscopically identified, we classify the system provisionally using the morphology of the other component. A spectroscopically identified galaxy with an extended companion is classified as a G+G candidate, whereas one with a point-like companion is classified as a G+Q candidate. A spectroscopically identified quasar with an extended companion is also classified as a G+Q candidate. In principle, a quasar with a point-like companion could likewise be labelled a Q+Q candidate. In practice, the imaging searches from which our sample was drawn favour pairs with similar colours across multiple bands. We therefore assess such pairs as lensed-quasar or dual-quasar candidates, depending on the image configuration and the evidence for a deflector. In all cases, spectroscopy of the companion is needed to confirm its nature.
	
	\subsection{Sample selection}
	\label{subsec:sample_selection}
	
	After cross-matching, {\tt CS-DESI} contains 677 systems with 937 spectra (Sect.~\ref{subsec:desi_dr1}). To keep the DESI selection separate from the targeted P200 follow-up, we assigned all three overlapping systems, J1743+3744, J1747+2212, and J1800+5305, to the P200 sample (Sect.~\ref{subsec:p200_targets}). Removing these systems and their five DESI spectra left 932 spectra associated with 674 systems, denoted {\tt CS-DESI-noP200}. We then excluded every system containing at least one spectrum classified as a star by the DESI pipeline. This produced {\tt CS-DESI-noP200-noSTAR}, containing 596 systems with 817 spectra: 486 quasar spectra and 331 galaxy spectra.
	
	We next grouped repeated observations to distinguish the number of spectra from the number of independently sampled positions. Before grouping, 411 systems had one spectrum, 155 had two, and 30 had more than two. Using \texttt{MEAN\_FIBER\_RA} and \texttt{MEAN\_FIBER\_DEC}, we transitively linked fibre centres separated by less than $0.5\arcsec$ within each system and counted each connected group as one spectroscopic position. All original spectra were retained without coaddition. The resulting counts were 419 systems with one position, 160 with two, 16 with three, and one with four.
	
	The 579 systems with one or two positions form the main sample, {\tt CS-DESI-noP200-noSTAR-VI}, classified in Sect.~\ref{subsec:one_two_positions}. The remaining 17 systems form the multiple-position sample discussed in Sect.~\ref{subsec:multi_positions}.
	
	\subsection{P200 targets}
	\label{subsec:p200_preliminary}
	
	We observed seven systems with P200/DBSP (Table~\ref{tab:p200}). Four have spectra from P200 alone, and three also have DESI spectra.
	
	We assign grade A to three of the seven systems and retain them for the final assessment in Sect.~\ref{sec:final_classification}. We classify the other four as non-lensed systems and exclude them from that assessment. Their spectra are presented in Appendix~\ref{app:P200_other}.
	
	Of the four systems with P200 spectra alone, J0225+0513 and J1652+0608 show clear evidence of stellar contamination. J1753+7333 shows characteristic quasar spectral features. Subtracting the two point sources reveals a prominent galaxy between them, as expected for a double-image lens. On the basis of the spectroscopy and image configuration, we assign J1753+7333 grade A following the criteria in Sect.~\ref{subsec:definition_used_in_pre_class}. In J2225+0409, two point sources flank a prominent galaxy. This configuration, together with the broad emission lines in the spectra, supports a grade-A classification.
	
	The three systems with both P200 and DESI spectra are J1743+3744, J1747+2212, and J1800+5305.
	
	J1743+3744 resembles a double-image lens, with a prominent galaxy between the two apparent images. Component A shows a typical AGN spectrum at \(z_A \simeq 1.916\). Both the P200 and DESI spectra of component B lack broad emission lines and favour its identification as a galaxy at \(z_B \simeq 0.507\). We therefore classify J1743+3744 as a G+Q alignment. Absorption features in the spectrum of the background quasar A could provide a probe of the circumgalactic gas around the foreground galaxies B and C.
	
	In J1747+2212, P200 resolves all three components spectroscopically, and DESI also provides spectra of B and C. Both datasets identify B and C as quasars at \(z_B \simeq 1.012\) and \(z_C \simeq 0.357\), respectively. Their redshift difference establishes them as a projected quasar pair (PQ). The P200 spectrum identifies component A as a foreground star.
	
	In J1800+5305, P200 resolves components A and B spectroscopically, and DESI also provides a spectrum of B. Both components show characteristic quasar spectra with three prominent broad emission lines, which we tentatively identify as Ly\(\alpha\), C\,\textsc{iv}, and C\,\textsc{iii}]. A possible foreground galaxy is visible between the two quasar images. Together, the spectra and image configuration support a grade-A classification.

	\subsection{Systems with one or two spectroscopic positions}
	\label{subsec:one_two_positions}
	
\begin{figure*}[!t]
		\centering
		\includegraphics[width=1\linewidth]{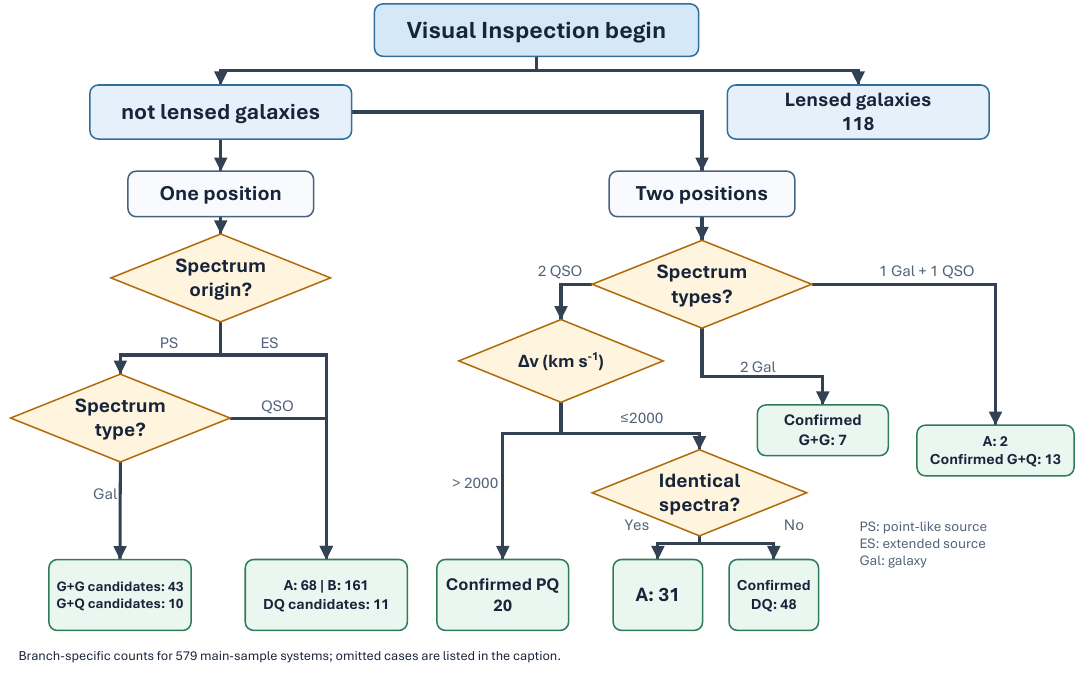}
		\caption{Preliminary classification workflow (Sect.~\ref{subsec:one_two_positions}). Each classification outcome is labelled with the number of systems in that branch. The remaining 47 systems comprise eight two-position grade-B systems, one single-position confirmed G+G system, five spectral failures, and 33 unclassified systems. PS and ES follow the definitions in the text. Gal denotes a galaxy. See footnote~\ref{fn:vi_images} for the visual-inspection images.}
		\label{fig:reclassification_flowchart}
	\end{figure*}

	We followed the classification workflow shown in Fig.~\ref{fig:reclassification_flowchart}. At the start of the assessment, we attempted to identify strong-lensing features (e.g., multiple images of the quasar, an arc, or a ring) and any possible foreground galaxy. If these could not be identified, we assigned the label \emph{Unclassified}.\footnote{\label{fn:vi_images}In some systems, the DESI fibre falls on a feature unrelated to strong lensing. In others, the imaging is consistent with a single quasar and shows no clear evidence of multiple images. The visual-inspection images, including those of unclassified systems, are grouped by sample and classification for release with this paper at \mbox{\url{https://github.com/EigenHermit/Lqso_DESI_26}}.} If they could be identified, the first step was to determine whether the system is a lens galaxy. Systems showing arc-like features or non-quasar spectra of fibre-targeted background sources were classified as lensed galaxies; all others followed the branches described below.
	
	\subsubsection{Systems with one spectroscopic position}
	\label{subsubsec:single_position}
	
	With only one independently sampled position, the classification depended on the spectral identity and morphology of the observed component:
	\begin{enumerate}
		\item[(i)] If the spectrum sampled an ES identified as the putative lens galaxy, the system entered the A/B branch defined above.
		\item[(ii)] If the spectrum identified a PS as a quasar, the system also entered the A/B branch, with one exception. When the source redshift and image separation implied, under the lensing hypothesis, a deflector bright enough to be detected in the available DESI-LS or HSC imaging, but no such galaxy was visible, we classified the system as a DQ candidate.
		\item[(iii)] If a PS had a galaxy spectrum, the morphology of the companion determined the candidate classification: an extended companion led to a G+G candidate, and a point-like companion to a G+Q candidate.
	\end{enumerate}
	
	\subsubsection{Systems with two spectroscopic positions}
	\label{subsubsec:two_positions}
	
	For two independently sampled positions, we distinguished three combinations of spectral types:
	\begin{enumerate}
		\item[(i)] \textit{Two galaxy spectra.} Both components were identified as galaxies, giving a spectroscopically confirmed G+G classification.
		
		\item[(ii)] \textit{One galaxy spectrum and one quasar spectrum.} The classification depended on which image component produced the galaxy spectrum. If the two putative quasar images were identified as a galaxy and a quasar, the system was classified as G+Q. If the galaxy spectrum instead sampled a separate putative deflector, with the quasar spectrum sampling a possible lensed image, we assigned grade A when the overall configuration supported lensing.
		
		\item[(iii)] \textit{Two quasar spectra.} We first compared the redshifts through the line-of-sight velocity difference, $\Delta v$. Pairs with $\Delta v > 2000~\mathrm{km\,s^{-1}}$ were classified as PQs. For $\Delta v \leq 2000~\mathrm{km\,s^{-1}}$, we compared the continuum shapes and emission-line profiles and examined the positions of the quasar components and any visible galaxy. Closely matching spectra and a configuration consistent with lensing led to grade A. Discrepant spectral shapes despite the small velocity difference led to the spectroscopically confirmed DQ category.
	\end{enumerate}
	
	\subsection{Systems with multiple spectroscopic positions}
	\label{subsec:multi_positions}
	
	The 17 systems with more than two independent spectroscopic positions required individual assessment to distinguish neighbouring sources from multiple images or different regions of the same object. We matched each spectrum to the imaging using its fibre position before applying the classification criteria in Sect.~\ref{subsec:one_two_positions}.
	
	In six systems, an additional spectrum samples a neighbouring galaxy or quasar outside the candidate configuration and does not affect its classification. Two of these systems, J0422-0447 and J0422-1427 (Fig.~\ref{fig:spec_image}), received grade A and proceed to the final assessment in Sect.~\ref{sec:final_classification}.
	
	The sample also contains six galaxy--galaxy lenses with fibres sampling different regions of the lens system. For example, in J0224-0336, three independent positions sample the foreground deflector and two images of the background source (Fig.~\ref{fig:lgal_new_pairs} in Appendix~\ref{app:lensed_gal}). In two other systems, multiple fibres sample a single galaxy. 
	
	\subsection{Reference-lens recovery and classification results}
	\label{subsec:desi_classification_results}
	Table~\ref{tab:classification_comparison} presents the preliminary classifications for the two DESI samples and the reference lenses. The reference sample provides a test of how effectively our selection recovers known lenses. Of its 51 systems, 44 were assigned grade A (\(86.3\%\)), all in the main sample. Among the five remaining reference lenses in that sample, J0911+0418 and J1411-0102 were classified as lensed galaxies, J0734+1915 had failed spectroscopy, and J1255+0737 and J1332+0347 remained unclassified. The two reference lenses in the multiple-position sample, J1012-0307 and J1527+0141, were classified as PQ and G+Q, respectively.
	
	The four misclassifications resulted from limitations in spectral quality and errors in associating spectra with individual components. In J0911+0418 and J1411-0102, the pipeline classified the low-S/N DESI spectra centred on the quasar images as galaxy spectra. These errors were not recognised during visual inspection, resulting in lensed-galaxy classifications. In \mbox{J1012-0307}, three fibres sampled the two lensed quasar images at \(z_s\simeq2.7\) and an unrelated neighbouring quasar at \(z_{\rm QSO}\simeq1.6\). The redshift mismatch between the lensed images and the unrelated quasar led to the erroneous PQ classification. In J1527+0141, three fibres sampled the lens galaxy, the brighter quasar image, and a position between them. The fainter quasar image was overlooked during the imaging inspection, resulting in the G+Q classification.
	
	Overall, the reference-sample results support the effectiveness of our preliminary classification, with an estimated completeness of approximately \(86\%\) for grade-A selection within this sample.
	
	Together, the main and multiple-position samples contain 103 grade-A systems, 169 grade-B systems, and 124 lensed galaxies. The remaining categories comprise galaxy and quasar pairs, five systems with spectral failures, and 34 unclassified systems.
	
	Including the three grade-A candidates from the P200 observations (Sect.~\ref{subsec:p200_targets}) brings the total number of grade-A systems to 106. Excluding the 44 known lenses in the reference sample leaves 62 grade-A candidates. We focus on these 62 systems as the most promising candidates for newly confirmed lensed quasars and assess them further in Sect.~\ref{sec:final_classification}.
	
	\begin{table*}[!t]
		\centering
		\caption{Preliminary classifications of the DESI samples and reference lenses.}
		\label{tab:classification_comparison}
		\begin{threeparttable}
			\small
			\setlength{\tabcolsep}{3pt}
			\renewcommand{\arraystretch}{1.08}
			\begin{tabular}{@{}lrrrrrr@{}}
				\toprule
				& \multicolumn{2}{c}{Main sample}
				& \multicolumn{2}{c}{Multiple-position sample}
				& \multicolumn{2}{c}{Reference lenses} \\
				& \multicolumn{2}{c}{($N=579$)}
				& \multicolumn{2}{c}{($N=17$)}
				& \multicolumn{2}{c}{($N=51$)} \\
				\cmidrule(lr){2-3}\cmidrule(lr){4-5}\cmidrule(l){6-7}
				Classification & Number & Fraction (\%) & Number & Fraction (\%) & Number & Fraction (\%) \\
				\midrule
				Grade A & 101 & 17.4 & 2 & 11.8 & 44 & 86.3 \\
				Grade B & 169 & 29.2 & 0 & 0.0 & 0 & 0.0 \\
				Lensed galaxies & 118 & 20.4 & 6 & 35.3 & 2 & 3.9 \\
				G+G (candidate) & 8(43) & 8.8 & 1 & 5.9 & 0 & 0.0 \\
				G+Q (candidate) & 13(10) & 4.0 & 1 & 5.9 & 1 & 2.0 \\
				DQ (candidate) & 48(11) & 10.2 & 0 & 0.0 & 0 & 0.0 \\
				DQ + galaxy (candidate) & 0 & 0.0 & 2 & 11.8 & 0 & 0.0 \\
				PQ & 20 & 3.5 & 2 & 11.8 & 1 & 2.0 \\
				Single galaxy & 0 & 0.0 & 2 & 11.8 & 0 & 0.0 \\
				Spectral failures & 5 & 0.9 & 0 & 0.0 & 1 & 2.0 \\
				Unclassified & 33 & 5.7 & 1 & 5.9 & 2 & 3.9 \\
				\midrule
				Total                       & 579 & 100.0 & 17 & 100.0 & 51 & 100.0 \\
				\bottomrule
			\end{tabular}
			\begin{tablenotes}[flushleft]
				\footnotesize
				\item Counts for the pair categories are $N_{\rm confirmed}(N_{\rm candidate})$, with zero candidate counts omitted; fractions include both and use the sample sizes in the column headings.
				\item The reference lenses are a subset of the two disjoint DESI samples and must not be added to their totals.
			\end{tablenotes}
	\end{threeparttable}\end{table*}
	
	\section{Final classification of DESI and P200 targets}
	\label{sec:final_classification}
	
	For the 62 candidates retained in Sect.~\ref{sec:pre_classification}, we combine the spectroscopic evidence with photometric and astrometric diagnostics to assess the lens interpretation. Multiband image decomposition provides component photometry and diagnostics of image-colour similarity, foreground-galaxy detection, and fit quality (Sect.~\ref{subsec:psf_subtraction}). Optical--infrared colours and Gaia proper motions help identify possible stellar contaminants (Sect.~\ref{subsec:colour_astrometry}). We use these results to define the confirmed and likely classes (Sect.~\ref{subsec:lq_classification_criteria}), then describe source-redshift estimation and mass modelling for the confirmed systems (Sect.~\ref{subsec:parameter_estimation}). The resulting samples are presented in Sects.~\ref{sec:confirmed_lq} and~\ref{sec:likely_lq}.
	
	\subsection{Photometry under the lensing hypothesis}
	\label{subsec:psf_subtraction}
	
	We modelled the compact sources with PSFs. When an extended foreground component was present, we added a PSF-convolved elliptical S\'ersic profile \citep[e.g.][]{He2025-AA}. Under the lensing hypothesis, these components represent the quasar images and the deflector, respectively. We used the survey PSF models for these fits. The image model in band $b$ is
	\begin{equation}
		I_{{\rm model},b}=\sum_{i=1}^{N_{\rm img}}I_{P,i,b}+I_{G,b}+B_b,
		\label{eq:photometry_model}
	\end{equation}
	where $N_{\rm img}$ is the number of compact components and $I_{P,i,b}$ is the PSF model of component $i$, scaled by its flux. The term $I_{G,b}$ describes the PSF-convolved galaxy light; for systems without a foreground galaxy, we omitted this component ($I_{G,b}=0$). The local background $B_b$ consists of a constant and two linear gradients.
	
	The Sérsic profile \(I_{G,b}\) is specified by its centre \((x_{\rm G},y_{\rm G})\), index \(n_s\), semi-major-axis half-light radius \(r_s\), axis ratio \(q_s=b/a\), position angle \(\phi_s\), and total flux in band \(b\). We fixed \(n_s=4\) by default and adopted \(n_s=1\) in all fitted bands for galaxies with a clear disk-like morphology. We measured \(\phi_s\) counterclockwise from the \(+Y\) axis of the reference image, modulo \(180^\circ\). In the joint fits to the HSC \(griz\) or DESI-LS \(grz\) images, we fitted a single half-light radius \(r_s\) for each galaxy using all available bands. The galaxy axis ratio and position angle, as well as the sky position of each component, were also shared across bands. For each configuration, we fitted the component fluxes and local background independently in each band.
	
	The best-fitting geometry minimized the noise-weighted sum of squared residuals. Galaxy magnitudes were derived from the S\'ersic total flux extrapolated to infinite radius. In Tables~\ref{tab:lens_gal} and~\ref{tab:psfs}, \texttt{nodata} identifies an unavailable band, whereas \texttt{below detection limit} denotes a fitted component without a finite magnitude. 
	
\begin{figure}
		\centering
		\includegraphics[width=0.9\linewidth]{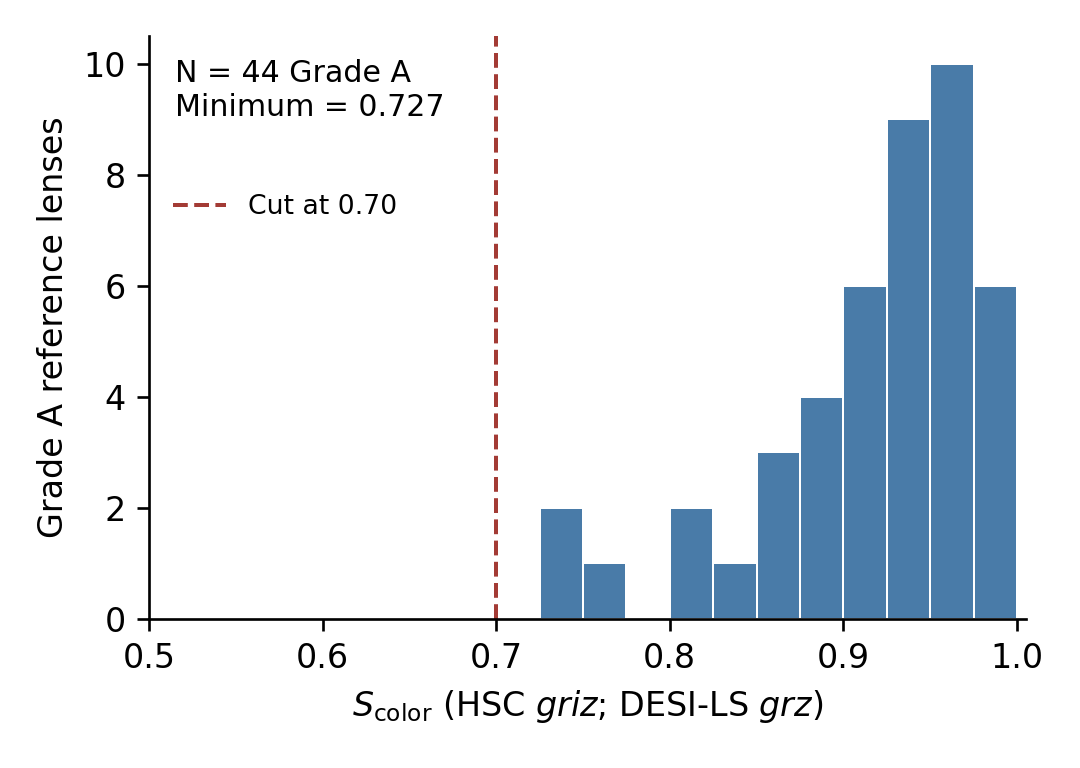}
		\caption{Distribution of $S_{\rm color}$ for all 44 Grade~A reference lenses. The scores use HSC $griz$ or DESI Legacy Survey $grz$ photometry. The dashed line marks the adopted threshold, $S_{\rm color}=0.7$.}
		\label{fig:scolor_hist}
	\end{figure}

	From the deblended optical fluxes, following the methodology used in \cite{He2023}, we quantified image-colour similarity as
	\begin{equation}
		S_{\rm colour}
		=
		\max\left[0,\,
		1-\frac{1}{N_{\rm col}}
		\sum_{k=1}^{N_{\rm col}}\sigma_k\right],
		\label{eq:scolor}
	\end{equation}
	where $\sigma_k$ is the population standard deviation of colour $k$ across the compact components and $N_{\rm col}$ counts all pairwise colours formed from the adopted bands, requiring finite magnitudes for every compact component. Thus, HSC $griz$ gives $N_{\rm col}=6$, whereas DESI-LS $grz$ gives $N_{\rm col}=3$. For a double, $\sigma_k=|C_{A,k}-C_{B,k}|/2$. The score approaches unity for similar image colours and is not weighted by photometric uncertainties. The fitted galaxy light is excluded. We also excluded WISE $W1/W2$ measurements because tests on reference lenses showed that uncertainties in the deblended component fluxes prevented a reliable determination of $S_{\rm color}$. Figure~\ref{fig:scolor_hist} shows the $S_{\rm color}$ distribution of 44 reference lenses, which motivates our adopted threshold of $S_{\rm color}\geq0.7$.
	
	The fitted photometry also quantifies the significance of the foreground-galaxy detection, while the image residuals allow us to assess the fit quality. We defined the galaxy S/N in band $b$ as
	\begin{equation}
		\nu_{{\rm G},b}=\frac{F_{{\rm G},b}}{\sigma_{F_{{\rm G},b}}},
	\end{equation}
	where $F_{{\rm G},b}$ is the fitted S\'ersic flux and $\sigma_{F_{{\rm G},b}}$ is its conditional uncertainty from the linear flux fit. For confirmed systems, we required
	\begin{equation}
		\max_b\nu_{{\rm G},b}\geq14,
	\end{equation}
	in at least one valid adopted band, together with an extended component whose morphology and residuals support the lens geometry. This S/N threshold retains 97.1\% of the reference lensed quasars with fitted foreground-galaxy components; reference systems without galaxy fits are excluded from this calculation. Table~\ref{tab:lens_gal} reports the maximum S/N and its corresponding band.
	
	Using the image model defined in Eq.~\ref{eq:photometry_model}, we assessed the fit quality through
	\begin{equation}
		\chi^2_{\nu,\rm img}=
		\frac{\displaystyle\sum_b\sum_{j\in\mathcal{V}_b}
			\left[(I_{{\rm data},bj}-I_{{\rm model},bj})w_{bj}\right]^2}
		{N_{\rm valid}-p},
		\label{eq:image_fit_diagnostic}
	\end{equation}
	where $I_{{\rm data},bj}$ and $I_{{\rm model},bj}$ are the observed and model intensities at pixel $j$ in band $b$, and $w_{bj}$ is the inverse-noise weight. The sum includes all fitted optical bands, retaining the original weights and each band's native pixel sampling. The set $\mathcal{V}_b$ contains valid, unmasked pixels within the common $5.5\arcsec\times5.5\arcsec$ cutout, or $10\arcsec\times10\arcsec$ for J1215-0058. Here $N_{\rm valid}$ is the total number of included pixels, and $p$ counts the fitted parameters, including registration offsets, with shared parameters counted once. Noise estimates come from inverse-variance maps where available, or from the robust median absolute deviation otherwise. The $\chi^2_{\nu,\rm img}$ column Table~\ref{tab:tab1} reports this local fit diagnostic.
	
	\subsection{Colour and astrometry examination}
	\label{subsec:colour_astrometry}
	We applied two complementary diagnostics to assess possible stellar contamination when assigning final classifications to the 62 grade-A systems. For components without spectroscopy, the colour examination (Sect.~\ref{subsubsec:colour}) uses optical--infrared photometry to assess whether the putative quasar images are consistent with AGN. The astrometry examination (Sect.~\ref{subsubsec:astrometry}) uses Gaia proper motions to identify possible stellar counterparts. Systems that otherwise remained plausible lenses but failed either examination were classified as likely lensed quasars. Missing or inconclusive measurements provided no independent constraint.
	
	\subsubsection{Colour examination}
	\label{subsubsec:colour}
	We required at least one spectrum showing quasar emission in each confirmed system. The putative images must also satisfy $S_{\rm color}\geq0.7$, our quantitative optical-colour similarity criterion (Sect.~\ref{subsec:psf_subtraction}). When the fibre or slit covers both images, spectroscopy provides a check on stellar contamination within the pair. However, some companions lie outside the available spectroscopic apertures and lack any spectroscopic observation. We used optical--infrared colours to assess their possible stellar identity, with particular attention to these unobserved companions.
	
	WISE does not resolve the compact components and the foreground galaxy individually. For a system containing two PSs and a galaxy, the blended flux can be expressed schematically as
	\begin{equation}
		F_{\mathrm{WISE}}^{\mathrm{obs}}
		= F_{\mathrm{Galaxy}} + F_{\mathrm{PSF1}} + F_{\mathrm{PSF2}},
		\label{eq:wise_flux_decomposition}
	\end{equation}
	where PSF1 denotes the spectroscopically identified quasar and PSF2 denotes the component without spectroscopy. This relation applies separately to the $W1$ and $W2$ bands.
	
	We first predicted the galaxy's $W1$ and $W2$ fluxes from its deblended DESI-LS or HSC $grz$ photometry using the colour matching described in Sect.~\ref{subsubsec:reference_galaxies}. Similarly, we matched PSF1 to reference quasars with comparable redshifts, optical magnitudes, and colours (Sect.~\ref{subsubsec:reference_agn}). Their $W1/W2$ fluxes were scaled to the $r$-band flux of PSF1 to predict its WISE contribution.
	
	With the galaxy and PSF1 contributions constrained, the remaining WISE emission constrains the flux of PSF2. We implemented this decomposition by jointly fitting the $W1$ and $W2$ images using the component positions and galaxy profile measured from the optical images. Neighbouring sources and the local background were included in the fit, and source fluxes were constrained to be non-negative. The inferred WISE fluxes of PSF2 were then combined with its optical photometry to assess whether its colours are consistent with an AGN.
	
	We propagated uncertainties in both predicted contributions through repeated fits, preserving the paired $W1/W2$ predictions. The resulting colour intervals are conditional on the adopted optical photometry and image geometry and exclude additional image-model and template-mismatch uncertainties.
	
	Following \citet{He2022}, we combined the optical and WISE fluxes on a common AB scale as
	\begin{equation}
		\begin{aligned}
			F_{grz}&=(F_g+0.8F_r+0.5F_z)/2.3,\\
			F_W&=0.75F_{W1}+0.25F_{W2}.
		\end{aligned}
		\label{eq:optical_wise_fluxes}
	\end{equation}
	The resulting colour is $m_{grz}-m_W=-2.5\log_{10}(F_{grz}/F_W)$, and we adopted median $m_{grz}-m_W>\max[-1,(m_g-m_z)-1]$ as the quasar-selection region. Colours below this boundary flag possible stellar contamination.
	
	\subsubsection{Astrometry examination}
	\label{subsubsec:astrometry}
	
	We also examined Gaia proper motions of all putative quasar images following \citet{Lemon2019}, using the proper-motion significance (PMSIG), defined as
	\begin{equation}
		{\rm PMSIG}=\left[\left(\frac{\mu_{\alpha *}}{\sigma_{\mu_{\alpha *}}}\right)^2+
		\left(\frac{\mu_\delta}{\sigma_{\mu_\delta}}\right)^2\right]^{1/2}.
		\label{eq:pmsig}
	\end{equation}
	Here $\mu_{\alpha *}=\mu_\alpha\cos\delta$ and $\mu_\delta$ are the proper-motion components, with uncertainties $\sigma_{\mu_{\alpha *}}$ and $\sigma_{\mu_\delta}$. We adopted ${\rm PMSIG}>10$ to flag possible stellar contamination. Astrometric-quality flags were considered because blending can produce spurious proper motions. Missing counterparts and position-only Gaia solutions provide no proper-motion constraint.

	\subsection{Definition of confirmed and likely lensed quasars}
	\label{subsec:lq_classification_criteria}
	\label{subsec:confirmed_lq_definition}
	\label{subsec:likely_lq_definition}
	
	We combined spectroscopy with the photometric and stellar-contamination diagnostics in Sects.~\ref{subsec:psf_subtraction} and~\ref{subsec:colour_astrometry}. A system was classified as \textit{confirmed} when it met the following conditions simultaneously:
	\begin{enumerate}
		\item[(1)] At least one spectrum associated with the putative images identifies quasar emission, either individually or in a blend.
		\item[(2)] The compact components satisfy $S_{\rm color}\geq0.7$ in the valid deblended optical photometry.
		\item[(3)] Where reliable constraints exist, the components pass the colour and astrometric checks in Sect.~\ref{subsec:colour_astrometry}. Missing or inconclusive measurements provide no independent constraint.
		\item[(4)] An extended component consistent with a deflector satisfies $\max_b\nu_{{\rm G},b}\geq14$ in the valid adopted bands, with supporting morphology and residuals.
		\item[(5)] The image positions relative to the deflector are consistent with an SIE model without external shear (Sect.~\ref{subsec:lens_modelling}).
	\end{enumerate}
	Independent spectra of every image and a separate deflector spectrum were not required.
	
	A system was retained as \textit{likely} when lensing remained plausible but the evidence was insufficient for confirmation. We grouped these systems by the outstanding uncertainty:
	\begin{enumerate}
		\item[(i)] \textit{Incomplete imaging evidence.} The deflector or multiple-image structure is not securely established. An undetected deflector is allowed only if its predicted brightness is near or below the survey limit.
		\item[(ii)] \textit{Uncertain background-source identity.} The identity of the background quasar or a putative companion remains unresolved, including cases with ambiguous spectroscopy or possible stellar contamination.
		\item[(iii)] \textit{Inconclusive image consistency.} A quasar and a galaxy are detected, but discrepant colours or uncertain photometry or spectra prevent a secure comparison of the putative images.
		\item[(iv)] \textit{Atypical lens configuration.} The component positions, morphology, or source--deflector redshift ordering leave the lens interpretation uncertain.
	\end{enumerate}
	These categories can overlap, and candidates with possible stellar contamination remain in the likely sample pending spectroscopic verification. Representative systems are discussed in Sect.~\ref{sec:likely_lq}.
	
	\subsection{Parameter estimation}
	\label{subsec:parameter_estimation}
	
	We characterized the confirmed systems selected in Sect.~\ref{subsec:confirmed_lq_definition} by refining source redshifts where needed and fitting simple lens mass models.
	
	\subsubsection{Spectroscopic redshift estimation}
	\label{subsec:redshift_estimation}
	
	For galaxy-dominated spectra, we subtracted the supplied DESI galaxy model at its fixed redshift and normalization. We fitted the residual with the low- and high-redshift DESI QSO template sets, each containing four eigenspectra \citep{Brodzeller2023,Lyke2020}.\footnote{We used version 0.8 of \texttt{redrock-templates}: \url{https://github.com/desihub/redrock-templates/tree/0.8}.} After removing boxcar-smoothed continua from both residual and templates, we optimized the template coefficients by inverse-variance weighting. The search spanned $0.05\leq z_s\leq6$ on a grid uniform in $\ln(1+z_s)$, with solutions ranked by improvement over a zero-signal model. Gaussian emission-line fits checked the solutions; fewer than two candidate features or comparably strong alternative redshifts left the identification ambiguous.
	
	When several spectra detected the same source, we jointly fitted their unsmoothed pixels at a shared $z_s$ by minimizing the summed inverse-variance-weighted squared residuals. Each spectrum had independent template coefficients and a quadratic continuum, with galaxy subtraction applied where required. We searched within $|z_s-z_{s,0}|\leq0.06$ of the established redshift $z_{s,0}$, using fixed masks for sky and telluric features, strong absorption, and the Ly$\alpha$ forest where applicable. Template mismatch, galaxy subtraction, and unavailable resolution matrices introduce systematics beyond the supplied pixel uncertainties.
	
	\subsubsection{Lens modelling}
	\label{subsec:lens_modelling}
	
	We fitted an SIE model with \texttt{lenstronomy} \citep{BirrerAmara2018_lenstronomy,Birrer2021_lenstronomy}, using the image positions and deblended flux ratios from Sect.~\ref{subsec:psf_subtraction}. The mass centre was fixed to the measured galaxy position, and no external shear was included. The five free parameters in $\boldsymbol{p}$ were the two source-plane coordinates, ellipticity components $(e_{1,\rm SIE},e_{2,\rm SIE})$, and Einstein radius $\theta_{\rm E}$. We minimized
	\begin{equation}
		\chi^2_{\rm SIE}(\boldsymbol{p})
		=\chi^2_{\rm pos}(\boldsymbol{p})+0.01\,\chi^2_{\rm flux}(\boldsymbol{p}),
	\end{equation}
	with
	\begin{align}
		\chi^2_{\rm pos}
		&=\sum_{i=1}^{N}\left[
		\frac{(x_i^{\rm mod}-x_i)^2}{\sigma_{x,i}^2}
		+\frac{(y_i^{\rm mod}-y_i)^2}{\sigma_{y,i}^2}\right],\\
		\chi^2_{\rm flux}
		&=\sum_{i=2}^{N}\frac{(R_i^{\rm mod}-R_i^{\rm obs})^2}{\sigma_{R_i}^2}.
	\end{align}
	Here $N=2$, $(x_i,y_i)$ are the measured image positions, and the superscript $\rm mod$ denotes model predictions. The observed and model flux ratios are $R_i^{\rm obs}=f_i/f_1$ and $R_i^{\rm mod}=|\mu_i|/|\mu_1|$, where $\mu_i$ is the magnification. We assumed $0.05\arcsec$ positional uncertainties per coordinate and used the photometric flux-ratio errors $\sigma_{R_i}$. The factor 0.01 reduces sensitivity to microlensing, differential extinction, and deblending errors.
	
	The four measured coordinates of the two images, $(x_1,y_1)$ and $(x_2,y_2)$, together with the observed flux ratio $R_2^{\rm obs}=f_2/f_1$, provide five constraints on the five SIE model parameters $(\theta_{\rm E},q,\phi,x_{\rm s},y_{\rm s})$, leaving nominally zero degrees of freedom. A good fit is therefore expected, whereas a poor fit flags a configuration difficult to reconcile with a simple SIE lens, thus making the lensing interpretation suspicious. We interpret only $\theta_{\rm E}$: the image separation sets the lensing scale, while different combinations of lens shape, orientation, and source position can reproduce similar observables. Table~\ref{tab:sie} lists the best fits without formal confidence intervals; $\chi^2_{\rm SIE}$ is the unnormalized objective defined above. Position angles are measured counterclockwise from north in the east--north tangent plane, modulo $180^\circ$.
	
	\section{Confirmed lensed quasars}
	\label{sec:confirmed_lq}
	
	This section presents the 15 confirmed lensed quasars selected from the grade-A sample in Sect.~\ref{sec:pre_classification}. We excluded the 44 grade-A reference lenses before applying the final criteria in Sect.~\ref{subsec:confirmed_lq_definition}. Sect.~\ref{subsec:confirmed_overview} summarizes the sample, and Sect.~\ref{subsec:confirmed_individual_notes} discusses the evidence for individual systems.
	
	\subsection{Overview}
	\label{subsec:confirmed_overview}
	
	The confirmed sample comprises 11 systems from the main DESI sample, two from the multiple-position sample, and two P200 targets. Figure~\ref{fig:cutouts} presents a montage of all 15 systems. Figure~\ref{fig:spec_image} shows the spectra, optical images, point-source models, and images after point-source subtraction for each system.
	
	Table~\ref{tab:confirmed_astrometry} lists the Gaia proper motions of components A and B and their PMSIG values, defined in Sect.~\ref{subsec:colour_astrometry}. Table~\ref{tab:tab1} summarizes the system coordinates, source redshifts, spectroscopic data sources and identifiers, and diagnostic statistics. These include the SIE and image-fitting statistics, $\chi^2_{\rm SIE}$ and $\chi^2_{\nu,\rm img}$, and the colour-similarity score, $S_{\rm color}$.
	
	Tables~\ref{tab:lens_gal} and~\ref{tab:psfs} report the photometric results for the lens galaxies and quasar images, respectively. Table~\ref{tab:lens_gal} gives the fitted S\'ersic parameters, total model magnitudes, and maximum conditional flux S/N across the valid bands, together with the corresponding band. Table~\ref{tab:psfs} lists the deblended magnitudes and uncertainties of components A and B in the available optical bands. Table~\ref{tab:sie} presents the SIE modelling results, including the Einstein radius, axis ratio, and position angle, alongside the input coordinates of the lens galaxy and quasar images.
	
	\begin{table*}[!tp]
		\centering
		\caption{Gaia DR3 proper motions of the 15 confirmed systems.}
		\label{tab:confirmed_astrometry}
		\begin{threeparttable}
			\small
			\setlength{\tabcolsep}{5pt}
			\renewcommand{\arraystretch}{1.12}
			\begin{tabular}{@{}lrrrrrr@{}}
				\toprule
				& \multicolumn{3}{c}{Component A} & \multicolumn{3}{c}{Component B} \\
				\cmidrule(lr){2-4}\cmidrule(l){5-7}
				System & $\mu_{\alpha *}$ & $\mu_\delta$ & PMSIG & $\mu_{\alpha *}$ & $\mu_\delta$ & PMSIG \\
				& \multicolumn{2}{c}{(mas\,yr$^{-1}$)} & & \multicolumn{2}{c}{(mas\,yr$^{-1}$)} & \\
				\midrule
				J0047+0322 & \multicolumn{3}{c}{No match} & \multicolumn{3}{c}{No match} \\
				J0138+0315 & \multicolumn{3}{c}{No match} & \multicolumn{3}{c}{No match} \\
				J0141+0039 & \multicolumn{3}{c}{No match} & \multicolumn{3}{c}{No match} \\
				J0422$-$0447 & \multicolumn{3}{c}{Position only} & $-0.33\pm0.31$ & $-1.63\pm0.22$ & 7.48 \\
				J0422$-$1427 & \multicolumn{3}{c}{Position only} & $0.29\pm0.24$ & $0.02\pm0.25$ & 1.25 \\
				J0642+5617 & $0.94\pm0.38$ & $1.28\pm0.36$ & 4.29 & \multicolumn{3}{c}{Position only} \\
				J0911$-$0044 & \multicolumn{3}{c}{No match} & \multicolumn{3}{c}{No match} \\
				J0954+0323 & \multicolumn{3}{c}{No match} & \multicolumn{3}{c}{No match} \\
				J1215$-$0058 & \multicolumn{3}{c}{No match} & \multicolumn{3}{c}{No match} \\
				J1220+0112 & \multicolumn{3}{c}{No match} & \multicolumn{3}{c}{No match} \\
				J1307+0011 & \multicolumn{3}{c}{Position only} & $2.12\pm0.94$ & $-1.87\pm0.71$ & 3.45 \\
				J1531+3724 & \multicolumn{3}{c}{Position only} & $0.37\pm0.27$ & $0.56\pm0.31$ & 2.30 \\
				J1753+7333 & $0.08\pm0.19$ & $-0.28\pm0.18$ & 1.59 & $-0.95\pm1.39$ & $-1.48\pm1.19$ & 1.41 \\
				J1800+5305 & \multicolumn{3}{c}{Position only} & $0.27\pm0.39$ & $0.17\pm0.42$ & 0.81 \\
				J2348+1530 & $0.69\pm0.78$ & $-0.39\pm0.66$ & 1.07 & \multicolumn{3}{c}{Position only} \\
				\bottomrule
			\end{tabular}
			\begin{tablenotes}[flushleft]
				\footnotesize
				\item Errors are formal $1\sigma$ uncertainties; PMSIG is defined in Sect.~\ref{subsec:colour_astrometry}, while \textit{Position only} and \textit{No match} denote unavailable proper-motion measurements.
			\end{tablenotes}
		\end{threeparttable}
	\end{table*}
	
	\begin{figure*}[!tp]
		\centering
		\includegraphics[width=0.83\textwidth]{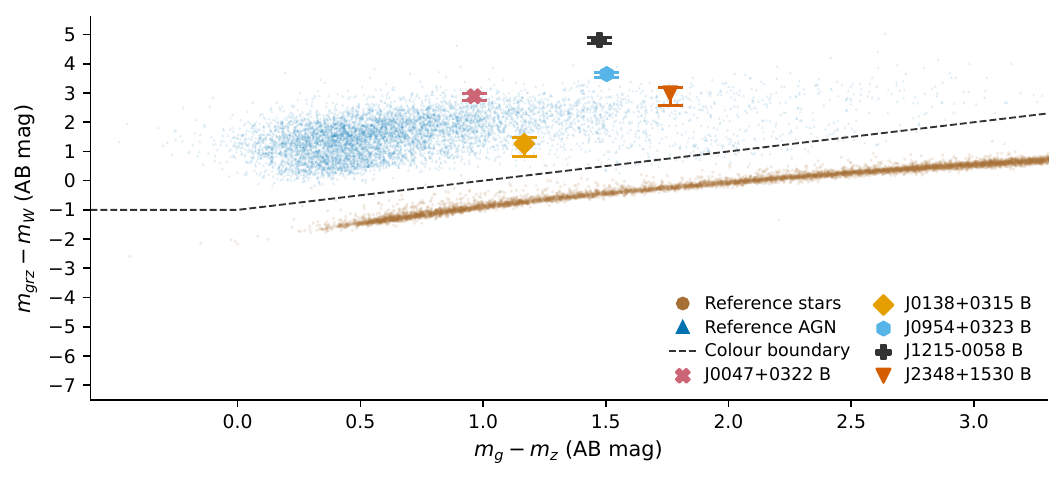}
		\caption{Optical--infrared colours of component B in the five indicated confirmed systems. The colours combine deblended optical photometry (Sect.~\ref{subsec:psf_subtraction}) with WISE fluxes inferred by constraining the contributions of the spectroscopically identified quasar and the lens galaxy (Sect.~\ref{subsubsec:colour}). Symbols show fits at the median predicted AGN and galaxy fluxes; bars span the conditional $1\sigma$ intervals (15.9th--84.1st percentiles) obtained by propagating both prediction uncertainties. Brown circles show reference stars, blue triangles show reference AGN, and the dashed line marks the adopted colour boundary. These conditional intervals exclude image-model and template-mismatch uncertainties. J1215-0058 and J2348+1530 retain image-fit quality flags; the J1215-0058 colour also uses a flagged HSC $z$-band measurement.}
		\label{fig:confirmed_colours}
	\end{figure*}

	\begin{figure*}
		\centering
		\includegraphics[width=0.9\linewidth]{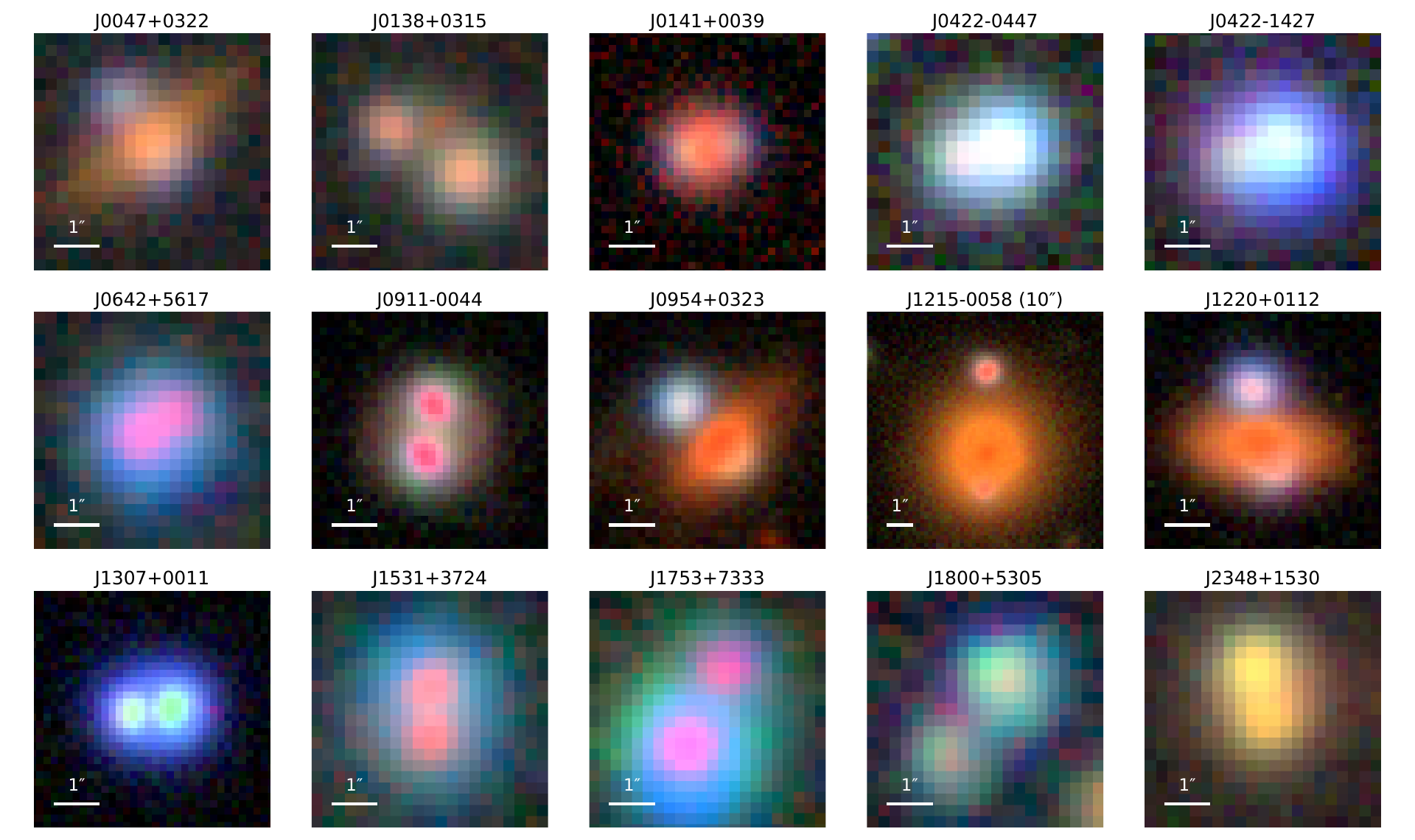}
		\caption{Cutouts of the 15 confirmed systems. The field of view (FOV) is $5.5\arcsec\times5.5\arcsec$, except for the $10\arcsec\times10\arcsec$ field of J1215-0058; scale bars indicate $1\arcsec$.}
		\label{fig:cutouts}
	\end{figure*}
	
	\begin{systemfigurepage}
		\includegraphics[width=0.96\linewidth,trim=0 30bp 0 0,clip]{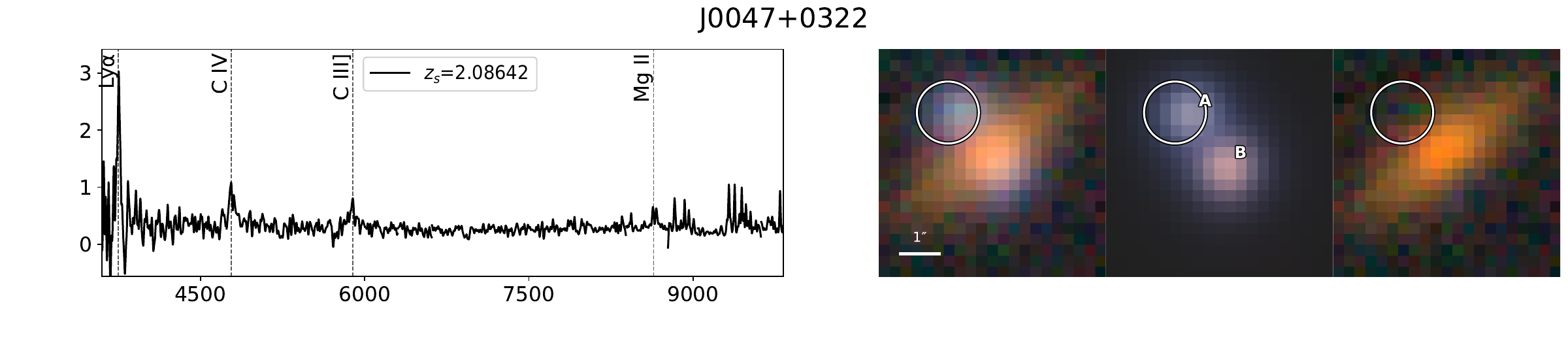}\par
		\vspace{2mm}
		\includegraphics[width=0.96\linewidth,trim=0 30bp 0 0,clip]{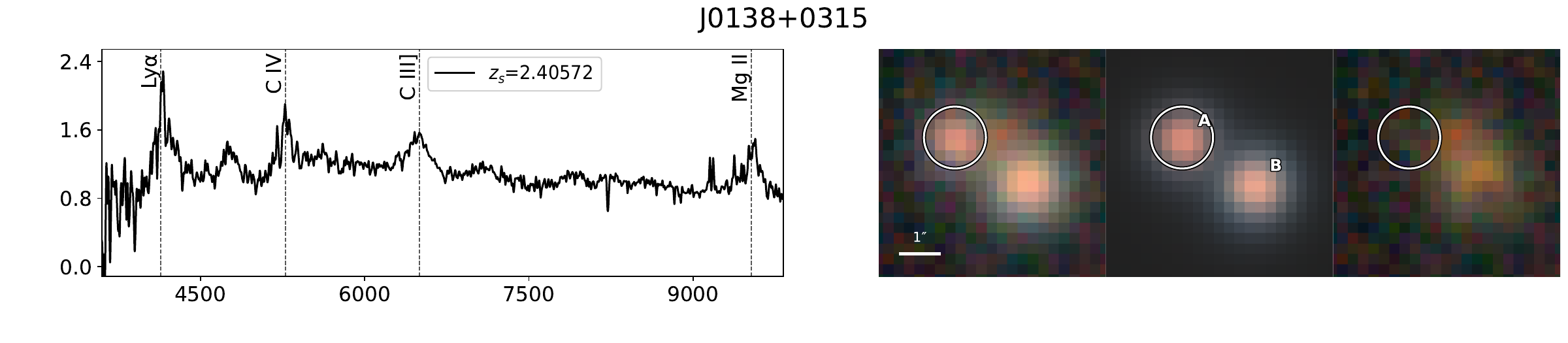}\par
		\vspace{2mm}
		\includegraphics[width=0.96\linewidth,trim=0 30bp 0 0,clip]{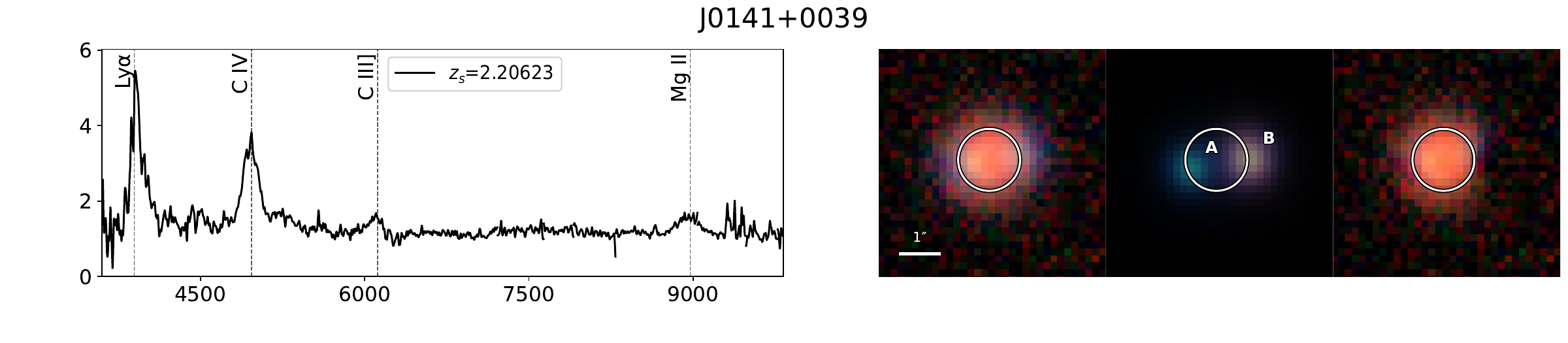}\par
		\vspace{2mm}
		\includegraphics[width=0.96\linewidth,trim=0 30bp 0 0,clip]{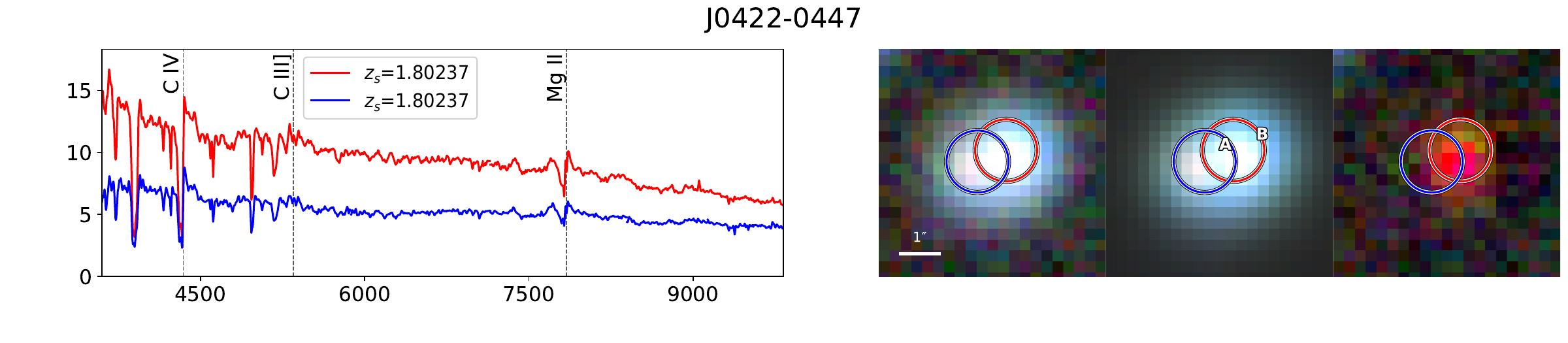}\par
		\vspace{2mm}
		\includegraphics[width=0.96\linewidth,trim=0 30bp 0 0,clip]{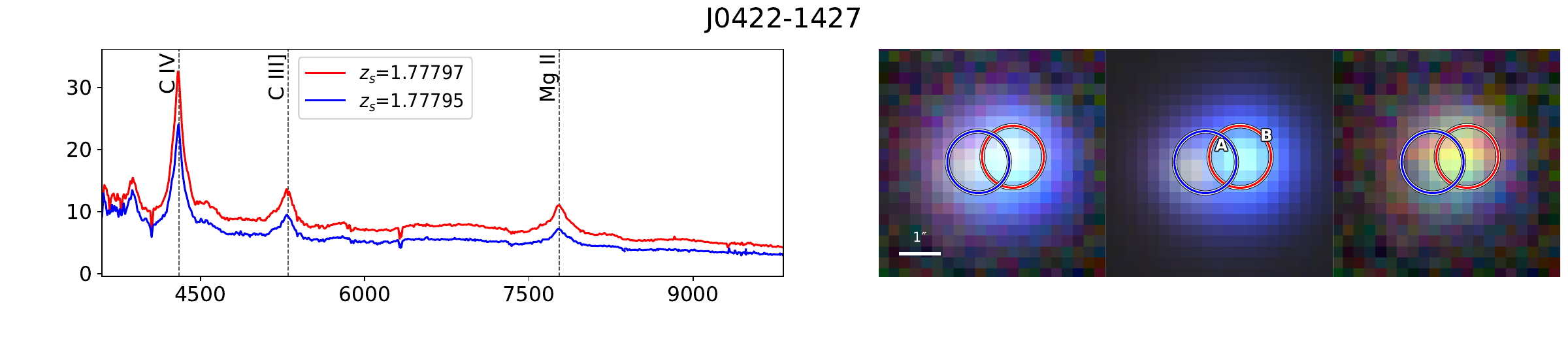}\par
		\vspace{2mm}
		\includegraphics[width=0.96\linewidth,trim=0 30bp 0 0,clip]{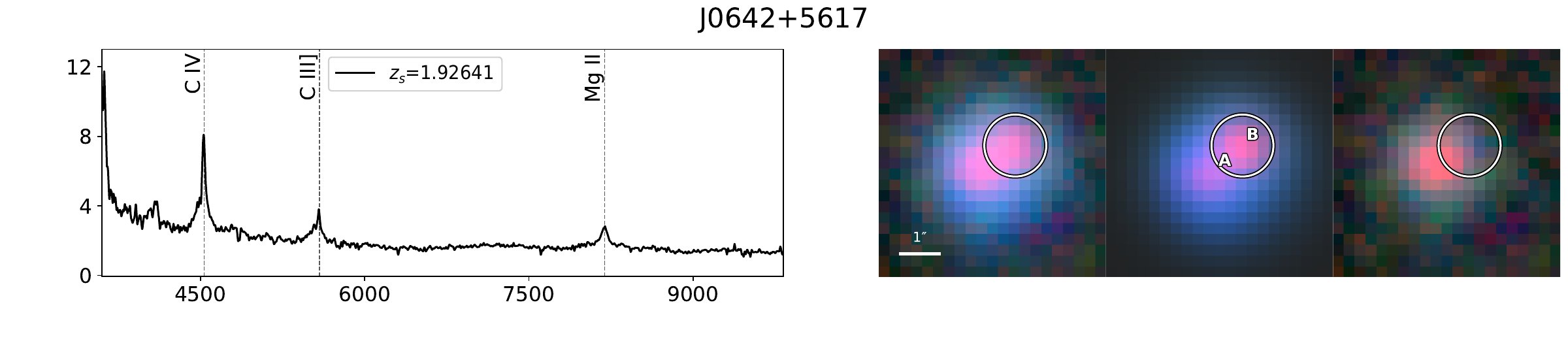}\par
		\caption{Spectra (left) and imaging decompositions (right: data, PSF model, and data minus PSF) of the 15 confirmed systems. Image fields span $5.5\arcsec\times5.5\arcsec$, except for the $10\arcsec\times10\arcsec$ field of J1215-0058; scale bars indicate $1\arcsec$. Circles and parallel lines mark DESI fibres and schematic P200 slits, respectively. In the P200 spectra of J1753+7333 and J1800+5305, a circled cross marks absorption by molecular oxygen in Earth's atmosphere near 7600\,\AA.}
		\label{fig:spec_image}
	\end{systemfigurepage}
	
	\begin{systemfigurepage}
		\ContinuedFloat
		\includegraphics[width=0.96\linewidth,trim=0 30bp 0 0,clip]{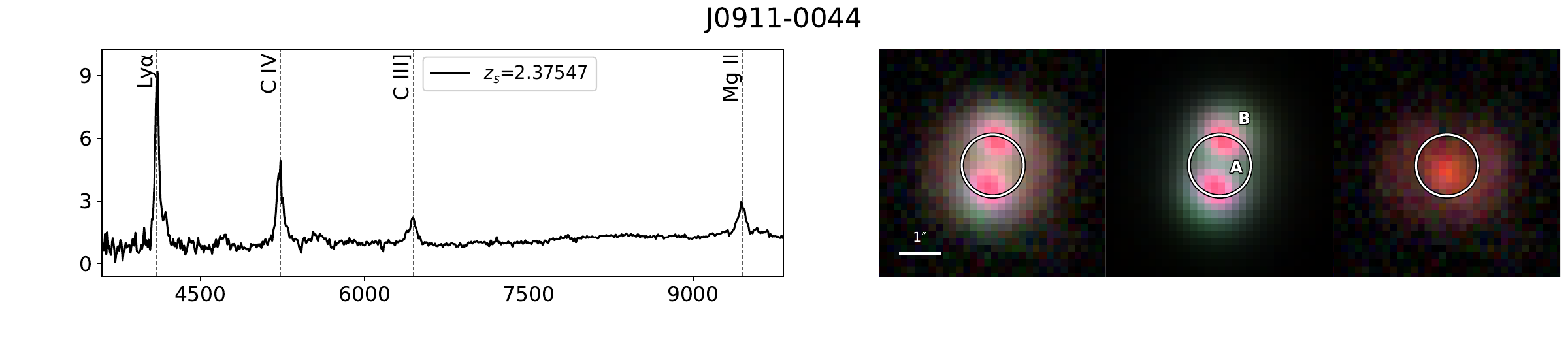}\par
		\includegraphics[width=0.96\linewidth,trim=0 30bp 0 0,clip]{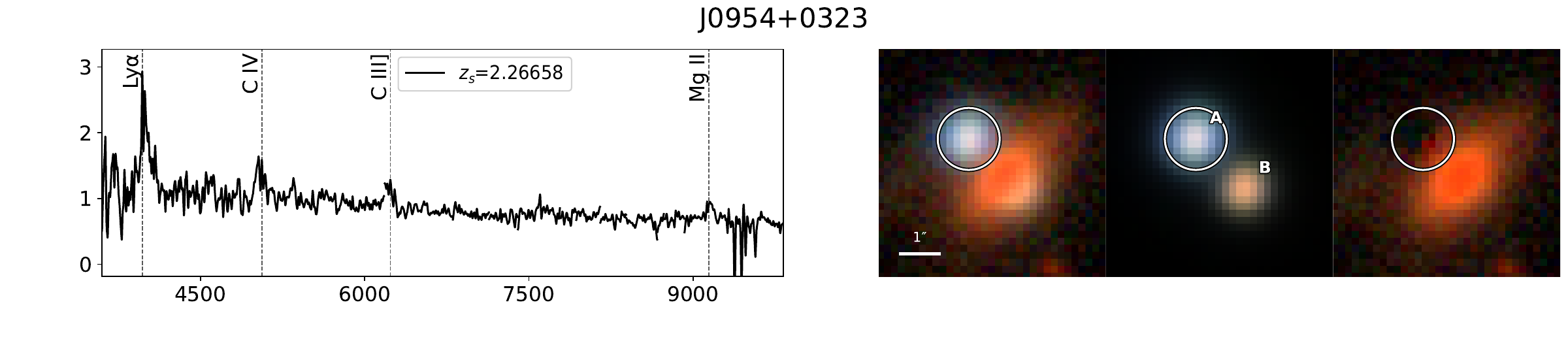}\par
		\includegraphics[width=0.96\linewidth,trim=0 30bp 0 0,clip]{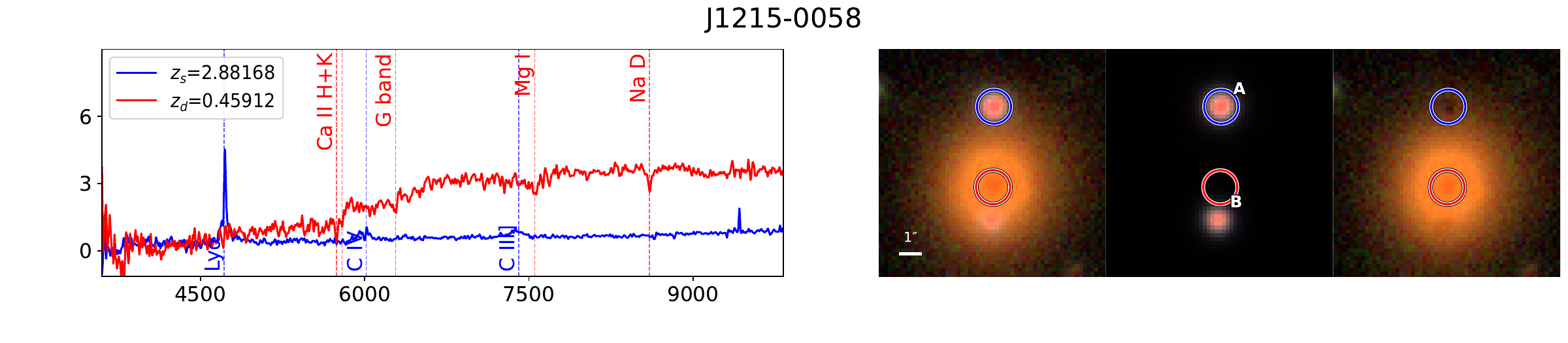}\par
		\includegraphics[width=0.96\linewidth,trim=0 30bp 0 0,clip]{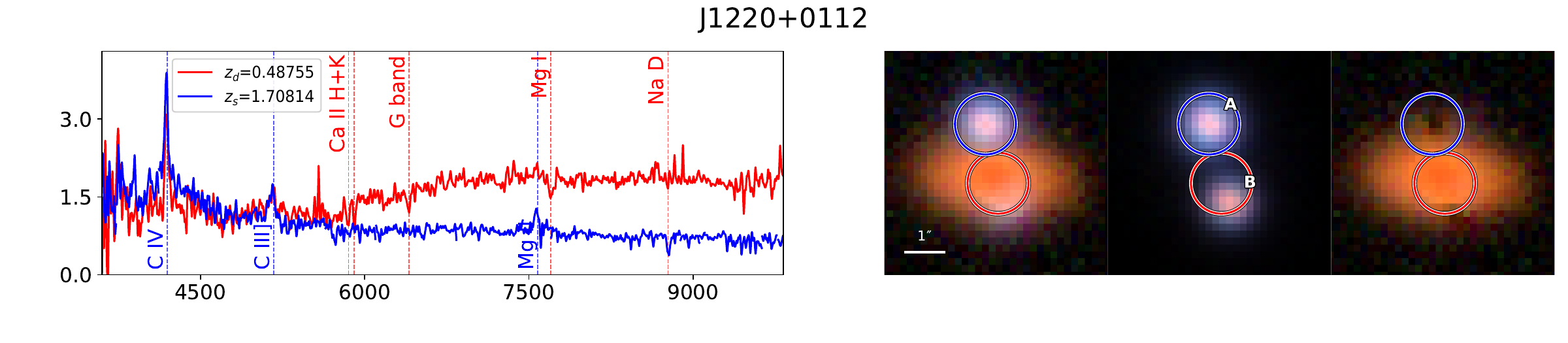}\par
		\includegraphics[width=0.96\linewidth,trim=0 30bp 0 0,clip]{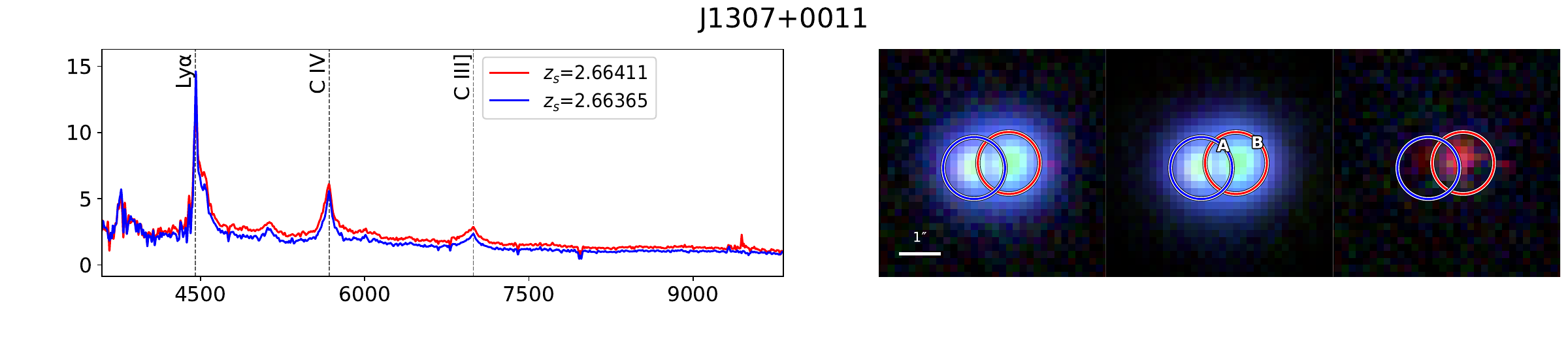}\par
		\includegraphics[width=0.96\linewidth,trim=0 30bp 0 0,clip]{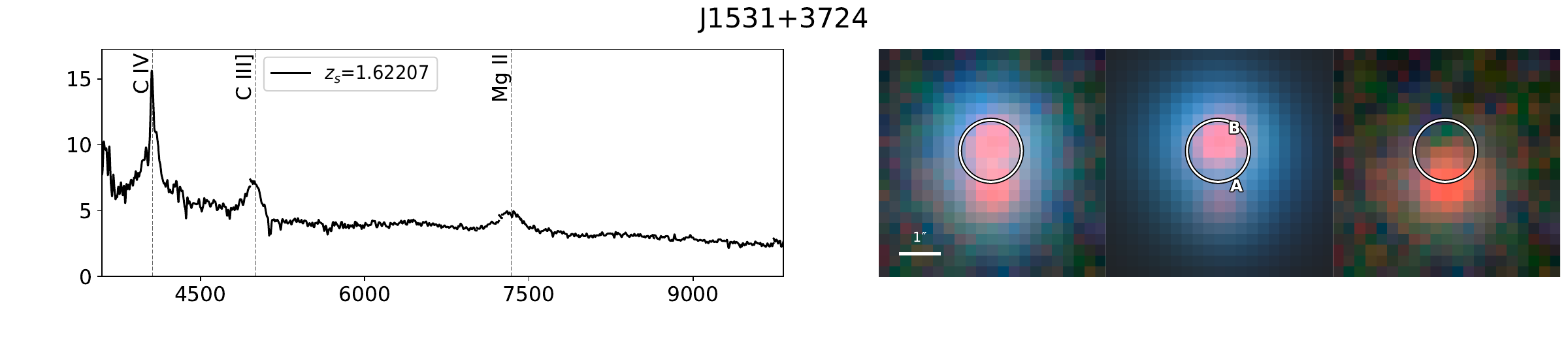}\par
		\includegraphics[width=0.96\linewidth,trim=0 30bp 0 0,clip]{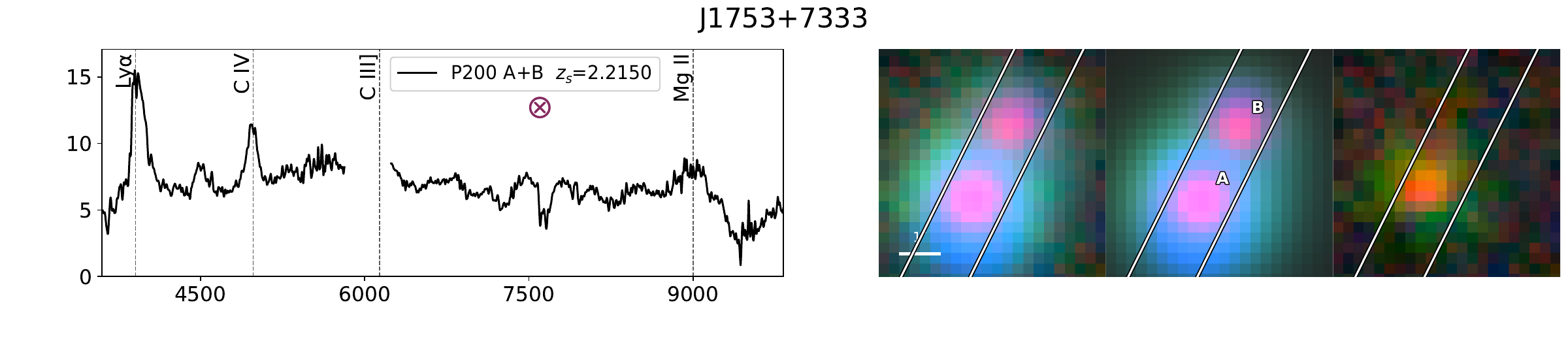}\par
		\caption{Continued.}
	\end{systemfigurepage}
	
	\begin{figure*}[!t]
		\centering
		\ContinuedFloat
		\includegraphics[width=0.96\linewidth,trim=0 30bp 0 0,clip]{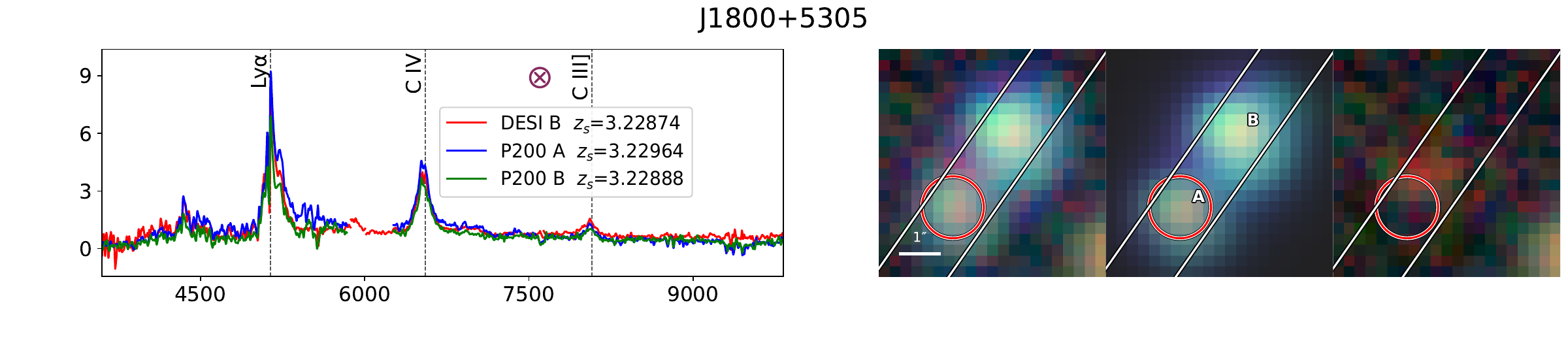}\par
		\vspace{2mm}
		\includegraphics[width=0.96\linewidth,trim=0 30bp 0 0,clip]{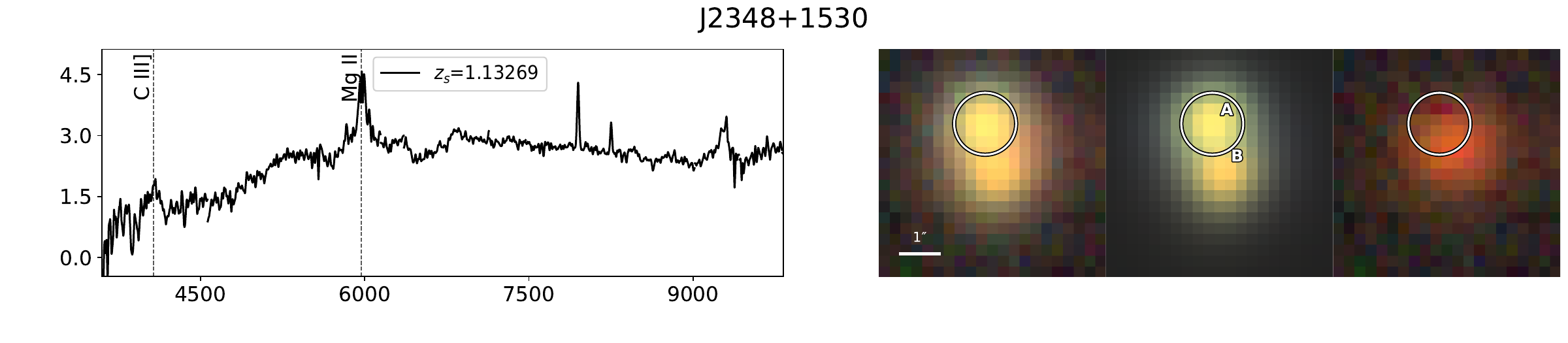}\par
		\caption{Continued.}
	\end{figure*}
	
	\begin{table*}[t]
		\centering
		\small
		\setlength{\tabcolsep}{2pt}
		\renewcommand{\arraystretch}{1.10}
		\caption{Properties and diagnostic modelling statistics of the 15 confirmed lensed quasars. The image-fit statistic combines all available fitted optical bands, including the HSC $z$ band, as defined in Sect.~\ref{subsec:psf_subtraction}. Reference codes are H23: \citet{He2023}; D23: \citet{Dawes2023}; C23: \citet{Chan2023}; A23: \citet{Andika2023}; and S25: \citet{Shu2025}.}
		\label{tab:tab1}
		\begin{tabular*}{\textwidth}{@{\extracolsep{\fill}}llcllcccc@{}}
			\toprule
			Name & \makecell{RA, Dec\\{}[deg]} & $z_s$ & Spectra & DESI SPECID & Refs. & $\chi^2_{\rm SIE}$ & $\chi^2_{\nu,\rm img}$ & $S_{\rm color}$ \\
			\midrule
			J0047+0322 & \makecell[l]{11.91232\\ 3.37556} & 2.0864 & DESI & \makecell[l]{2789831897776129} & C23;S25 & 1.14e-26 & 1.065 & 0.860 \\
			J0138+0315 & \makecell[l]{24.61357\\ 3.25511} & 2.4057 & DESI & \makecell[l]{39627863657286622} & C23 & 5.7e-26 & 0.876 & 0.958 \\
			J0141+0039 & \makecell[l]{25.46863\\ 0.66559} & 2.2062 & DESI & \makecell[l]{39627803271895916} & C23 & 1.71e-28 & 1.713 & 0.914 \\
			J0422-0447 & \makecell[l]{65.58130\\ -4.79716} & 1.810 & DESI & \makecell[l]{39627671142927980\\ 39627671142927987} & D23;H23 & 4.73e-24 & 1.517 & 0.856 \\
			J0422-1427 & \makecell[l]{65.65655\\ -14.46659} & 1.7787 & DESI & \makecell[l]{39627439000783472\\ 39627439000783478} & D23;H23 & 3.77e-26 & 2.195 & 0.739 \\
			J0642+5617 & \makecell[l]{100.65294\\ 56.28642} & 1.9264 & DESI & \makecell[l]{39633338255806502} & D23;H23 & 1.03e-26 & 0.729 & 0.707 \\
			J0911-0044 & \makecell[l]{137.96489\\ -0.73524} & 2.3755 & DESI & \makecell[l]{39627768920544632} & C23;A23 & 1.12e-23 & 1.106 & 0.957 \\
			J0954+0323 & \makecell[l]{148.68976\\ 3.39569} & 2.2666 & DESI & \makecell[l]{39627871764878042} & A23 & 1.48e-24 & 1.733 & 0.719 \\
			J1215-0058 & \makecell[l]{183.88950\\ -0.97856} & 2.8817 & DESI & \makecell[l]{39627763652496396\\ 39627763652496397} & A23;S25 & 2.43e-25 & 2.969 & 0.947 \\
			J1220+0112 & \makecell[l]{185.07900\\ 1.21517} & 1.7069 & DESI & \makecell[l]{39627818031646108\\ 39627818031646110} & C23;A23;S25 & 1.42e-25 & 2.844 & 0.926 \\
			J1307+0011 & \makecell[l]{196.88880\\ 0.18966} & 2.6669 & DESI & \makecell[l]{39627794069589256\\ 39627794069589261} & C23 & 8.95e-26 & 1.060 & 0.928 \\
			J1531+3724 & \makecell[l]{232.94340\\ 37.40290} & 1.6221 & DESI & \makecell[l]{39633030372919474} & D23 & 4.01e-26 & 0.706 & 0.955 \\
			J1753+7333 & \makecell[l]{268.48410\\ 73.56490} & 2.2150 & P200 & -- & H23 & 8.92e-26 & 0.979 & 0.842 \\
			J1800+5305 & \makecell[l]{270.01470\\ 53.08359} & 3.2292 & \makecell[l]{DESI;\\ P200} & \makecell[l]{39633294265944616} & D23;H23 & 1.19e-26 & 0.879 & 0.968 \\
			J2348+1530 & \makecell[l]{357.04209\\ 15.50576} & 1.1327 & DESI & \makecell[l]{39628160865667843} & D23;H23 & 1.17e-23 & 0.889 & 0.921 \\
			\bottomrule
		\end{tabular*}
	\end{table*}
	
	\begin{table*}[t]
		\centering
		\caption{Lens-galaxy S\'ersic parameters, total magnitudes, and conditional flux S/N. The S\'ersic index was fixed to 1 for J0047+0322 and J0954+0323 and to 4 for the other systems.}
		\label{tab:lens_gal}
		\small
		\setlength{\tabcolsep}{1.5pt}
		\renewcommand{\arraystretch}{1.18}
		\begin{tabular*}{\textwidth}{@{\extracolsep{\fill}}lccccccccc@{}}
			\toprule
			& \multicolumn{4}{c}{Shared S\'ersic parameters} & \multicolumn{4}{c}{Total model magnitudes} & \\
			\cmidrule(lr){2-5}\cmidrule(lr){6-9}
			Name & $n_s$ & $r_s$ [\arcsec] & $q_s$ & $\phi_s$ [$^\circ$] & $m_g$ & $m_r$ & $m_i$ & $m_z$ & $\max_b\nu_{{\rm G},b}$ \\
			\midrule
			J0047+0322 & 1 & 1.151 & 0.182 & 127.19 & $23.024\pm0.080$ & $20.968\pm0.014$ & nodata & $19.819\pm0.013$ & 80.6 ($z$) \\
			J0138+0315 & 4 & 1.674 & 0.575 & 37.05 & $22.119\pm0.065$ & $20.854\pm0.022$ & nodata & $20.105\pm0.029$ & 49.3 ($r$) \\
			J0141+0039 & 4 & 0.288 & 0.609 & 114.71 & $22.200\pm0.012$ & $21.575\pm0.007$ & $20.651\pm0.011$ & $20.273\pm0.005$ & 149.8 ($r$) \\
			J0422-0447 & 4 & 0.197 & 0.493 & 177.66 & $23.804\pm0.438$ & $24.071\pm0.928$ & nodata & $20.536\pm0.042$ & 25.9 ($z$) \\
			J0422-1427 & 4 & 0.318 & 0.639 & 150.02 & $20.392\pm0.027$ & $19.716\pm0.011$ & nodata & $19.212\pm0.015$ & 102.8 ($r$) \\
			J0642+5617 & 4 & 0.030 & 0.150 & 172.30 & $21.136\pm0.042$ & $20.613\pm0.041$ & nodata & $20.083\pm0.028$ & 38.3 ($z$) \\
			J0911-0044 & 4 & 2.839 & 0.745 & 84.79 & $22.085\pm0.031$ & $21.427\pm0.032$ & $20.506\pm0.010$ & $19.823\pm0.013$ & 107.3 ($i$) \\
			J0954+0323 & 1 & 0.653 & 0.472 & 137.36 & $23.660\pm0.038$ & $21.832\pm0.010$ & $20.614\pm0.005$ & $20.206\pm0.006$ & 239.1 ($i$) \\
			J1215-0058 & 4 & 2.135 & 0.917 & 171.45 & $20.573\pm0.004$ & $18.840\pm0.001$ & $17.967\pm0.001$ & $17.594\pm0.001$ & 1577.5 ($i$) \\
			J1220+0112 & 4 & 1.239 & 0.499 & 83.40 & $21.892\pm0.010$ & $20.489\pm0.004$ & $19.572\pm0.002$ & $19.173\pm0.002$ & 613.4 ($i$) \\
			J1307+0011 & 4 & 0.093 & 0.150 & 102.57 & $23.731\pm0.113$ & $24.163\pm0.096$ & $22.678\pm0.031$ & $23.130\pm0.144$ & 34.7 ($i$) \\
			J1531+3724 & 4 & 0.256 & 0.456 & 89.16 & $21.282\pm0.083$ & $20.706\pm0.050$ & nodata & $19.964\pm0.022$ & 50.5 ($z$) \\
			J1753+7333 & 4 & 0.015 & 0.150 & 178.81 & \makecell{below detection\\ limit} & $21.088\pm0.094$ & nodata & $20.711\pm0.043$ & 25.5 ($z$) \\
			J1800+5305 & 4 & 1.447 & 0.150 & 75.92 & $23.500\pm0.194$ & $22.517\pm0.100$ & nodata & $20.917\pm0.071$ & 15.3 ($z$) \\
			J2348+1530 & 4 & 1.428 & 0.720 & 95.48 & $22.576\pm0.159$ & $21.044\pm0.043$ & $20.180\pm0.027$ & $19.591\pm0.021$ & 52.7 ($z$) \\
			\bottomrule
		\end{tabular*}
	\end{table*}
	
	\begin{table*}[t]
		\centering
		\caption{Deblended photometry of the 15 confirmed lensed quasars.}
		\label{tab:psfs}
		\small
		\setlength{\tabcolsep}{4pt}
		\renewcommand{\arraystretch}{1.05}
		\begin{tabular*}{\textwidth}{@{\extracolsep{\fill}}llcccc@{}}
			\toprule
			Name & Survey & $m_g$ & $m_r$ & $m_i$ & $m_z$ \\
			\midrule
			J0047+0322 A & DESI-LS & $21.976\pm0.014$ & $21.815\pm0.013$ & nodata & $21.432\pm0.025$ \\
			J0047+0322 B & DESI-LS & $21.874\pm0.016$ & $21.559\pm0.013$ & nodata & $20.911\pm0.020$ \\\addlinespace[2pt]
			J0138+0315 A & DESI-LS & $22.119\pm0.016$ & $21.517\pm0.010$ & nodata & $20.828\pm0.014$ \\
			J0138+0315 B & DESI-LS & $21.662\pm0.016$ & $21.058\pm0.010$ & nodata & $20.494\pm0.014$ \\\addlinespace[2pt]
			J0141+0039 A & HSC & $23.765\pm0.035$ & $23.888\pm0.041$ & $25.481\pm0.629$ & $24.245\pm0.108$ \\
			J0141+0039 B & HSC & $23.241\pm0.017$ & $23.193\pm0.019$ & $22.817\pm0.046$ & $22.136\pm0.014$ \\\addlinespace[2pt]
			J0422-0447 A & DESI-LS & $20.834\pm0.009$ & $20.190\pm0.008$ & nodata & $19.797\pm0.007$ \\
			J0422-0447 B & DESI-LS & $19.169\pm0.005$ & $18.791\pm0.005$ & nodata & $18.564\pm0.005$ \\\addlinespace[2pt]
			J0422-1427 A & DESI-LS & $21.144\pm0.018$ & $20.846\pm0.011$ & nodata & $20.330\pm0.016$ \\
			J0422-1427 B & DESI-LS & $19.601\pm0.009$ & $19.642\pm0.007$ & nodata & $19.568\pm0.014$ \\\addlinespace[2pt]
			J0642+5617 A & DESI-LS & $20.640\pm0.023$ & $20.848\pm0.043$ & nodata & $20.730\pm0.046$ \\
			J0642+5617 B & DESI-LS & $21.010\pm0.014$ & $20.829\pm0.019$ & nodata & $20.222\pm0.016$ \\\addlinespace[2pt]
			J0911-0044 A & HSC & $21.678\pm0.004$ & $21.382\pm0.006$ & $20.942\pm0.002$ & $20.577\pm0.004$ \\
			J0911-0044 B & HSC & $21.880\pm0.004$ & $21.460\pm0.005$ & $21.045\pm0.002$ & $20.616\pm0.003$ \\\addlinespace[2pt]
			J0954+0323 A & HSC & $21.810\pm0.003$ & $21.514\pm0.003$ & $21.590\pm0.005$ & $21.315\pm0.008$ \\
			J0954+0323 B & HSC & $23.089\pm0.014$ & $22.416\pm0.010$ & $22.146\pm0.010$ & $21.585\pm0.014$ \\\addlinespace[2pt]
			J1215-0058 A & HSC & $22.367\pm0.006$ & $21.893\pm0.005$ & $21.255\pm0.003$ & $20.762\pm0.004$ \\
			J1215-0058 B & HSC & $23.038\pm0.010$ & $22.603\pm0.010$ & $22.085\pm0.005$ & $21.564\pm0.008$ \\\addlinespace[2pt]
			J1220+0112 A & HSC & $21.660\pm0.003$ & $21.613\pm0.004$ & $21.433\pm0.003$ & $21.370\pm0.005$ \\
			J1220+0112 B & HSC & $22.553\pm0.007$ & $22.417\pm0.008$ & $22.103\pm0.006$ & $22.018\pm0.011$ \\\addlinespace[2pt]
			J1307+0011 A & HSC & $20.978\pm0.003$ & $20.793\pm0.002$ & $20.885\pm0.002$ & $20.669\pm0.005$ \\
			J1307+0011 B & HSC & $20.227\pm0.004$ & $19.964\pm0.002$ & $20.331\pm0.003$ & $19.955\pm0.007$ \\\addlinespace[2pt]
			J1531+3724 A & DESI-LS & $22.052\pm0.122$ & $21.714\pm0.094$ & nodata & $21.696\pm0.079$ \\
			J1531+3724 B & DESI-LS & $19.994\pm0.010$ & $19.791\pm0.008$ & nodata & $19.695\pm0.008$ \\\addlinespace[2pt]
			J1753+7333 A & DESI-LS & $18.859\pm0.004$ & $18.805\pm0.011$ & nodata & $18.552\pm0.006$ \\
			J1753+7333 B & DESI-LS & $20.767\pm0.009$ & $20.530\pm0.019$ & nodata & $19.985\pm0.011$ \\\addlinespace[2pt]
			J1800+5305 A & DESI-LS & $21.374\pm0.015$ & $20.952\pm0.013$ & nodata & $20.845\pm0.032$ \\
			J1800+5305 B & DESI-LS & $20.435\pm0.006$ & $19.956\pm0.005$ & nodata & $19.945\pm0.014$ \\\addlinespace[2pt]
			J2348+1530 A & DESI-LS & $21.175\pm0.012$ & $20.133\pm0.005$ & $19.786\pm0.005$ & $19.692\pm0.006$ \\
			J2348+1530 B & DESI-LS & $21.787\pm0.023$ & $20.627\pm0.008$ & $20.175\pm0.008$ & $20.025\pm0.009$ \\\addlinespace[2pt]
			\bottomrule
		\end{tabular*}
	\end{table*}
	
	\begin{table*}[t]
		\centering
		\caption{SIE parameters and input positions of the lens galaxy (G) and quasar images (A, B).}
		\label{tab:sie}
		\small
		\setlength{\tabcolsep}{3pt}
		\begin{tabular}{lccccccccc}
			\specialrule{\heavyrulewidth}{0pt}{\doublerulesep}
			\specialrule{\heavyrulewidth}{0pt}{\belowrulesep}
			Name
			& $\theta_{\rm E}$
			& $\phi_{\rm SIE}$
			& $q_{\rm SIE}$
			& $\mathrm{RA}_{\rm G}$
			& $\mathrm{Dec}_{\rm G}$
			& $\mathrm{RA}_{\rm A}$
			& $\mathrm{Dec}_{\rm A}$
			& $\mathrm{RA}_{\rm B}$
			& $\mathrm{Dec}_{\rm B}$ \\
			& [\arcsec] & [$^\circ$] & [--]
			& [deg] & [deg] & [deg] & [deg] & [deg] & [deg] \\
			\midrule
			J0047+0322 & 0.729 & 49.59 & 0.726 & 11.912349 & 3.375599 & 11.912523 & 3.375833 & 11.912284 & 3.375485 \\
			J0138+0315 & 1.002 & 19.23 & 0.329 & 24.613640 & 3.255209 & 24.614052 & 3.255382 & 24.613571 & 3.255080 \\
			J0141+0039 & 0.755 & 66.89 & 0.766 & 25.468645 & 0.665596 & 25.468794 & 0.665535 & 25.468407 & 0.665598 \\
			J0422-0447 & 0.431 & 166.30 & 0.446 & 65.581057 & -4.797364 & 65.581242 & -4.797410 & 65.580992 & -4.797341 \\
			J0422-1427 & 0.536 & 157.01 & 0.447 & 65.656518 & -14.466574 & 65.656735 & -14.466654 & 65.656421 & -14.466588 \\
			J0642+5617 & 0.494 & 162.66 & 0.542 & 100.653118 & 56.286383 & 100.653158 & 56.286284 & 100.652819 & 56.286458 \\
			J0911-0044 & 0.578 & 85.60 & 0.685 & 137.964883 & -0.735315 & 137.964920 & -0.735409 & 137.964866 & -0.735082 \\
			J0954+0323 & 0.810 & 47.20 & 0.737 & 148.689788 & 3.395715 & 148.690040 & 3.395973 & 148.689710 & 3.395639 \\
			J1215-0058 & 2.489 & 54.74 & 0.949 & 183.889495 & -0.978573 & 183.889495 & -0.977613 & 183.889533 & -0.978997 \\
			J1220+0112 & 0.993 & 116.44 & 0.892 & 185.078978 & 1.214999 & 185.079039 & 1.215359 & 185.078905 & 1.214827 \\
			J1307+0011 & 0.414 & 165.28 & 0.212 & 196.888823 & 0.189676 & 196.889012 & 0.189636 & 196.888780 & 0.189657 \\
			J1531+3724 & 0.736 & 165.95 & 0.772 & 232.943434 & 37.402749 & 232.943411 & 37.402633 & 232.943428 & 37.403021 \\
			J1753+7333 & 0.892 & 118.94 & 0.339 & 268.484847 & 73.564551 & 268.485005 & 73.564430 & 268.484164 & 73.564908 \\
			J1800+5305 & 1.202 & 3.17 & 0.824 & 270.014985 & 53.083517 & 270.015116 & 53.083268 & 270.014504 & 53.083783 \\
			J2348+1530 & 0.606 & 24.90 & 0.917 & 357.042045 & 15.505822 & 357.042137 & 15.505980 & 357.042064 & 15.505669 \\
			\bottomrule
		\end{tabular}
	\end{table*}
	
	\subsection{Notes on individual systems}
	\label{subsec:confirmed_individual_notes}
	
	For J0047+0322, J0138+0315, J0954+0323, J1215$-$0058, and J2348+1530, we used the optical--infrared colour of component B to check for stellar contamination (Fig.~\ref{fig:confirmed_colours}).
	
	\paragraph{J0047+0322.}
	Figure~\ref{fig:spec_image} shows an elongated, disc-like deflector in this system, which was reported by \citet{Chan2023} and subsequently by \citet{Shu2025}. We fixed the deflector's S\'ersic index at $n_s=1$ in the photometric fit (Table~\ref{tab:lens_gal}). The DESI spectrum centred on A gives \zs{2.0864}, and the two compact components have \Sc{0.860}. Component B lies in the quasar-selection region of Fig.~\ref{fig:confirmed_colours} ($m_{grz}-m_W=2.89$), although the reference disc galaxies used to predict the deflector's WISE flux are an imperfect match to its optical colours. Their positions around the galaxy support a double-image configuration. The fitted galaxy magnitudes are \mg{23.02}, \mr{20.97}, and \mz{19.82}, and the SIE model gives \TE{0.729}.
	
	\paragraph{J0138+0315.}
	In this candidate from \citet{Chan2023}, DESI identifies a quasar at \zs{2.4057} (Fig.~\ref{fig:spec_image}). The decomposition separates a galaxy with \mg{22.12}, \mr{20.85}, and \mz{20.11} from two compact images with closely matched colours (\Sc{0.958}). For component B, $m_{grz}-m_W=1.26$, and the full conditional $1\sigma$ interval lies above the quasar-selection boundary in Fig.~\ref{fig:confirmed_colours}. The SIE fit gives \TE{1.002}.
	
	\paragraph{J0141+0039.}
	The main limitation in this \citet{Chan2023} candidate is blending of the faint compact component with the galaxy (Fig.~\ref{fig:spec_image}). DESI identifies a quasar at \zs{2.2062}, and the galaxy has \mg{22.20}, \mr{21.57}, and \mi{20.65}. The faint component's $i$-band flux has a conditional S/N of only 1.7 and is not reliably recovered, preventing a full $gri$ colour comparison. The valid $g$ and $r$ measurements give \Sc{0.914}; this restricted score, together with the quasar spectrum, galaxy detection, and component configuration, supports its inclusion in the confirmed sample, with the photometric uncertainty explicitly retained. The SIE model gives \TE{0.755}.
	
	\paragraph{J0422-0447.}
	\label{subsec:j0422_spectrum}
	The two DESI spectra of this candidate from \citet{Dawes2023} and \citet{He2023} show similar continua and absorption features (Fig.~\ref{fig:spec_image}). An absorption-masked joint QSO-template fit gives $z_s\simeq1.810$, but varying the trough masks by $8$--$16$\,\AA\ changes the result by up to $\Delta z_s\simeq0.0011$. The source redshift is therefore template- and mask-dependent. Narrow Mg\,II and Fe\,II absorption establishes $z_{\rm abs}\simeq0.775$ without identifying the deflector redshift; the galaxy-template solution at $z_{\rm fit}=0.1614$ lacks supporting features. Broad troughs suggest a possible broad absorption line (BAL) quasar interpretation \citep{Gibson2009BAL}.
	
	The fibre centres are only $0\farcs74$ apart, so blended light can contribute to the spectral similarity. An unrelated third quasar at $z_{\rm QSO}=3.3502$ is excluded from Fig.~\ref{fig:spec_image}. The imaging evidence consists of a significantly detected galaxy with $m_z=20.54$, similar compact-component colours ($S_{\rm color}=0.856$), and a plausible double-image configuration. The SIE fit gives $\theta_{\rm E}=0\farcs431$.
	
	\paragraph{J0422-1427.}
	For this candidate selected by \citet{Dawes2023} and \citet{He2023}, the DESI fibres centred on the two compact components yield consistent quasar spectra (Fig.~\ref{fig:spec_image}). A joint fit to the two spectra following the procedure in Sect.~\ref{subsec:redshift_estimation} gives a source redshift of \zs{1.7787}. A third DESI source, classified as a galaxy at $z_{\rm gal}=1.2604$, lies away from the fitted deflector centre and is treated as unrelated. Its spectrum and fibre marker are omitted from Fig.~\ref{fig:spec_image}. The detected galaxy has \mg{20.39}, \mr{19.72}, and \mz{19.21}, and the compact components have \Sc{0.739}. The SIE model gives \TE{0.536} and $q_{\rm SIE}=0.447$.
	
	\paragraph{J0642+5617.}
	Subtracting the compact components in this subarcsecond pair reveals a red galaxy with \mg{21.14}, \mr{20.61}, and \mz{20.08} (Fig.~\ref{fig:spec_image}). The system was selected by both \citet{Dawes2023} and \citet{He2023}. A single DESI fibre encompasses both images, whose blended spectrum contains broad C\,IV and C\,III] emission at \zs{1.9264}; it does not provide separate image redshifts. The deblended colours (\Sc{0.707}) and component positions support the double-image interpretation, with \TE{0.494} from the SIE fit.
	
	\paragraph{J0911-0044.}
	For this candidate from \citet{Andika2023} and \citet{Chan2023}, a single DESI fibre covers both compact images, yielding a blended quasar spectrum at \zs{2.3755} (Fig.~\ref{fig:spec_image}). The HSC decomposition reveals a galaxy with \mg{22.09}, \mr{21.43}, and \mi{20.51}, and two compact images with \Sc{0.957}. The fitted Einstein radius is \TE{0.578}.
	
	\paragraph{J0954+0323.}
	A red galaxy with \mg{23.66}, \mr{21.83}, and \mi{20.61} lies in the double-image configuration of this \citet{Andika2023} candidate (Fig.~\ref{fig:spec_image}). We fixed the deflector's S\'ersic index at $n_s=1$ in the photometric fit (Table~\ref{tab:lens_gal}). The DESI quasar redshift is \zs{2.2666}, and the HSC decomposition gives \Sc{0.719}. The inferred colour of B ($m_{grz}-m_W=3.65$) lies above the quasar-selection boundary in Fig.~\ref{fig:confirmed_colours}. As in J0047+0322, the reference disc galaxies only approximately match the deflector's optical colours. The fitted Einstein radius is \TE{0.810}.
	
	\paragraph{J1215-0058.}
	DESI obtained separate spectra of the foreground galaxy at $z_d=0.45912$ and the background quasar at $z_s=2.8817$, making J1215-0058 one of only two systems in our 15-lens sample with spectroscopic redshifts for both the deflector and source; the other is J1220+0112 (Fig.~\ref{fig:spec_image}). Selected as a candidate by \citet{Andika2023}, it was subsequently reported as confirmed by \citet{Shu2025} on the basis of these DESI redshifts and inspection of the spectra and imaging. Our analysis complements that identification with new deblended photometry, an explicit test of image-colour consistency ($S_{\rm color}=0.947$), and lens modelling. The nominal colour of B ($m_{grz}-m_W=4.83$) lies in the quasar-selection region of Fig.~\ref{fig:confirmed_colours}, although both the HSC $z$-band photometry and the central WISE image fit carry quality flags.
	
	The image separation of $4\farcs983$ is the largest among our 15 confirmed systems. The SIE model gives $\theta_{\rm E}=2\farcs489$ and $q_{\rm SIE}=0.949$, reproducing the two observed image positions. Combining this Einstein radius with the measured deflector and source redshifts gives a projected enclosed mass of $M_{\rm E}\simeq1.31\times10^{12}\,M_\odot$ within $R_{\rm E}\simeq14.9$~kpc, consistent with a massive foreground deflector. We estimate this mass as $M_{\rm E}\simeq[c^2/(4G)](D_dD_s/D_{ds})\theta_{\rm E}^2$ in the adopted cosmology, where the $D$ terms are angular-diameter distances and $\theta_{\rm E}$ is in radians.
	
	\paragraph{J1220+0112.}
	Together with J1215-0058, J1220+0112 is one of only two systems in our 15-lens sample with DESI spectroscopic detections of both the foreground galaxy and the background quasar (Fig.~\ref{fig:spec_image}). Previously selected by \citet{Andika2023} and \citet{Chan2023}, it was reported as confirmed by \citet{Shu2025}, who identified the deflector at $z_d=0.48755$ and the source at $z_s\simeq1.7081$. Our deblended photometry, image-colour comparison, and SIE modelling provide complementary tests of this interpretation. The compact images have closely matched colours (\Sc{0.926}), and the fitted Einstein radius is \TE{0.993}.
	
	The galaxy-centred fibre also substantially overlaps image B, so its spectrum contains broad quasar emission in addition to the foreground galaxy light. Following the procedure in Sect.~\ref{subsec:redshift_estimation}, we subtracted the foreground galaxy model from this spectrum and jointly fitted the residual and the independent spectrum centred on image A. The fit used the quasar signal in both spectra at a shared source redshift, yielding our adopted value of $z_s=1.7069$. Applying the same method as for J1215-0058, we estimate a projected enclosed mass of $M_{\rm E}\simeq2.56\times10^{11}\,M_\odot$ within $R_{\rm E}\simeq6.17$~kpc.
	
	\paragraph{J1307+0011.}
	Both DESI spectra identify the background quasar in this \citet{Chan2023} candidate, giving a shared redshift of \zs{2.6669} (Fig.~\ref{fig:spec_image}). The HSC decomposition yields galaxy magnitudes of \mg{23.73}, \mr{24.16}, and \mi{22.68}, together with closely matched image colours (\Sc{0.928}). The image separation of $0\farcs836$ is the smallest among our 15 confirmed systems. The compact double has an SIE Einstein radius of \TE{0.414}.
	
	\paragraph{J1531+3724.}
	In this candidate from \citet{Dawes2023}, DESI spectroscopy places the source at \zs{1.6221} (Fig.~\ref{fig:spec_image}). The DESI-LS decomposition separates a galaxy with \mg{21.28}, \mr{20.71}, and \mz{19.96} from the compact components, whose colours give \Sc{0.955}. The SIE model gives \TE{0.736}.
	
	\paragraph{J1753+7333.}
	P200 provides the only spectroscopy for this candidate from \citet{He2023}, establishing \zs{2.215} (Fig.~\ref{fig:spec_image}). The adopted DESI-LS $grz$ fit gives galaxy magnitudes of \mr{21.09} and \mz{20.71} and zero fitted $g$-band galaxy flux. The two compact components have \Sc{0.842} and form a double-image configuration around the galaxy.
	
	The fitted galaxy centre lies only $0\farcs46$ from image A. When we allowed the S\'ersic index to vary over $0.5\leq n_s\leq8$, the best $grz$, $rz$, and $z$-only fits all gave $n_s\simeq0.5$. The $grz$ half-light radius remained at its lower bound of $0\farcs015$, whereas the $rz$ fit gave $0\farcs79$ and assigned zero $r$-band flux to the galaxy. In the $z$ band alone, half-light radii of $0\farcs76$ and $1\farcs82$ yielded fits differing by only $\Delta\chi^2=1.25$. Given the blending and the sensitivity to band choice, the galaxy radius, axis ratio, and position angle in Table~\ref{tab:lens_gal} are poorly constrained.
	
	Using the fitted component positions and fluxes, the SIE solution gives \TE{0.892}, $q_{\rm SIE}=0.339$, and $\phi_{\rm SIE}=118.94^\circ$. At the adopted flux-ratio weight of 0.01, it reproduces both image positions and the $r$-band ratio $f_B/f_A=0.2043$ without additional images. A flux-ratio anomaly is therefore not required by this fit.
	
	\paragraph{J1800+5305.}
	For this candidate selected by \citet{Dawes2023} and \citet{He2023}, we combined a $3000\,\mathrm{s}$ P200/DBSP observation with a DESI spectrum of image B. Ly$\alpha$ and N\,V appear in the DBSP spectra of both components and in the DESI spectrum, although N\,V is detected at lower S/N in the DESI spectrum (Fig.~\ref{fig:spec_image}). Following the procedure in Sect.~\ref{subsec:redshift_estimation}, we jointly fitted the three spectra at a shared source redshift and obtained $z_s=3.2292$, the highest source redshift among our 15 confirmed systems. Subtracting the two PSFs reveals the deflector, with $m_g=23.50$, $m_r=22.52$, and $m_z=20.92$. The consistent source spectra and resolved galaxy light provide complementary evidence for the lens interpretation.
	
	\paragraph{J2348+1530.}
	This candidate, independently selected by \citet{Dawes2023} and \citet{He2023}, has a DESI source redshift of \zs{1.1327}, the lowest among our 15 confirmed systems (Fig.~\ref{fig:spec_image}). Its red galaxy has \mg{22.58}, \mr{21.04}, \mi{20.18}, and \mz{19.59}. The WISE fit places component B above the quasar-selection boundary in Fig.~\ref{fig:confirmed_colours} ($m_{grz}-m_W=2.99$), although residuals in the central image limit the accuracy of this colour. The two compact components agree in colour across all four bands (\Sc{0.921}), with a double-image geometry and fitted Einstein radius of \TE{0.606}.
	
	\section{Likely lensed quasars}
	\label{sec:likely_lq}
	
	The 47 likely systems retain a possible lens interpretation but lack sufficient evidence for confirmation. We first describe their catalogue (Sect.~\ref{subsec:likely_status}), then examine the imaging, source-identification, and configuration limitations defined in Sect.~\ref{subsec:likely_lq_definition}, using representative systems in Sects.~\ref{subsec:likely_typei}--\ref{subsec:likely_typeiii_iv}.
	
	\subsection{Overview}
	\label{subsec:likely_status}
	
	We compiled \texttt{likely\_catalogue.csv} to summarize the measurements and classification of all 47 likely lensed-quasar systems.\footnote{\raggedright The catalogue and the colour diagnostics for Figs.~\ref{fig:confirmed_colours} and~\ref{fig:transferred_diagnostics} are available at \mbox{\url{https://github.com/EigenHermit/Lqso_DESI_26}}.} The catalogue contains 127 records: 126 components in the 46 DESI-selected systems and one P200 target J2225+0409. Its 47 columns are organized into four groups, as detailed in Table~\ref{tab:likely_catalogue_columns}. We refitted all 47 systems with the same procedure as the confirmed sample: 34 include a S\'ersic component and 13 use PSFs only. The accompanying component-photometry and diagnostic tables give these flags, colour scores, galaxy S/N, and combined image-fit statistics. These diagnostics do not by themselves revise the classifications.
	
	Columns~1--9 provide names, coordinates, DESI target identifiers, redshifts, and spectral types. Column~10 records the Type~I--IV limitations that prevent confirmation; multiple types are listed where more than one applies. Columns~11--36 contain the available HSC and DESI-LS $griz$ photometry, WISE $W1/W2$ measurements, composite magnitudes $m_{grz}$ and $m_W$ (Sect. \ref{subsec:colour_astrometry}), and associated uncertainties and measurement notes. Missing measurements are left blank. The WISE columns retain the direct-fit measurements with their original geometry and uncertainty conventions; values based on a superseded component inventory are withheld. The AGN- and galaxy-constrained colours shown in Figs.~\ref{fig:confirmed_colours} and~\ref{fig:transferred_diagnostics} are supplied separately in \texttt{colour\_prediction\_diagnostics.csv}. Columns~37--47 provide Gaia astrometry, including parallax, proper motion, and indicators of astrometric quality and image structure.
	
\begin{figure*}[!p]
		\centering
		\includegraphics[width=0.985\linewidth]{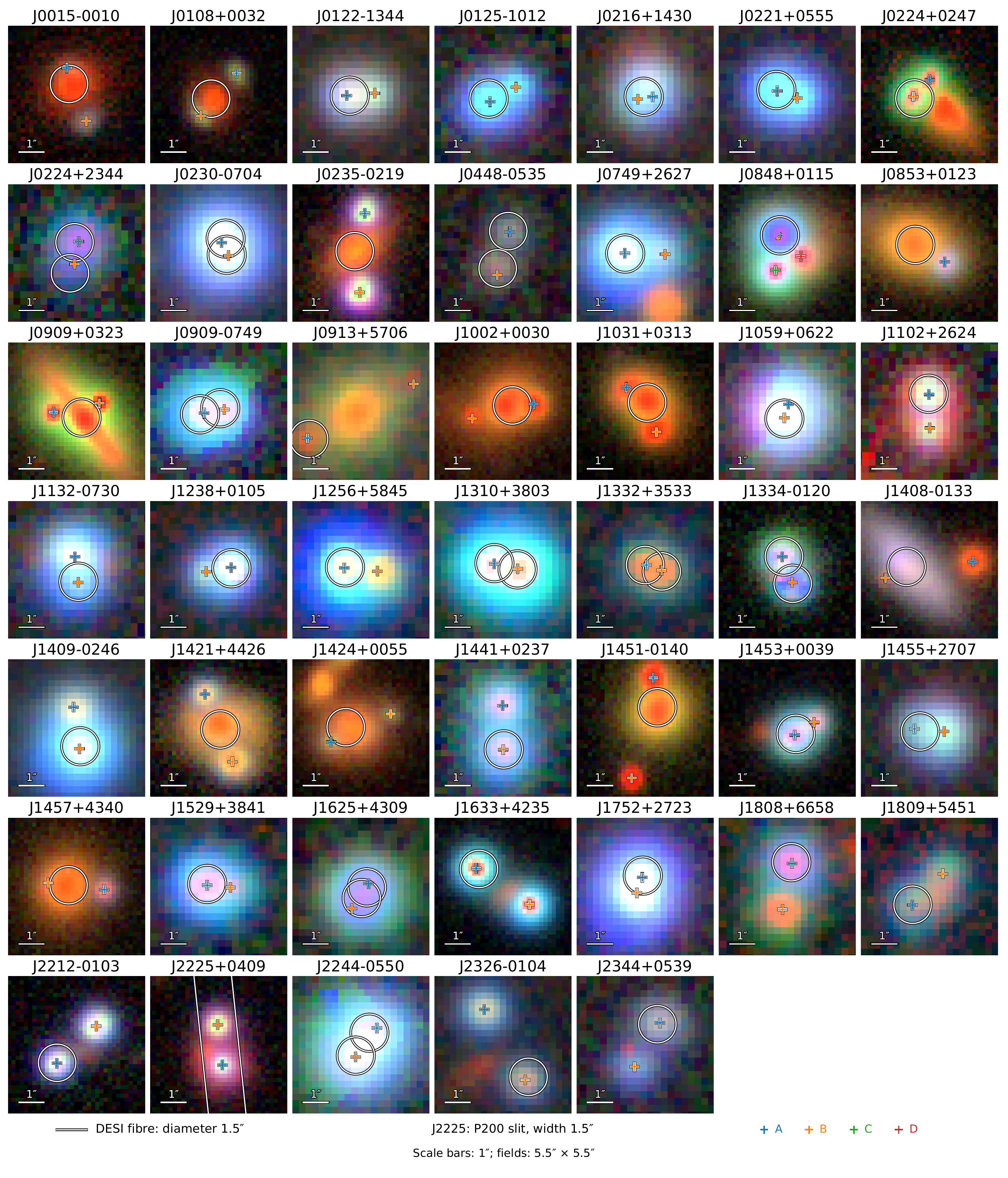}
		\caption{Cutouts of the 47 likely systems, ordered by name, with $5.5\arcsec\times5.5\arcsec$ fields and $1\arcsec$ scale bars. Outlined white circles mark DESI fibres and matching lines mark the schematic P200 slit. Blue, orange, green, and red crosses mark compact components A, B, C, and D, respectively.}
		\label{fig:likely_cutouts}
	\end{figure*}

	Incomplete imaging evidence (Type~I) and uncertain background-source identity (Type~II) affect 46.8\% and 42.6\% of the sample, respectively. Types~III and IV account for one and five systems and are discussed together. Figure~\ref{fig:likely_cutouts} shows all 47 systems, and Fig.~\ref{fig:likely_examples} presents spectra and image decompositions for selected examples discussed below.
	
	\subsection{Incomplete imaging evidence}
	\label{subsec:likely_typei}
	
	Type~I includes 22 systems: 21 have this limitation alone and one also has a Type~II limitation. The available images do not securely establish the deflector or multiple-image structure. The four examples below illustrate the difficulty of detecting a deflector in compact double-image candidates.
	
	\paragraph{J0448-0535.}
	The two DESI spectra in Fig.~\ref{fig:likely_examples} identify quasars at $z_{\rm QSO}=0.97356$ and $0.97497$. The compact components have similar colours ($S_{\rm color}=0.870$). However, the fitted extended component lies close to B and has similar colours, raising the possibility that it is the host galaxy of B rather than a foreground deflector. A S\'ersic component in the image decomposition therefore does not by itself establish the presence of a lens galaxy. Higher-resolution imaging and spectroscopy of the extended light are needed to distinguish these interpretations.
	
	\paragraph{J1059+0622.}
	Gaia resolves two sources separated by $0\farcs592$, with $G=17.95$ and $17.47$; the brighter source has \texttt{ipd\_frac\_multi\_peak}$=27$, consistent with an unresolved or blended image structure \citep{Holl2023Gaia}. We fitted the DESI-LS images shown in Fig.~\ref{fig:likely_examples} with two PSFs, initializing their centres at the Gaia positions and allowing each to move by up to $0\farcs3$. The fitted separation is $0\farcs572$, and the component colours are similar ($S_{\rm color}=0.923$). Both centres lie within the DESI aperture, whose spectrum identifies a quasar at $z_{\rm QSO}=2.19524$. This blended spectrum establishes the presence of quasar emission but does not identify both components independently. Neither Gaia source has a proper-motion or parallax measurement, so astrometry cannot exclude a stellar companion.
	
	The two-source morphology and similar optical colours support a double-image lens interpretation, but no deflector is securely detected. If the two components instead form a DQ at the measured redshift, their fitted angular separation corresponds to a projected physical separation of approximately $4.84$~kpc. Under the lens hypothesis, matching the fitted separation and source redshift to the mock catalogue of \citet{Oguri2010}, within $0\farcs1$ and $\Delta z_s=0.1$, yields 164 systems with a median deflector magnitude of $m_i=23.0$ and a 16th--84th percentile range of 22.1--23.2. Deeper, higher-resolution imaging is needed to search for such a faint deflector, while spatially resolved spectroscopy would test whether both compact components are quasar images.
	
	\paragraph{J1238+0105.}
	A DESI spectrum identifies a quasar at $z_{\rm QSO}=3.13905$ (Fig.~\ref{fig:likely_examples}). The two compact components, separated by $1\farcs009$, have similar DESI-LS colours ($S_{\rm color}=0.936$), but the adopted decomposition yields no significant deflector detection. Without a measured galaxy component, we do not apply the prediction-constrained WISE test. We retain the Type~I classification; deeper, higher-resolution imaging is needed to search for the deflector.
	
	\paragraph{J1752+2723.}
	Gaia resolves two sources separated by $0\farcs704$, with $G=18.65$ and $17.43$ and ${\rm PMSIG}=0.50$ and 4.27, respectively (Fig.~\ref{fig:likely_examples}). Neither exceeds the adopted proper-motion rejection threshold. The brighter source, associated with B, has \texttt{ipd\_frac\_multi\_peak}$=30$, consistent with blending \citep{Holl2023Gaia}. Using the same Gaia-initialized procedure as for J1059+0622, we modelled the DESI-LS images with two PSFs and allowed their centres to move by up to $0\farcs3$. The fitted separation is $0\farcs659$, and the optical colours satisfy the adopted similarity criterion ($S_{\rm color}=0.850$). The local image-fit diagnostic is $\chi^2_{\nu,\rm img}=60.93$ in the common $5.5\arcsec$ cutout (Eq.~\ref{eq:image_fit_diagnostic}), indicating substantial residual structure despite the similar colours.
	
	A single DESI spectrum identifies a quasar at $z_{\rm QSO}=2.05403$, but blends the two compact components. Together, the imaging and astrometry support a possible double-image configuration without securely establishing a deflector or independent quasar identities. If the two components instead form a DQ at the measured redshift, their fitted angular separation corresponds to a projected physical separation of approximately $5.64$~kpc. Applying the same mock-catalogue matching as for J1059+0622 yields 197 systems with a median deflector magnitude of $m_i=22.6$ and a 16th--84th percentile range of 22.0--22.9. Higher-resolution imaging would test for faint extended light between the compact sources; separate spectra would determine whether they share the same quasar spectrum.
	
	\subsection{Uncertain background-source identity}
	\label{subsec:likely_typeii}
	
	Type~II includes 20 systems: 19 have this limitation alone and one also has a Type~I limitation. Two recurring limitations are possible stellar contamination and spectra dominated by foreground galaxy light.
	
	\subsubsection{Possible stellar contamination}
\begin{figure}[!tp]
		\centering
		\includegraphics[width=0.9\linewidth]{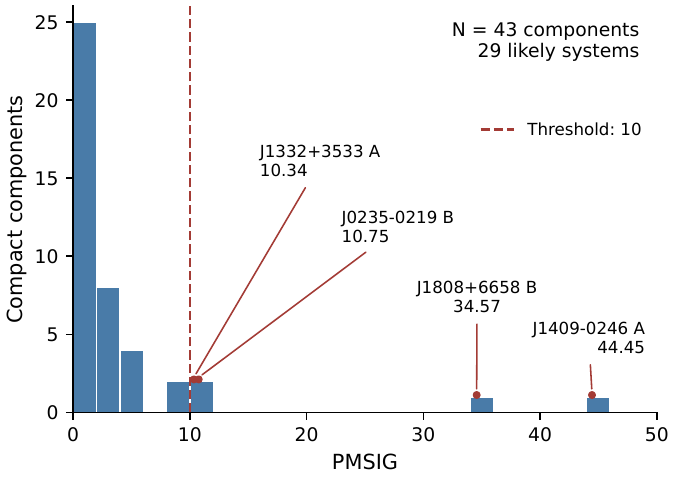}
		\caption{Distribution of Gaia PMSIG values for 43 compact components in 29 likely systems. The dashed line marks ${\rm PMSIG}=10$; the four components above it are labelled.}
		\label{fig:likely_pmsig}
	\end{figure}

\begin{figure*}[!t]
		\centering
		\includegraphics[width=0.83\textwidth]{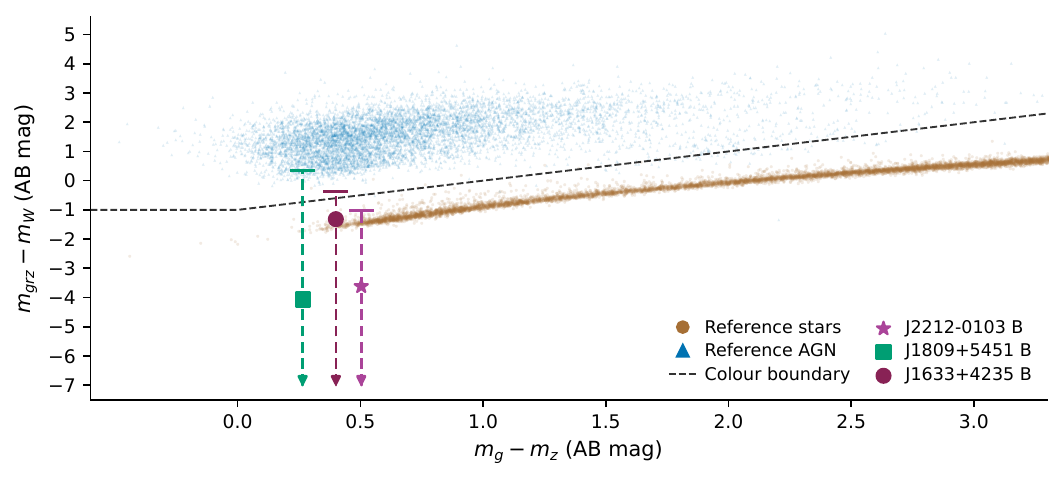}
		\caption{The same colour diagnostic as Fig.~\ref{fig:confirmed_colours} for component B in J2212-0103, J1809+5451, and J1633+4235. Symbols show fits at the median predicted AGN and galaxy fluxes. Upper caps and dashed downward arrows show the conditional $1\sigma$ intervals obtained by propagating both prediction uncertainties. All three lower bounds are $-\infty$; arrow lengths are graphical. The interval for J2212-0103 B lies below the quasar boundary, whereas those for the other two components cross it.}
		\label{fig:transferred_diagnostics}
	\end{figure*}

	Ten of the 20 Type~II systems show indications of possible stellar contamination. The optical--infrared colours flag four systems, and Gaia proper motions flag four others (Fig.~\ref{fig:likely_pmsig}). In the remaining two, J0122-1344 and J2225+0409, the blended spectra leave open the possibility that one compact component is a foreground star. Figure~\ref{fig:transferred_diagnostics} shows three of the four systems flagged by colour. We discuss these three and the two with blended spectra below.
	
	In J1809+5451 and J1633+4235, the conditional \(1\sigma\) colour interval for component B crosses the quasar-selection boundary, leaving its identity unresolved. In J2212-0103, the entire interval for B lies below the boundary, although a mismatch between the galaxy and its template limits this inference. A lensed-quasar interpretation is less plausible for the four systems flagged by Gaia. Each contains a compact component whose proper-motion significance exceeds our conservative threshold of \({\rm PMSIG}=10\), with measured values of 10.34, 10.75, 34.57, and 44.45.
	
	\paragraph{J0122-1344.}
	Broad C\,III] and Mg\,II emission establishes the presence of a quasar, but the identity of B remains uncertain (Fig.~\ref{fig:likely_examples}). The P200 spectrum analysed by \citet{He2025-AA} blended A, B, and the intervening galaxy. Although the DESI fibre is centred near A, the fitted A--B separation of $1\farcs127$, together with atmospheric seeing and the $1\farcs5$ fibre diameter, allows light from B and the galaxy to contribute to the spectrum. The DESI spectrum broadly resembles the earlier P200 spectrum and does not resolve the ambiguity identified in that work: B could be a star projected near a quasar and a galaxy, or a second image of the quasar. Spectroscopy that isolates B is therefore needed to distinguish these interpretations.
	
	\paragraph{J1633+4235.}
	DESI identifies A as a quasar at $z_{\rm QSO}=2.2928$ in this \citet{Chan2023} candidate, and the HSC decomposition detects a galaxy and compact components with similar optical colours ($S_{\rm color}=0.891$; Fig.~\ref{fig:likely_examples}). However, B's conditional optical--infrared colour interval crosses the quasar-selection boundary and does not securely establish an AGN identity. A stellar companion remains possible despite the plausible lens configuration.
	
	\paragraph{J1809+5451.}
	The DESI spectrum centred on A shows C\,III] and Mg\,II at $z_{\rm QSO}=1.1669$ in this candidate from \citet{Dawes2023} and \citet{He2023}. The detected galaxy and similar optical component colours ($S_{\rm color}=0.947$) support a possible lens configuration (Fig.~\ref{fig:likely_examples}), but B's conditional colour interval crosses the quasar-selection boundary, leaving its identity unresolved. Component-resolved spectroscopy is needed to distinguish a quasar image from a projected star in both systems.
	
	\paragraph{J2212-0103.}
	A DESI spectrum identifies A as a quasar at $z_{\rm QSO}=2.16980$, while B remains without spectroscopy (Fig.~\ref{fig:likely_examples}). For B, $m_g-m_z=0.504$ and the conditional $1\sigma$ colour range extends from $-\infty$ to $m_{grz}-m_W=-1.023$, below the quasar-selection boundary (Fig.~\ref{fig:transferred_diagnostics}). This is compatible with stellar contamination, although the galaxy-template mismatch limits a definitive interpretation.
	
	\paragraph{J2225+0409.}
	The broad emission feature in the blended P200 spectrum may be C\,III] (Fig.~\ref{fig:likely_examples}). The source redshift $z_s\simeq1.534$ remains tentative because Mg\,II is not securely detected at its expected wavelength near 7100\,\AA. The continuum declines and then rises towards longer wavelengths, unlike a typical quasar continuum. As in J0122-1344, this leaves two plausible interpretations: a lensed quasar blended with deflector light, or a projected combination of a quasar, a galaxy, and a star. The available colour and proper-motion diagnostics do not flag stellar contamination, but incomplete or blended component measurements prevent its exclusion. Spatially resolved spectroscopy of the compact sources is needed to establish their identities and distinguish these scenarios.
	
	\subsubsection{Unidentified background sources in foreground-centred spectra}
	In ten Type~II systems the DESI spectrum is assigned to the putative deflector, and the background source lacks a secure spectroscopic identification. One of these ten also carries a stellar-contamination flag. We discuss three examples below.
	
	\paragraph{J0108+0032.}
	Figure~\ref{fig:likely_examples} shows a prominent galaxy beside two compact components with similar $griz$ colours ($S_{\rm color}=0.940$). The DESI spectrum measures the galaxy redshift, $z_{\rm gal}=0.90188$, but leaves the compact-source identities unresolved. Spectroscopy centred on these components is needed to establish whether they are images of a background quasar.
	
	\paragraph{J1002+0030.}
	The main emission lines can be attributed to an AGN in the putative deflector at $z_{\rm gal}\simeq0.573$, so they do not establish the presence of a more distant source (Fig.~\ref{fig:likely_examples}). Component-resolved spectroscopy is needed to determine the nature and redshifts of the compact sources and test whether the foreground AGN host acts as a lens.
	
	\paragraph{J1421+4426.}
	The DESI fibre is centred on the galaxy at $z_{\rm gal}=0.46169$ (Fig.~\ref{fig:likely_examples}). A and B lie $1\farcs54$ and $1\farcs39$ from the fibre centre, respectively, outside its $0\farcs75$ radius. Their deblended HSC $griz$ colours are similar ($S_{\rm color}=0.782$). Separate spectra are needed to test whether both are images of the same background quasar.

	\subsection{Inconclusive image consistency and atypical lens configurations}
	\label{subsec:likely_typeiii_iv}
	
	Type~III contains one system (2.1\%), J0913+5706, whose image colours are discrepant. Five systems (10.6\%) have Type~IV limitations: J0224+0247, J0848+0115, J1451-0140, J1453+0039, and J2326-0104. Their source--deflector configurations remain uncertain, and all five carry Type~IV alone. We discuss these systems below.
	
	\paragraph{J0224+0247.}
	The DESI pipeline classified the spectrum as a galaxy at $z_{\rm DESI}=0.74899$. We refitted it with QSO templates and obtained $z_{\rm QSO}\simeq3.4664$, consistent with broad Ly$\alpha$ and C\,III] emission near 5430 and 8510\,\AA\ (Fig.~\ref{fig:likely_examples}). This identifies quasar emission in the blended spectrum but does not establish the nature of each compact component.
	
	The HSC decomposition contains two compact components and a clearly detected extended component G, but their relative positions are atypical of a simple double-image lens. Adding a third point source does not appreciably improve the image fit, and subtraction of the extended light reveals no clear fourth compact component. The only Gaia source lies near B, with $G=20.66$, ${\rm PMSIG}=0.93$, and no reported multi-peak windows. These measurements do not resolve the geometrical ambiguity. We therefore retain the system provisionally pending higher-resolution imaging and component-resolved spectroscopy.
	
	\paragraph{J0848+0115.}
	This system was listed as a contaminant by \citet{Chan2023}, who noted a quasar--galaxy--star interpretation. Our four-point-source decomposition can approximate the observed light distribution (Fig.~\ref{fig:likely_examples}), but an exploratory SIE fit favours a mass centre displaced from the fitted galaxy-light centre. The single DESI spectrum covering the close A/B pair identifies a quasar at $z_{\rm QSO}=0.64603$; it does not establish separate redshifts for A and B. Under the lens hypothesis, matching this redshift and the maximum image separation of approximately $1\farcs44$ to the mock catalogue of \citet{Oguri2010} gives a characteristic deflector magnitude of $i\simeq18.4$. The low source redshift alone does not exclude lensing, but the uncertain component identities and mass--light alignment leave the non-lens interpretation favoured. We retain the system provisionally because the available observations do not independently identify all compact components.
	
	\paragraph{J0913+5706.}
	A DESI spectrum identifies a quasar at $z_{\rm QSO}=2.08130$, but the $grz$ colour score of $S_{\rm color}=0.637$ falls below the adopted threshold of 0.7 (Fig.~\ref{fig:likely_examples}). A spectrum of the other compact component is needed to determine whether the colour difference occurs between images of the same quasar or between unrelated sources.
	
	\paragraph{J1451-0140.}
	DESI measures the prominent galaxy at $z_{\rm gal}=0.34690$, but neither compact component has a secure AGN identification (Fig.~\ref{fig:likely_examples}). The prediction-constrained WISE test therefore lacks a spectroscopic-quasar reference and is not applied. The available data do not establish a background source behind the galaxy. We retain the system as Type~IV pending spectroscopy of the compact components to test the source--deflector interpretation.
	
	\paragraph{J1453+0039.}
	A DESI spectrum centred on A identifies a quasar at $z_{\rm QSO}=1.09414$, while B remains without an independent spectrum (Fig.~\ref{fig:likely_examples}). The adopted $gri$ decomposition gives similar compact-component colours ($S_{\rm color}=0.791$) and reveals a yellow extended component between A and B, closer to A. Because the extended component lies close to A, it could instead be the host galaxy of A, leaving its identification as a foreground deflector uncertain. The relative component positions and fluxes are not readily explained by a simple double-image configuration. A neighbouring red galaxy could contribute to a more complex lens potential if it lies in the foreground; \citet{Chan2023} also noted the displaced galaxy position and possible influence of a nearby galaxy. The Gaia source assigned to A has ${\rm PMSIG}=3.17$, but provides no independent constraint on B. A spectrum of B would test whether the two compact sources share a redshift and spectral features. Agreement would strengthen the lens interpretation, which would then require a model consistent with the observed geometry and any contribution from the neighbouring galaxy.
	
	\paragraph{J2326-0104.}
	The two compact sources are separated by $3\farcs27$, but the extended light visible after PSF subtraction lies to one side of the line joining A and B (Fig.~\ref{fig:likely_examples}). The fitted centre of G is $1\farcs31$ from this line, giving an opening angle $\angle{\rm AGB}\simeq101^\circ$. This departure from the near alignment expected for a simple double centred on G motivates the Type~IV classification. The fitted centre of G reaches its allowed positional limit, so the offset should be checked with higher-resolution imaging. DESI identifies B as a quasar at $z_{\rm QSO}=2.53842$. A spectrum of A would test whether it shares this redshift.
	
	\begin{systemfigurepage}
		\includegraphics[width=0.96\linewidth,trim=0 30bp 0 0,clip]{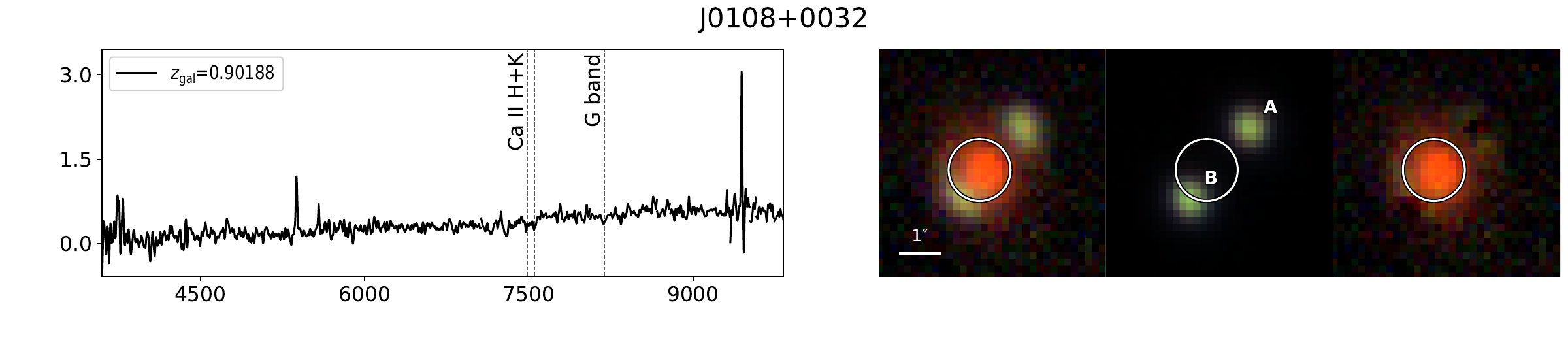}\par
		\vspace{2mm}
		\includegraphics[width=0.96\linewidth,trim=0 30bp 0 0,clip]{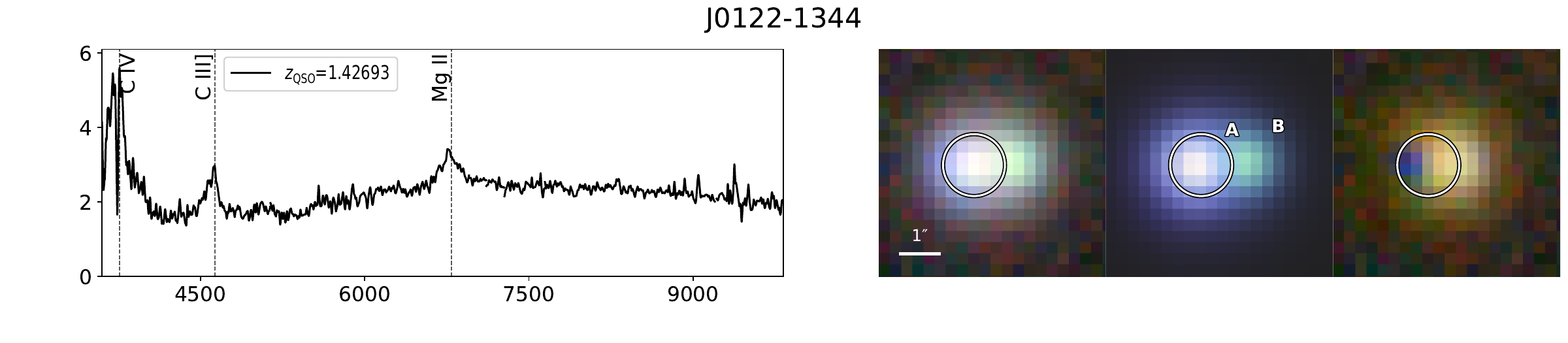}\par
		\vspace{2mm}
		\includegraphics[width=0.96\linewidth,trim=0 30bp 0 0,clip]{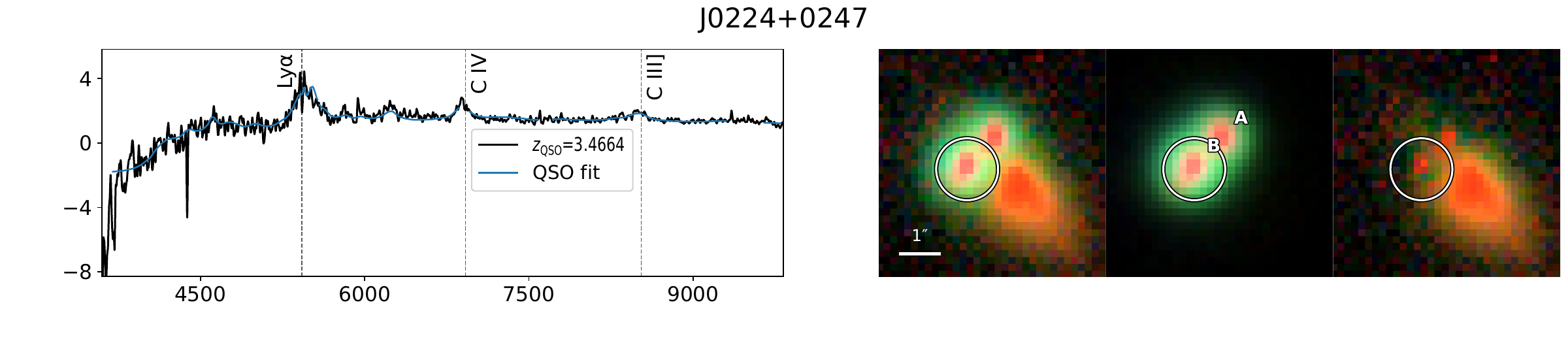}\par
		\vspace{2mm}
		\includegraphics[width=0.96\linewidth,trim=0 30bp 0 0,clip]{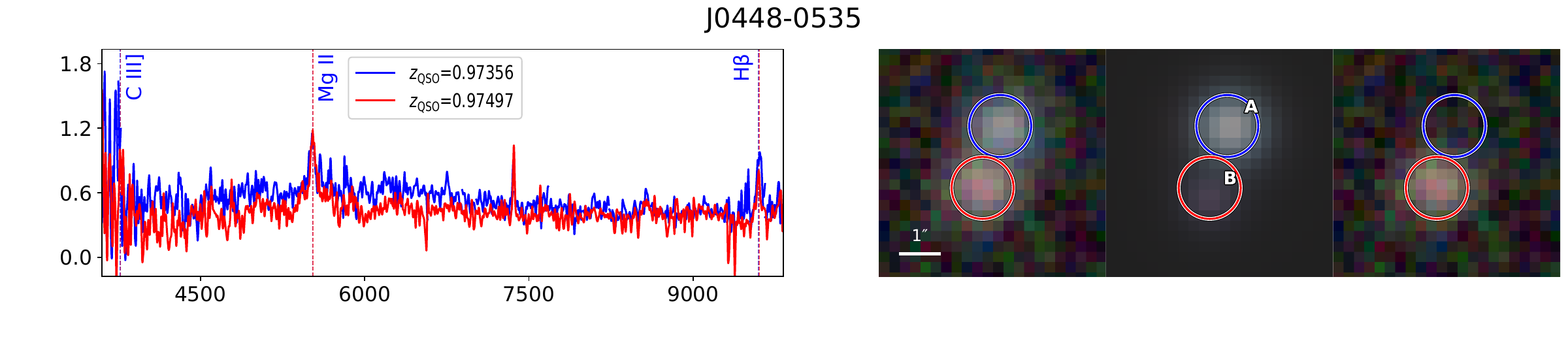}\par
		\vspace{2mm}
		\includegraphics[width=0.96\linewidth,trim=0 30bp 0 0,clip]{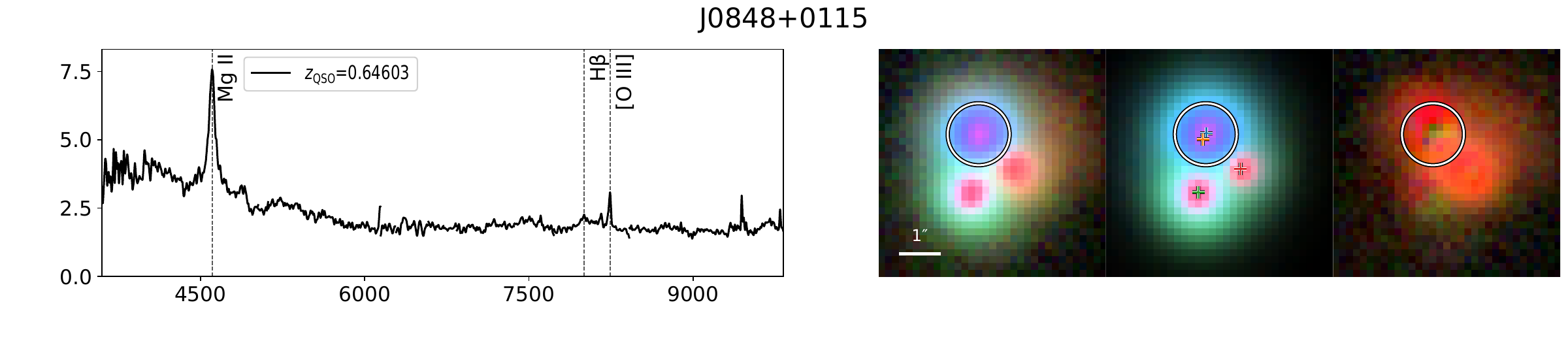}\par
		\vspace{2mm}
		\includegraphics[width=0.96\linewidth,trim=0 30bp 0 0,clip]{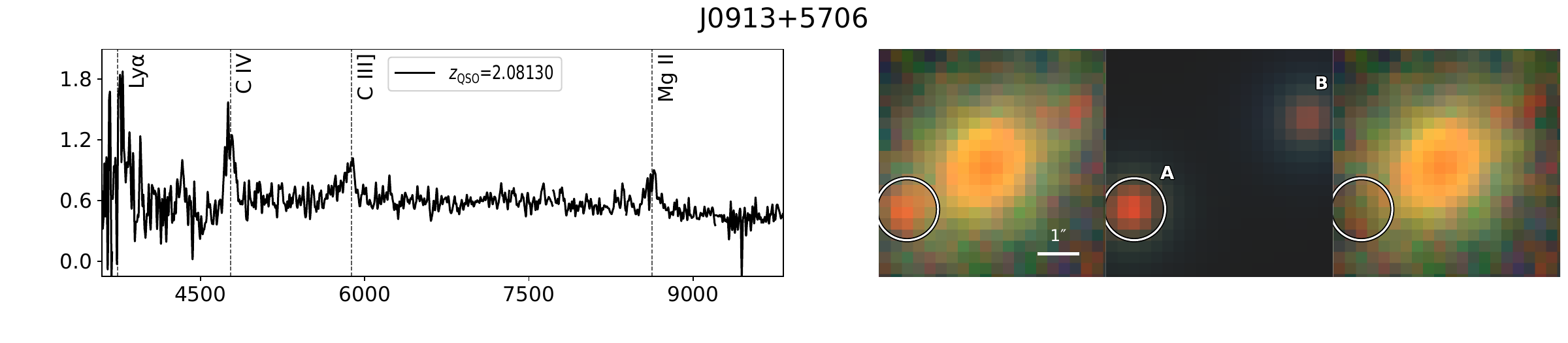}\par
		\caption{Spectra and optical-image decompositions of the 18 likely systems discussed individually in Sect.~\ref{sec:likely_lq}. Each row shows the spectrum or spectra, the optical data, the PSF model, and the data after PSF subtraction. Image fields span $5.5\arcsec\times5.5\arcsec$, with $1\arcsec$ scale bars. Circles mark DESI fibres and parallel lines mark the schematic P200 slit, as in Fig.~\ref{fig:spec_image}. The circled cross in the P200 spectrum of J2225+0409 marks absorption by molecular oxygen in Earth's atmosphere near 7600\,\AA.}
		\label{fig:likely_examples}
	\end{systemfigurepage}
	
	\begin{systemfigurepage}
		\ContinuedFloat
		\includegraphics[width=0.96\linewidth,trim=0 30bp 0 0,clip]{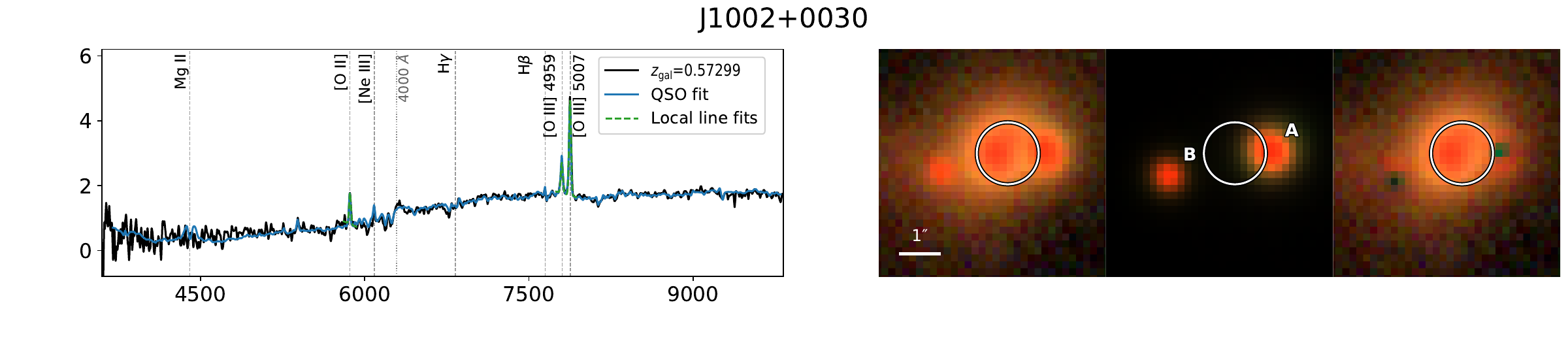}\par
		\includegraphics[width=0.96\linewidth,trim=0 30bp 0 0,clip]{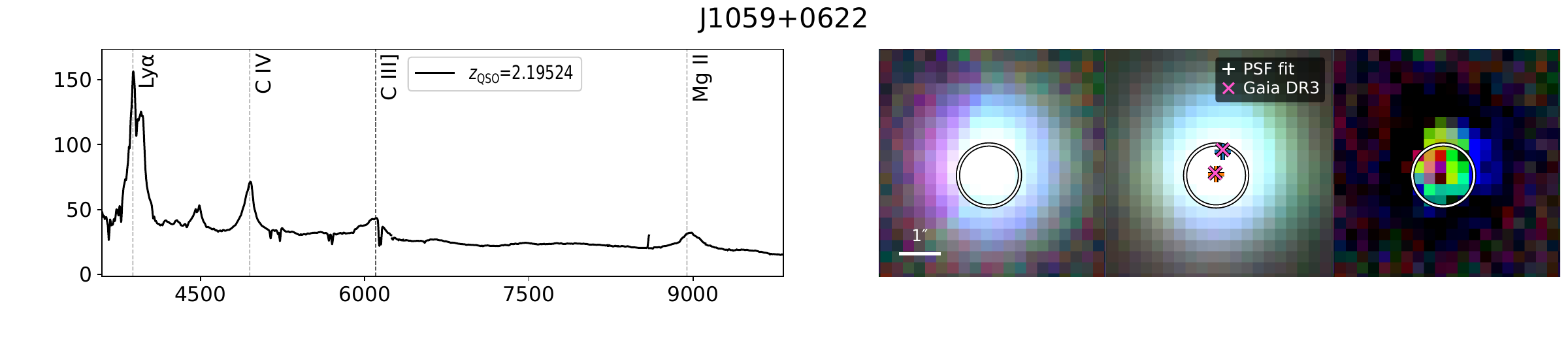}\par
		\includegraphics[width=0.96\linewidth,trim=0 30bp 0 0,clip]{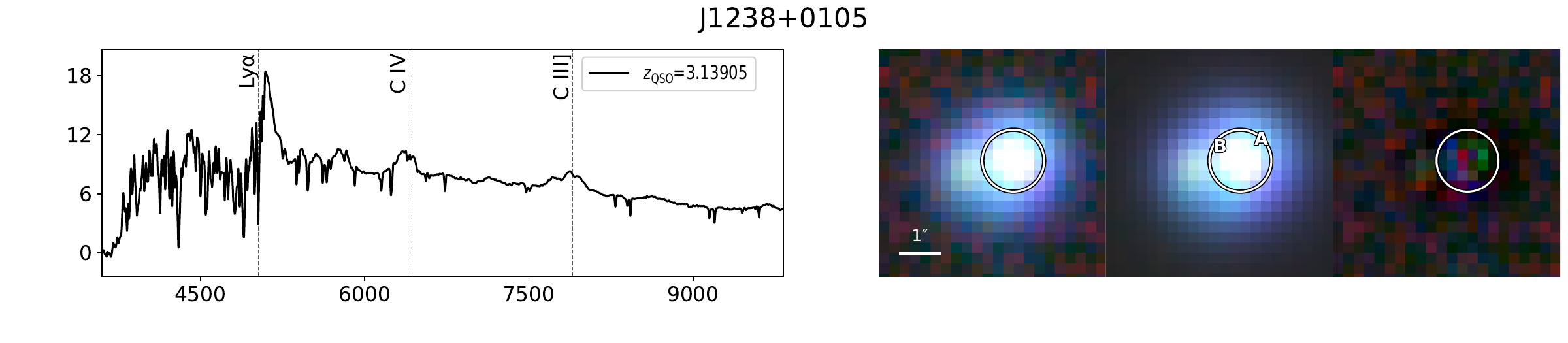}\par
		\includegraphics[width=0.96\linewidth,trim=0 30bp 0 0,clip]{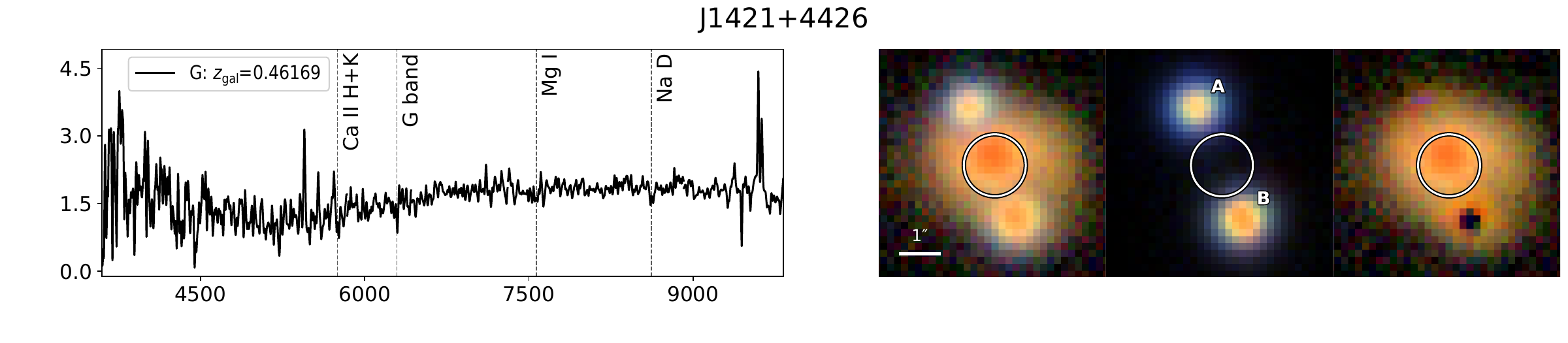}\par
		\includegraphics[width=0.96\linewidth,trim=0 30bp 0 0,clip]{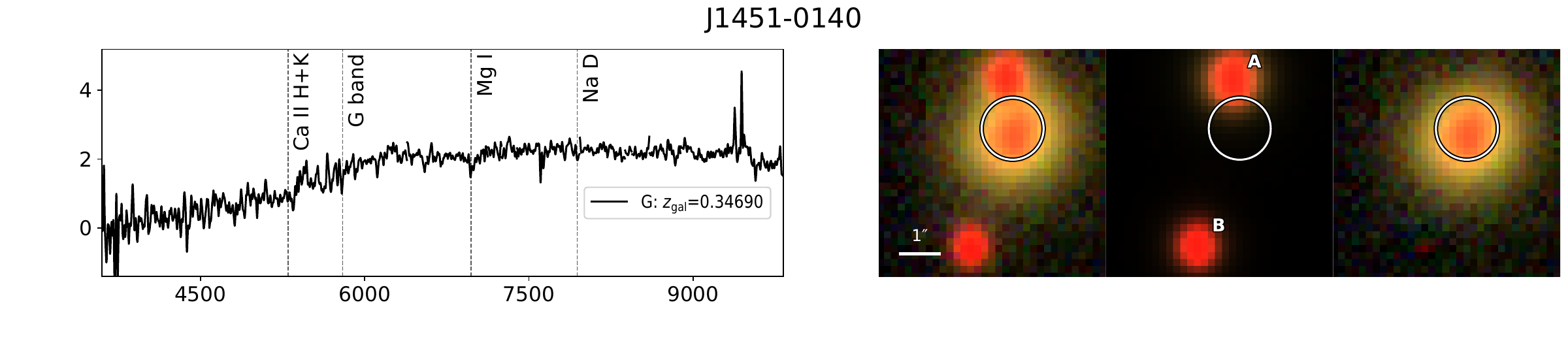}\par
		\includegraphics[width=0.96\linewidth,trim=0 30bp 0 0,clip]{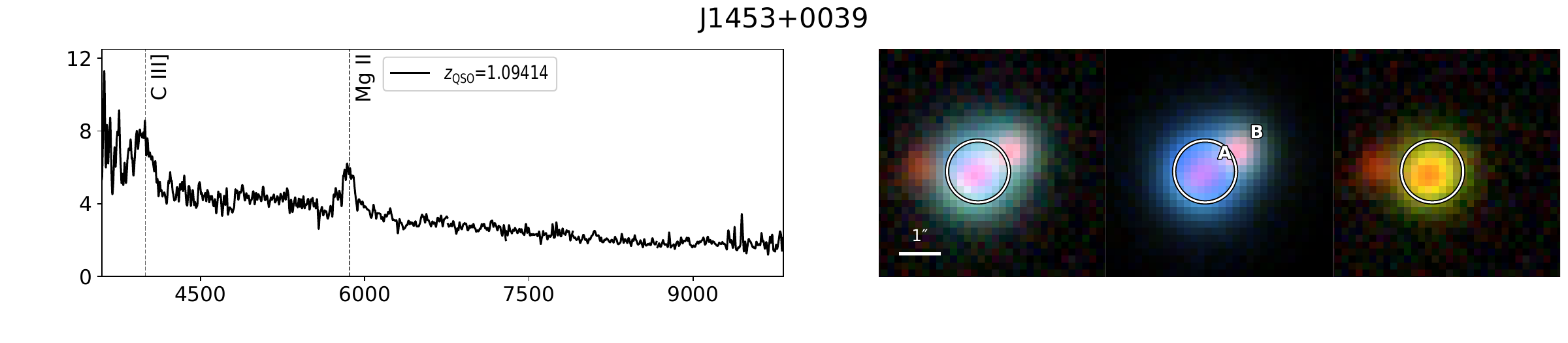}\par
		\includegraphics[width=0.96\linewidth,trim=0 30bp 0 0,clip]{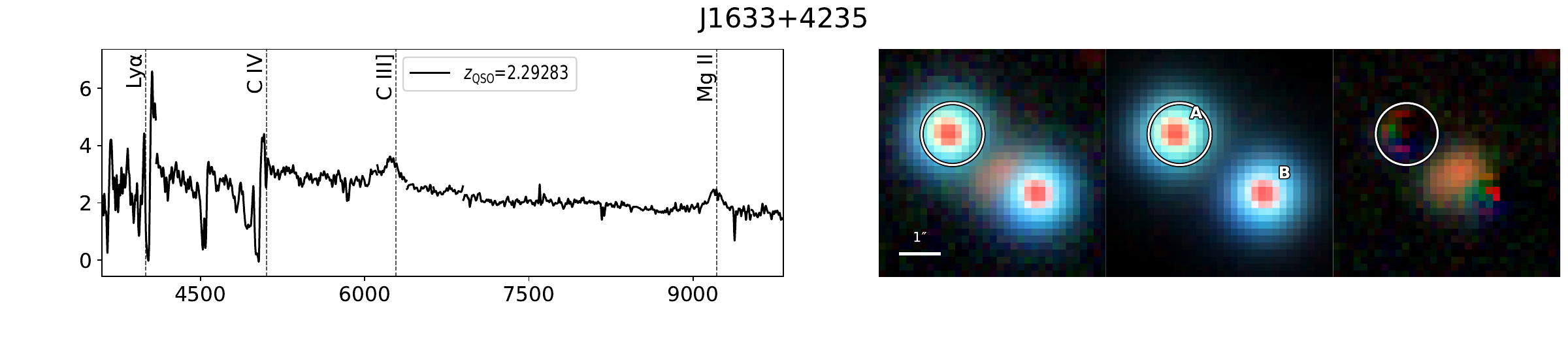}\par
		\caption{Continued.}
	\end{systemfigurepage}
	
	\begin{systemfigurepage}
		\ContinuedFloat
		\includegraphics[width=0.96\linewidth,trim=0 30bp 0 0,clip]{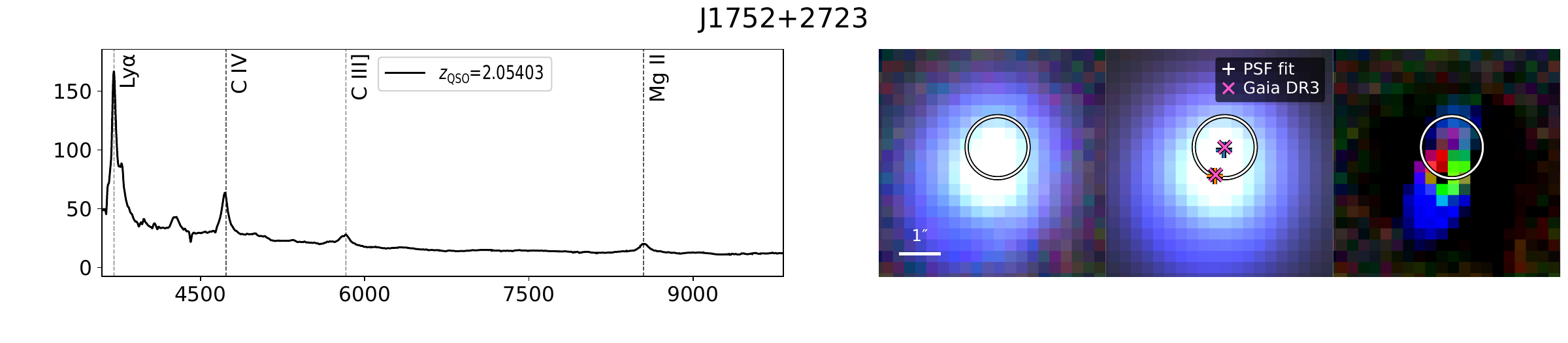}\par
		\vspace{2mm}
		\includegraphics[width=0.96\linewidth,trim=0 30bp 0 0,clip]{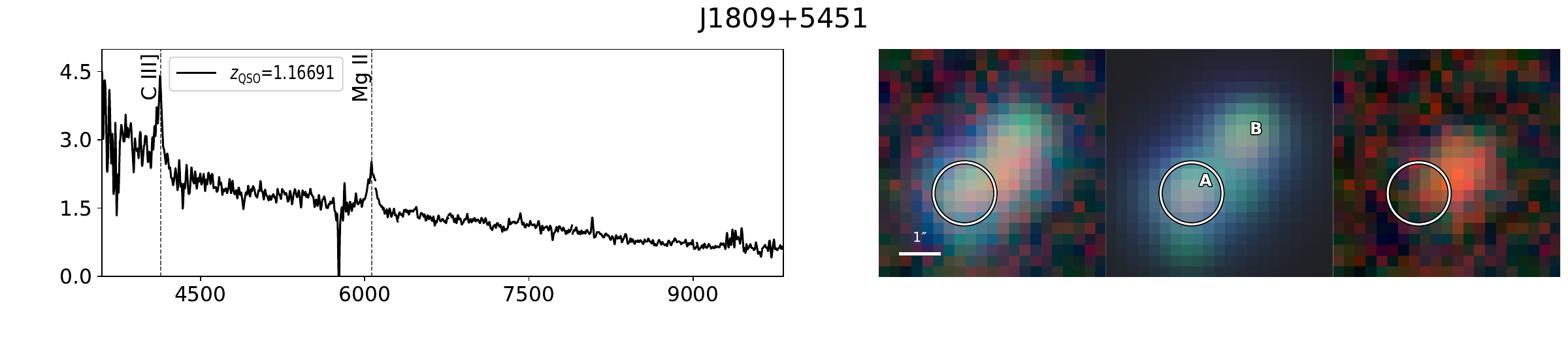}\par
		\vspace{2mm}
		\includegraphics[width=0.96\linewidth,trim=0 30bp 0 0,clip]{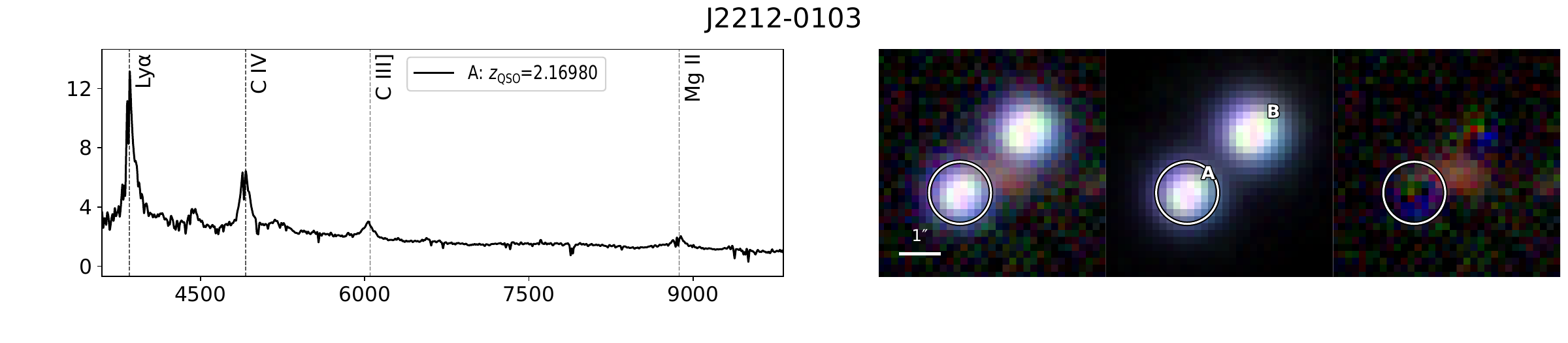}\par
		\vspace{2mm}
		\includegraphics[width=0.96\linewidth,trim=0 30bp 0 0,clip]{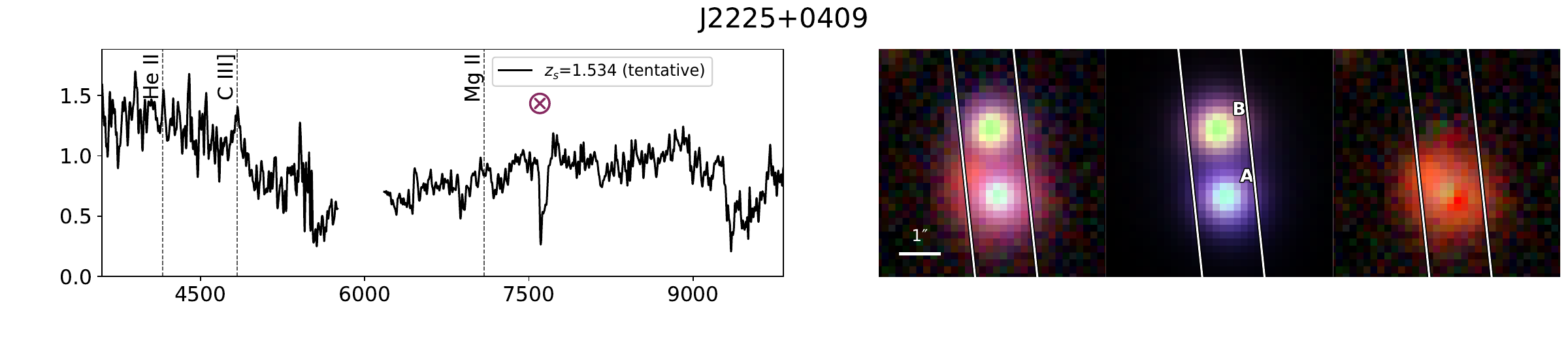}\par
		\vspace{2mm}
		\includegraphics[width=0.96\linewidth,trim=0 30bp 0 0,clip]{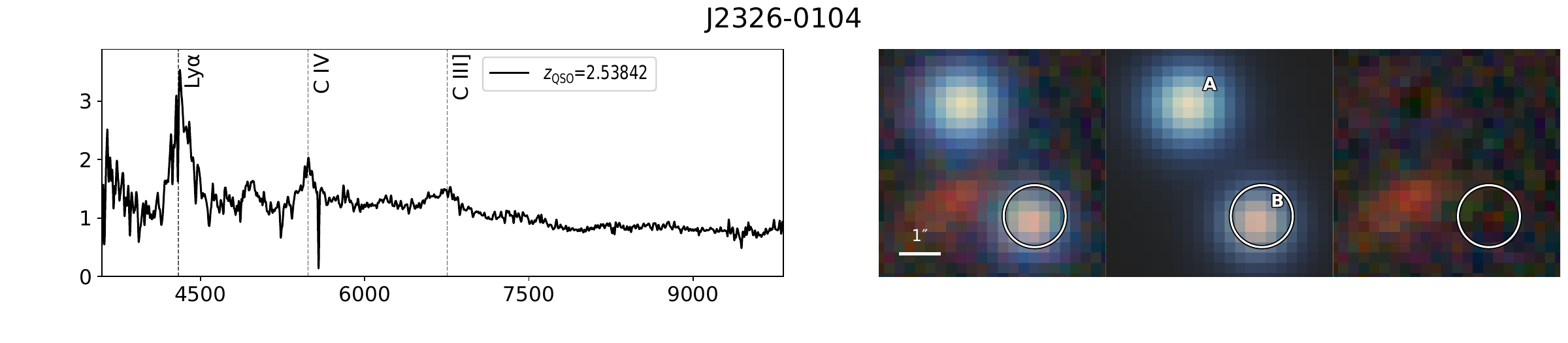}\par
		\caption{Continued.}
		\label{fig:j2225_spectrum_image}
	\end{systemfigurepage}
	
	\begin{table*}[!t]
		\centering
		\caption{Contents of \texttt{likely\_catalogue.csv}, organized into four groups. The catalogue contains 47 columns and 127 records covering 47 likely systems.}
		\label{tab:likely_catalogue_columns}
		\begin{threeparttable}
			\small
			\setlength{\tabcolsep}{5pt}
			\renewcommand{\arraystretch}{1.15}
			\begin{tabular}{@{}>{\raggedright\arraybackslash}p{0.18\textwidth}>{\raggedright\arraybackslash}p{0.30\textwidth}>{\raggedright\arraybackslash}p{0.47\textwidth}@{}}
				\toprule
				Group (columns) & Fields & Description \\
				\midrule
				Basic information (1--9) & \texttt{name}, \texttt{system}, \texttt{component}, \texttt{sample}, \texttt{RA\_deg}, \texttt{DEC\_deg}, \texttt{TARGETID}, \texttt{z\_DESI}, \texttt{DESI\_spectype} & Source and component names, DESI/P200 sample membership, J2000 coordinates in degrees, and the assigned DESI target identifier, pipeline redshift, and spectral type. \\
				Classification (10) & \texttt{lens\_type} & Type~I--IV limitations of the parent system. Multiple types are separated by semicolons, for example \texttt{I;II}. \\
				Photometry (11--36) & \texttt{HSC\_*}, \texttt{DESILS\_*}, \texttt{W1\_*}, \texttt{W2\_*}, \texttt{grz\_AB}, \texttt{W\_AB}, \texttt{WISE\_uncertainty}, \texttt{WISE\_note} & Native HSC and DESI-LS $griz$ magnitudes and errors, WISE $W1/W2$ magnitudes with lower and upper errors, and composite magnitudes $m_{grz}$ and $m_W$. The WISE error convention and necessary measurement cautions are retained. \\
				Astrometry (37--47) & \texttt{gaia\_source\_id}, \texttt{parallax}, \texttt{pmra}, \texttt{pmdec}, their \texttt{\_err} fields, \texttt{PMSIG}, \texttt{RUWE}, \texttt{astrometric\_excess\_noise}, \texttt{ipd\_frac\_multi\_peak} & Gaia identifier, parallax and proper motion with uncertainties, PMSIG, RUWE, astrometric excess noise, and the percentage of multi-peak image windows. Parallax and excess noise are in mas; proper motions are in mas\,yr$^{-1}$. \\
				\bottomrule
			\end{tabular}
			
		\end{threeparttable}
	\end{table*}


	\begin{figure*}
		\centering
		\includegraphics[width=\linewidth]{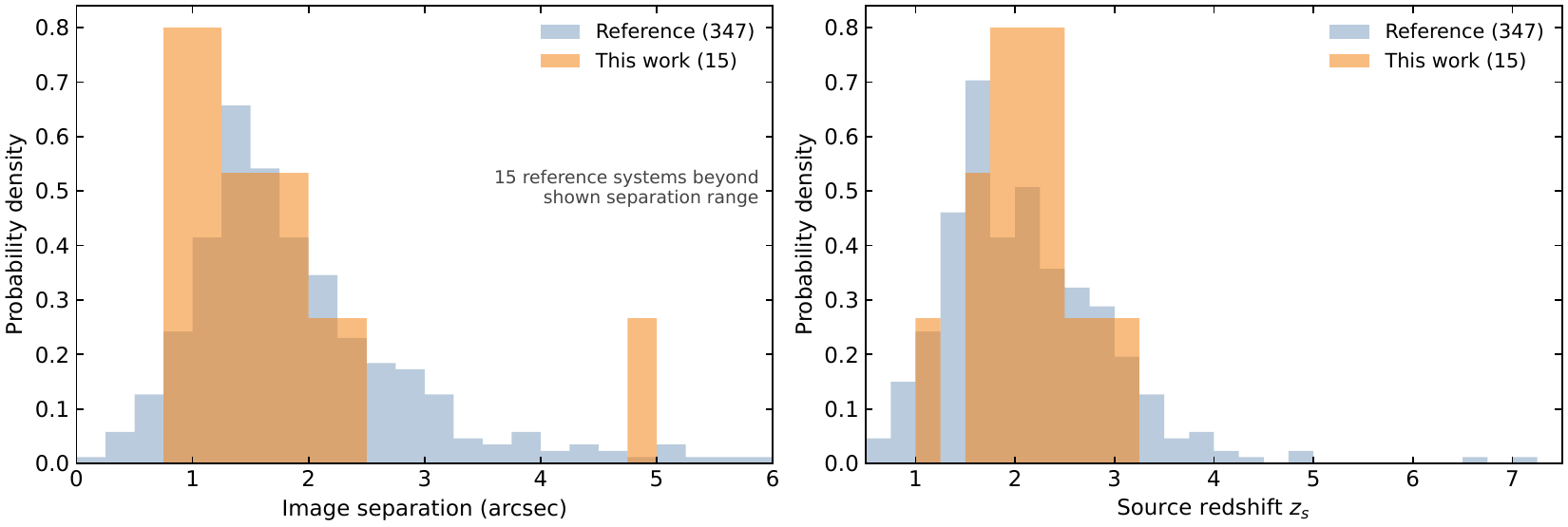}
		\caption{Image-separation and source-redshift distributions of the 15 confirmed lenses and the literature comparison sample. Histograms are normalized by the valid counts in the legends, including 15 reference systems with separations above the displayed $6\arcsec$ range.}
		\label{fig:compare_with_known_sample}
	\end{figure*}
	
	\raggedbottom
	\section{Discussion and conclusion}
	\label{sec:diss}
	
	We discuss the contributions of the confirmation workflow and lens samples, compare the outcomes of the input selection methods, and outline how imaging and forthcoming spectroscopy can extend this work.
	
	\subsection{Main contribution of this work}
	\label{subsec:discussion_small_separation}
	
	The first contribution is a practical workflow for confirming lensed quasars with general-purpose spectroscopic surveys. By associating fibre apertures with resolved image components, we combine the available spectra with image decomposition, deflector detection, and colour and astrometric checks. This makes use of survey spectra even when they cover only one image or blend several components, and identifies the observations needed to resolve the remaining ambiguity. The workflow is scalable to larger spectroscopic samples and can be adapted to the MUltiplexed Survey Telescope \citep[MUST,][]{Cai2025MUST}, the Prime Focus Spectrograph \citep[PFS,][]{Tamura2016HSCPFS}, and the 4-metre Multi-Object Spectroscopic Telescope \citep[4MOST,][]{deJong2019}, with component associations and quality criteria adjusted to each instrument's aperture and resolution.
	
	The second contribution is a sample of 15 confirmed lensed quasars with homogeneous deblended photometry and SIE models, together with 47 likely systems for further investigation. The confirmed sample is 4.3\% of the size of the 347 literature systems with measured image separations in Fig.~\ref{fig:compare_with_known_sample}. Twelve confirmed systems have image separations below $2\arcsec$, including three below $1\arcsec$. Small-separation lenses provide access to lower-mass or higher-redshift deflectors and their central mass distributions. Recovering them also addresses an important source of incompleteness in ground-based searches, where blending obscures the multiple-image structure \citep{Lemon2017SmallSep,Yue2022Mock}.
	
	Six confirmed systems have source redshifts $z_s<2$, including J2348+1530 at $z_s=1.133$. At the lower-redshift end, reduced surface-brightness dimming and lensing magnification favour resolved studies of quasar-host structure and molecular-gas kinematics, as demonstrated for RX~J1131-1231 \citep{Paraficz2018}. Combining reconstructed host properties with broad-line black-hole mass estimates can test the evolution of black-hole--host relations \citep{Ding2017}.
	
	J0422$-$0447 is a promising target for studying quasar outflows. For one of its two DESI spectra, the \href{https://data.desi.lbl.gov/doc/releases/dr1/vac/zlya/}{DESI DR1 Ly$\alpha$ quasar catalogue} \citep{Martini2025BAL} reports a C\,IV balnicity index of $461~\mathrm{km\,s^{-1}}$, supporting the BAL interpretation in Sect.~\ref{subsec:j0422_spectrum}. The DESI fibre centres are only $0\farcs74$ apart, so each spectrum can contain light from both images. Monitoring the images separately would allow us to test whether changes in trough depth at a given velocity recur in the second image after the lensing time delay, as expected for intrinsic variability of the outflow \citep{Misawa2018BAL}. If stellar microlensing magnifies the compact continuum more strongly than the broad-line region, comparing the two spectra could also reveal at which velocities the wind absorbs the continuum alone and at which it also absorbs broad-line emission \citep{ODowd2015BAL}.
	
	A third contribution concerns galaxy--galaxy lenses. Within the literature and catalogues checked, we obtain four newly completed spectroscopic deflector--source redshift pairs and the first spectroscopic deflector redshifts for 16 further systems. J0224$-$0336 and J1338$-$0109 are particularly promising: compact concentrations of neighbouring galaxies suggest possible group environments, while secure deflector and source redshifts provide a basis for lens modelling (Fig.~\ref{fig:lgal_new_pairs}). If these associations are physical, their lensing configurations could constrain how mass is partitioned between individual galaxies and a common group halo, and how the surrounding environment perturbs the lens potential \citep{Moller2002Groups,Limousin2010Group}. Spectroscopy of the remaining neighbours and resolved modelling of the lensed emission are needed to establish membership and measure these contributions.
	
	\subsection{Evaluation of different selection methods}
	\label{subsec:discussion_census}
	
	Table~\ref{tab:catalogue_selection_efficiency} summarizes the classification results, DESI DR1 spectroscopic coverage, and mean image separations of confirmed systems for the five input catalogues. We use the confirmed-lens and stellar-contamination fractions to characterize their yields and contamination, respectively; the confirmed samples include known reference lenses. These statistics apply to the classified DESI DR1 subsets and are affected by spectroscopic coverage and DESI target selection. They therefore do not fully represent the parent samples, but still provide a useful reference for understanding differences among the methods.
	
	The five studies can be grouped into three categories according to their main candidate-search procedure. The first comprises neural-network-based image searches, represented by \citet[][H25]{He2025-Apj} and \citet[][A23]{Andika2023}. Despite their similar approaches, they select different candidate populations: the lensed-galaxy fraction in A23 is much higher than the 4.1\% found for H25. This difference may partly reflect how the training samples were constructed. A23 modelled background quasars with Gaussian light profiles before applying lensing and PSF convolution. The authors noted that some simulated multiple images could resemble extended arcs, potentially causing the networks to select lensed galaxies. H25 instead constructed simulated images from multiple PSs, more directly targeting the multiple-point-source morphology of lensed quasars. These differences suggest that source morphology and the source-to-deflector brightness contrast in the training set may influence the networks' selection preferences.
	
	The second category searches for multiple images in point-source catalogues, as in \citet[][H23]{He2023} and \citet[][D23]{Dawes2023}. Both use the DESI-LS catalogue to identify potential lenses through the spatial proximity and colours of quasar candidates. Their main difference lies in the parent sample. H23 adopted a broader quasar-candidate catalogue without a hard magnitude cut, whereas D23 selected candidates satisfying $m_r<22.7$. Their candidate catalogues contain 971 and 436 systems, of which 310 and 159, respectively, receive classifications in this study. H23 has a slightly higher confirmed fraction than D23 (12.9\% versus 10.7\%), but also higher stellar contamination (12.9\% versus 6.3\%). These results are consistent with a trade-off between retaining more candidates through a broader parent sample and reducing contamination through stricter preselection. Both searches rely on the catalogue identifying individual images. Faint or blended images, or images misclassified as ESs, may therefore cause systems to be missed.
	
	The third category is the analytic image search used by \citet[][C23]{Chan2023}. This method constructs a sharpening operator from the Laplacians of the image and PSF to improve the separation of neighbouring PSs. It then detects, measures, and subtracts the PSs and examines the residuals for a foreground galaxy. The detection procedure has a relatively transparent physical interpretation and does not require simulated lenses to train the detection operator. It also depends less on whether the original catalogue has already deblended the images. This provides a complementary route to finding compact systems, although the results remain sensitive to the PSF, noise, and point-source subtraction residuals. In our sample, C23 has a confirmed fraction of 10.6\% and a stellar-contamination fraction of 13.6\%, both within the ranges found for the other methods.
	
	Overall, the confirmed lensed-quasar fractions are close to 10\% across all five studies (8.2--12.9\% for H23, D23, C23, and H25, and 10.3\% for A23 after excluding lensed galaxies). Thus, once lensed galaxies are distinguished, the methods have similar confirmed fractions, and the present results do not establish a clear ranking of their performance. The candidate populations and stellar-contamination fractions show more pronounced differences.
	
	H25 has the highest stellar-contamination fraction (24.7\%). Some stellar configurations resemble lensed quasars in the $gri$ images and survive H25's photometric and morphological preselection. The lower contamination fractions in D23 and A23 are consistent with their additional screening criteria. D23 applies a magnitude cut and stringent optical--infrared selection to its parent sample, while A23 uses multiband photometry and Gaia astrometry. Excluding lensed galaxies from the A23 sample raises its stellar-contamination fraction from 1.4\% to 5.1\%, close to the 6.3\% found for D23. These values suggest that residual stellar contamination can persist at around 5\%, although whether this represents a practical limit for colour--colour cuts and astrometric checks remains uncertain. Differences in contamination among the studies may depend more on how candidates enter the parent sample and which additional checks they undergo than on the choice of search method (neural networks, catalogue searches, or analytic algorithms). The comparison highlights the complementary roles of morphology, colour, and astrometry in candidate selection.
	
	\begin{table*}[t]
		\centering
		\caption{Classification fractions, DESI DR1 spectroscopic coverage, and mean confirmed-lens image separation by input catalogue and selection method. Coverage is $N_{\rm DR1}/N_{\rm all}$: the fraction of the adopted input systems with at least one DR1 spectrum within $6\arcsec$, before classification exclusions. Parent sizes refer to the lists used in CS (grades A--C for \citealt{Chan2023}). The five classification percentages use the labelled subset $N$. Mean image separations use exactly the confirmed samples, including reference lenses, and are arithmetic means of the maximum pairwise separations of fitted images. The $\dagger$ row excludes the 108 lensed galaxies from the \citet{Andika2023} sample ($N=39$); coverage is not repeated, and the lensed-galaxy fraction is omitted for this subset.}
		\label{tab:catalogue_selection_efficiency}
		\small
		\setlength{\tabcolsep}{1pt}
		\renewcommand{\arraystretch}{1.12}
		\begin{tabular*}{\textwidth}{@{\extracolsep{\fill}}llrrrrrrrr@{}}
			\toprule
			Reference & Method & \makecell[r]{Labelled\\$N$} & \makecell[r]{DR1 coverage\\$N_{\rm DR1}/N_{\rm all}$ (\%)} & \makecell[r]{Confirmed\\(\%)} & \makecell[r]{Confirmed\\+ likely\\(\%)} & \makecell[r]{DQ\\(\%)} & \makecell[r]{Star\\Cont.\\(\%)} & \makecell[r]{Lensed\\galaxies\\(\%)} & \makecell[r]{Mean image\\separation\\(arcsec)} \\
			\midrule
			\makecell[l]{\citeauthor{He2023}\\(\citeyear{He2023})} & Catalog-based & 310 & $329/971$ (33.9) & 12.9 & 20.3 & 16.1 & 12.9 & 1.3 & 1.55 \\
			\makecell[l]{\citeauthor{Dawes2023}\\(\citeyear{Dawes2023})} & Catalog-based & 159 & $166/436$ (38.1) & 10.7 & 21.4 & 22.0 & 6.3 & 0.6 & 1.70 \\
			\makecell[l]{\citeauthor{Chan2023}\\(\citeyear{Chan2023})} & \makecell[l]{Analytic\\image search} & 66 & $69/131$ (52.7) & 10.6 & 21.2 & 6.1 & 13.6 & 15.2 & 1.46 \\
			\makecell[l]{\citeauthor{Andika2023}\\(\citeyear{Andika2023})} & \makecell[l]{Neural-network\\image search} & 147 & $150/242$ (62.0) & 2.7 & 10.9 & 0.7 & 1.4 & 73.5 & 2.46 \\
			\makecell[l]{\citeauthor{Andika2023}\\(\citeyear{Andika2023})$^{\dagger}$} & \makecell[l]{Neural-network\\image search} & 39 & -- & 10.3 & 41.0 & 2.6 & 5.1 & -- & 2.46 \\
			\makecell[l]{\citeauthor{He2025-Apj}\\(\citeyear{He2025-Apj})} & \makecell[l]{Neural-network\\image search} & 97 & $105/272$ (38.6) & 8.2 & 13.4 & 5.2 & 24.7 & 4.1 & 1.50 \\
			\bottomrule
		\end{tabular*}
	\end{table*}
	
	\begin{figure*}[t]
		\centering
		\includegraphics[width=\textwidth]{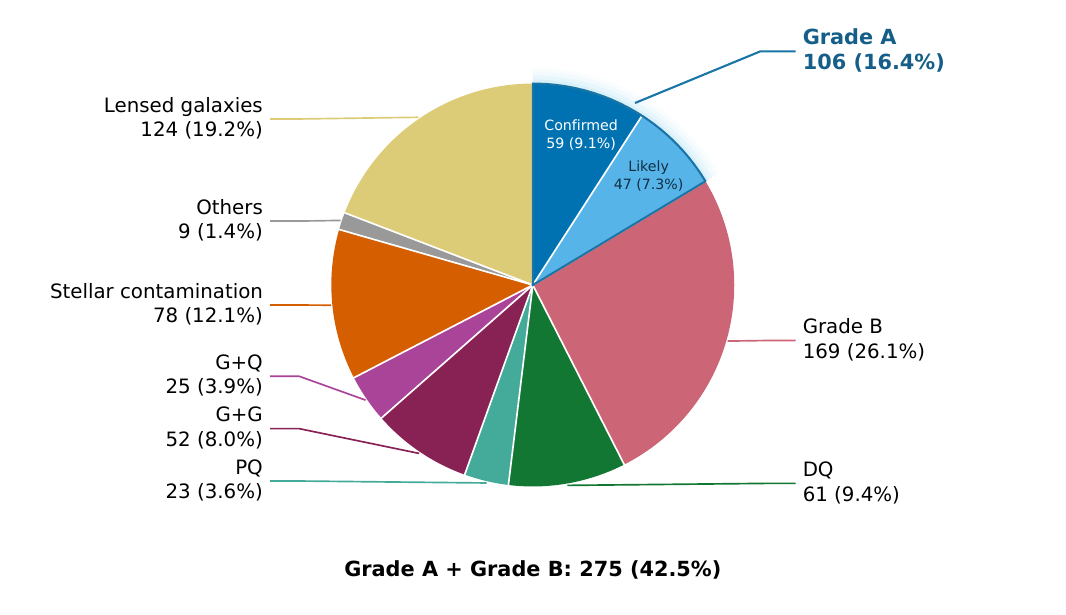}
		\caption{Classification census of 647 labelled systems, with percentages relative to this total; the 34 systems that remain unclassified are excluded. The highlighted Grade~A sector combines confirmed and likely lenses.}
		\label{fig:classification_census}
	\end{figure*}
	
	\subsection{Future work}
	\label{subsec:discussion_future_searches}
	
	The discussion in Sect.~\ref{subsec:discussion_census} highlights the different strengths of existing search methods, which future searches could combine. Once trained, machine-learning models can rapidly process large numbers of images, making them suitable for initial candidate selection in large surveys. In our evaluated subsets, H25 and A23 achieve confirmed lensed-quasar fractions comparable to those of the other methods, supporting their effectiveness in identifying candidates. However, their selection effects are difficult to quantify, and their results may be highly sensitive to the properties of the training samples. The background-source morphology, image separation, and source-to-deflector brightness contrast in the training set may all influence which systems a network preferentially selects. This may partly explain the high lensed-galaxy fraction in A23. Future work should therefore improve recognition performance while also establishing which systems the networks preferentially recover and which they may miss.
	
	One direction worth exploring is to combine rapid machine-learning selection with image analysis that has a clear physical interpretation. For example, a neural network could first select candidates with multiple-image morphology. Following the approach of C23, image sharpening, point-source fitting, and subtraction could then be used to examine individual images and foreground galaxies in the residuals. Deblended colours and Gaia astrometry could subsequently help identify stellar contaminants and distinguish lensed quasars, quasar pairs, and lensed galaxies.
	
	Such a combined search must also balance contamination control against sample completeness. The comparison in Sect.~\ref{subsec:discussion_census} is consistent with lower stellar contamination under stricter parent-sample preselection. However, such cuts may exclude faint or blended images, or images not identified as PSs in the catalogue. Catalogue searches could therefore be complemented by direct image searches for compact multiple-image systems. To quantify selection effects at each stage, simulated systems spanning image separation, flux ratio, source redshift, and deflector brightness could be injected into real survey images. Their recovery rates could then be measured through successive selection stages, with additional validation using independent known lenses and real contaminant samples. This would help assess the completeness cost of reducing contamination and test whether the combined approach recovers systems missed by individual methods.
	
	Beyond candidate searches, spectroscopic confirmation remains essential for expanding reliable samples. As subsequent DESI releases provide wider spectroscopic coverage and more repeat observations, our workflow can be applied to additional candidates and to reassess the likely systems. Future imaging releases from Euclid and the Chinese Space Station Survey Telescope (CSST, see \citealp{Wei2026} and the references therein) will also provide data to support lens-galaxy detection in systems that require it. Combining automated searches, survey-spectroscopy checks, and targeted follow-up observations could expand the lensed-quasar sample while improving our understanding of its contamination and selection effects.
	
	\subsection{Conclusion}
	\label{subsec:conclusion}
	
	Our joint analysis of DESI DR1 and P200/DBSP spectroscopy identifies 15 confirmed lensed quasars and 47 likely systems, with deblended photometry and SIE models for the confirmed sample. Twelve confirmed systems have image separations below $2\arcsec$, and six have source redshifts below two. The galaxy--galaxy lens analysis additionally supplies four newly completed deflector--source redshift pairs and first spectroscopic deflector redshifts for 16 systems, within the literature and catalogues checked.
	
	The methodological contribution is a transferable workflow that combines fibre-level spectral identifications with resolved imaging and complementary contamination checks. It makes existing survey spectra useful for lens confirmation while defining the follow-up required for ambiguous systems. Extending this approach to future DESI releases and surveys with MUST, PFS, and 4MOST will connect increasingly large imaging candidate lists to spectroscopic evidence. Wide-field imaging forecasts already anticipate thousands of discoverable lensed quasars \citep{Oguri2010,Yue2022Mock}; efficient selection and scalable confirmation offer a route to establishing such samples in a near future.
	
	\begin{acknowledgements}
		
		Z.H. acknowledges support from the National Natural Science Foundation of China (Grant No.~12403104). Q.T. acknowledges support from the Jiangxi Provincial Natural Science Foundation (Grant No. 20242BAB26012). This research uses data obtained through the Telescope Access Program (TAP), which has been funded by the TAP association, including the Centre for Astronomical Mega-Science of the Chinese Academy of Sciences, Xiamen University, Peking University, Tsinghua University, the University of Science and Technology of China, Nanjing University, Yunnan University, and Sun Yat-sen University. We thank Y. Shu for insightful discussions. \\

		This project used data obtained with DECam, which was constructed by the Dark Energy Survey (DES) collaboration. Funding for the DES Projects has been provided by the U.S. Department of Energy, the U.S. National Science Foundation (NSF), the Ministry of Science and Education of Spain, the Science and Technology Facilities Council of the United Kingdom, the Higher Education Funding Council for England, the National Center for Supercomputing Applications at the University of Illinois at Urbana-Champaign, the Kavli Institute of Cosmological Physics at the University of Chicago, Center for Cosmology and Astro-Particle Physics at the Ohio State University, the Mitchell Institute for Fundamental Physics and Astronomy at Texas A\&M University, Financiadora de Estudos e Projetos, Fundacao Carlos Chagas Filho de Amparo, Financiadora de Estudos e Projetos, Fundacao Carlos Chagas Filho de Amparo a Pesquisa do Estado do Rio de Janeiro, Conselho Nacional de Desenvolvimento Cientifico e Tecnologico and the Ministerio da Ciencia, Tecnologia e Inovacao, the Deutsche Forschungsgemeinschaft and the Collaborating Institutions in DES. The Collaborating Institutions are Argonne National Laboratory, the University of California at Santa Cruz, the University of Cambridge, Centro de Investigaciones Energeticas, Medioambientales y Tecnologicas-Madrid, the University of Chicago, University College London, the DES-Brazil Consortium, the University of Edinburgh, the Eidgenossische Technische Hochschule (ETH) Zurich, Fermi National Accelerator Laboratory, the University of Illinois at Urbana-Champaign, the Institut de Ciencies de l’Espai (affiliated with the Institut d'Estudis Espacials de Catalunya and the Consejo Superior de Investigaciones Cient\'ificas), the Institut de Fisica d’Altes Energies, Lawrence Berkeley National Laboratory, the Ludwig Maximilians Universitat Munchen and the associated Excellence Cluster Universe, the University of Michigan, NSF’s NOIRLab, the University of Nottingham, the Ohio State University, the University of Pennsylvania, the University of Portsmouth, SLAC National Accelerator Laboratory, Stanford University, the University of Sussex, and Texas A\&M University.
		
	\end{acknowledgements}
	
	%
	%
	
	\bibliographystyle{aa}
	\bibliography{cite-database}

	\begin{appendix} 
		\raggedbottom
		
		\section{P200 targets classified as non-lensed systems}
		\label{app:P200_other}
		
\begin{figure*}
			\centering
			\includegraphics[width=0.95\textwidth,height=0.87\textheight,keepaspectratio]{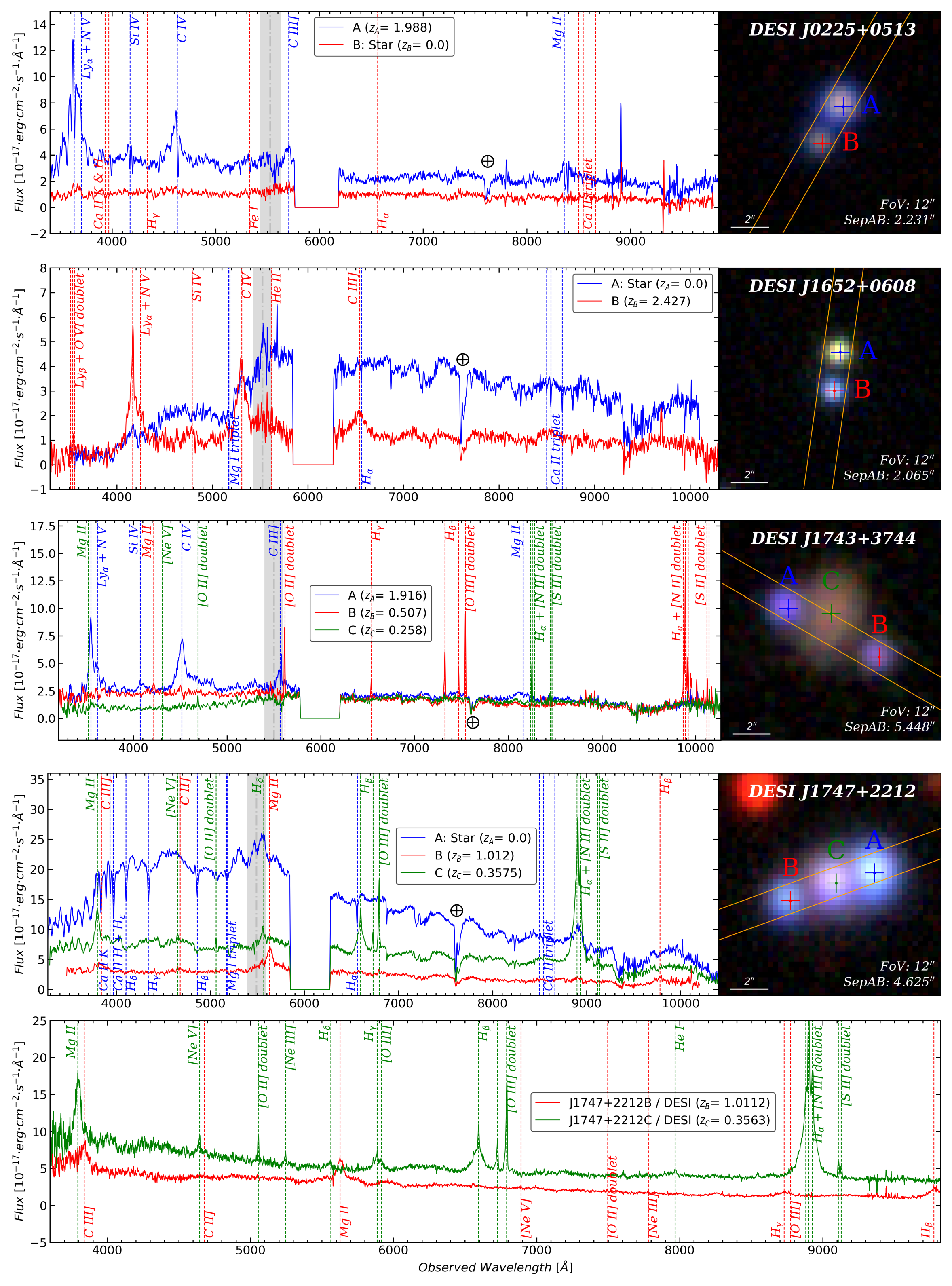}
			\caption{P200/DBSP spectra and images of J0225+0513, J1652+0608, J1743+3744, and J1747+2212 (top four rows, in order). Spectral colours match the labelled image components. The first two systems are Q+S alignments; J1743+3744 is a G+Q system, and J1747+2212 is a PQ with an additional foreground star (Table~\ref{tab:p200}). The bottom panel shows the independent DESI spectra of quasar components B and C in J1747+2212.}
			\label{fig:p200_nonlenses}
		\end{figure*}

		Of the seven targets in the P200/DBSP sample analysed here, two are classified as confirmed lensed quasars (Sect.~\ref{sec:confirmed_lq}) and one, J2225+0409, as a likely lensed quasar (Sect.~\ref{sec:likely_lq}). Figure~\ref{fig:p200_nonlenses} presents the spectra and images of the remaining four systems. As summarized in Table~\ref{tab:p200}, J0225+0513 and J1652+0608 are Q+S alignments, J1743+3744 is a G+Q system, and J1747+2212 is a PQ with an additional foreground star.
		
		\noindent\textit{J0225+0513 and J1652+0608.}
		The spectra identify component A of J0225+0513 as a quasar at $z_{\rm QSO}=1.988$ and component B as a star. In J1652+0608, component A is a star and component B is a quasar at $z_{\rm QSO}=2.427$. Both are therefore straightforward Q+S contaminants.
		
		\noindent\textit{J1743+3744.}
		Component A is a quasar at $z_A=1.916$, whereas B is a galaxy at $z_B=0.507$; the central galaxy C lies at $z_C=0.258$ (Fig.~\ref{fig:p200_nonlenses}). The system is therefore classified as a G+Q alignment rather than a PQ. Quasar A may provide a background sightline for absorption studies of galaxy C, although intervening absorption at its redshift remains to be tested.
		
		\noindent\textit{J1747+2212.}
		Component A is a foreground star, while components B and C are quasars at $z_B=1.012$ and $z_C=0.3575$, respectively. Their different redshifts are independently supported by the DESI spectra shown in the bottom panel of Fig.~\ref{fig:p200_nonlenses}. The two quasar components thus form a projected pair rather than multiple images of a single quasar. Table~\ref{tab:p200} records the system as a projected QSO pair; the star is an additional component.
		
		\section{Galaxy-galaxy lenses identified by DESI}
		\label{app:lensed_gal}
		
		We compile 124 galaxy--galaxy lens systems and candidates, comprising 118 objects in the main sample and six in the multiple-position sample (Table~\ref{tab:classification_comparison}). Following the catalogue presentation for the likely lensed quasars, we provide a machine-readable system catalogue, \texttt{LGAL124\_systems.csv}, together with a linked inventory of the individual DESI spectra, \texttt{LGAL124\_spectra.csv}. Table~\ref{tab:lgal_catalogue_columns} defines the system-catalogue columns. Discovery references, previously reported redshifts and their measurement types, and references for adopted redshifts are retained in the catalogue and its accompanying reference list, \texttt{LGAL124\_references.csv}.\footnote{\label{fn:appendix_catalogues}The galaxy--galaxy lens tables and the DQ catalogue are available at \mbox{\url{https://github.com/EigenHermit/Lqso_DESI_26}}.} The literature comparison includes the original discovery studies \citep[e.g.,][]{Andika2023,Chan2023,Shu2022}, Survey of Gravitationally-lensed Objects in HSC Imaging (SuGOHI) and lenscat records, \citet{Shu2025}, and available SDSS and Galaxy and Mass Assembly (GAMA) spectroscopy. We distinguish photometric estimates from spectroscopic measurements and do not assign a neighbouring object's redshift to the deflector or source solely on the basis of a positional match. The six newly reclassified systems are included in the expanded catalogue with their pipeline spectra; their deflector/source redshift assignments and literature novelty remain unassessed, so they do not alter the redshift contributions reported below.
		
\begin{figure*}[!tp]
			\centering
			\includegraphics{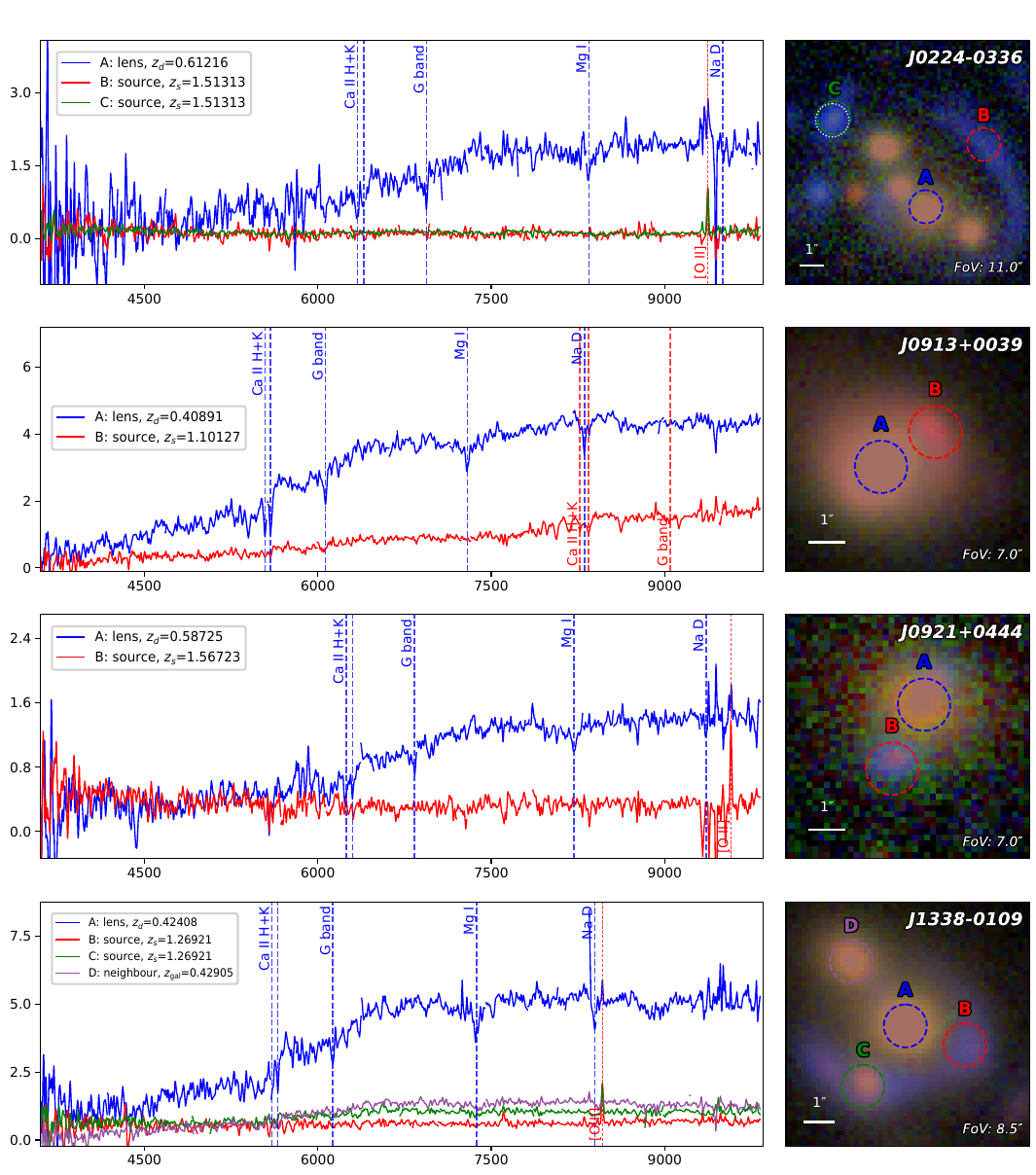}
			\caption{Newly completed spectroscopic deflector--source redshift pairs: J0224$-$0336, J0913+0039, J0921+0444, and J1338$-$0109, from top to bottom. Left: DESI spectra at distinct fibre positions, overlaid on a common flux scale without rescaling or vertical offsets; Gaussian smoothing with $\sigma=6$ native spectral pixels is applied only for display, as in Fig.~\ref{fig:spec_image}. The horizontal and vertical axes give observed wavelength in \AA\ and flux density in $10^{-17}\,\mathrm{erg\,s^{-1}\,cm^{-2}}\,\mathring{\mathrm{A}}^{-1}$, respectively; full-height dashed markers indicate expected spectral-feature positions at the adopted redshifts. Right: unsubtracted HSC $gri$ images, with square fields enclosing all fibre apertures; the system name, FOV, and a $1\arcsec$ scale bar are shown inside each cutout. Colour-matched labels link each spectrum to its $1\farcs5$-diameter DESI fibre aperture. Nearly coincident observations are represented by the higher-S/N spectrum; the white dotted aperture in J0224$-$0336 marks the omitted repeat. All four distinct positions in J1338$-$0109 are retained, including the neighbouring galaxy at $z_{\rm gal}=0.42905$, which is separate from the deflector at $z_d=0.42408$; the source-labelled spectrum adopts $z_s=1.26921$ rather than the rejected pipeline solution $z_{\rm DESI}=1.65351$.}
			\label{fig:lgal_new_pairs}
		\end{figure*}

		Our principal additions are four systems for which the present analysis supplies the first complete spectroscopic deflector--source redshift pair within the literature and catalogues checked. 
		Specifically, we add the source redshifts for J0224$-$0336 and J0913+0039, completing the pairs $(z_d,z_s)=(0.61216,1.51313)$ and $(0.40891,1.10127)$, respectively, and provide both deflector and source spectroscopic redshifts for J0921+0444, with $(z_d,z_s)=(0.58725,1.56723)$, and J1338$-$0109, with $(z_d,z_s)=(0.42408,1.26921)$, with the spectra and imaging of all four systems shown in Fig.~\ref{fig:lgal_new_pairs}.
		
		The higher-redshift components in J0224$-$0336, J0921+0444 and J1338$-$0109 show [O\,\textsc{ii}] features, while the two redshifts in J0913+0039 are supported by stellar absorption and continuum breaks. We additionally identify 16 systems for which this analysis provides a first spectroscopic redshift attributable to the putative deflector, with no previous spectroscopic measurement found in the sources checked.
		
		For the figure, spectra at distinct fibre positions are overlaid on the same flux scale. Observations with fibre centres separated by less than $0\farcs5$ are treated as one position, and the spectrum with the higher median per-pixel $f_\lambda\sqrt{\mathrm{IVAR}}$ over valid pixels between 4000 and 9000~\AA\ is displayed. In J0224$-$0336, this selects \texttt{39627700788274154} (S/N=1.12) over the nearly coincident \texttt{2789663211257856} (S/N=0.57); both records remain in the catalogue. All distinct positions in J1338$-$0109 are retained, including the neighbouring galaxy at $z_{\rm gal}=0.42905$, which is not averaged into $z_d$. The pipeline solution $z_{\rm DESI}=1.65351$ for \texttt{39627757960826773} is not adopted; the source features support $z_s\simeq1.26921$.
		
		\begin{table*}[t]
			\centering
			\caption{Column definitions for the 124-system galaxy--galaxy lens catalogue. The companion spectral inventory preserves one row per DESI \texttt{TARGETID}.}
			\label{tab:lgal_catalogue_columns}
			\begin{threeparttable}
				\small
				\setlength{\tabcolsep}{5pt}
				\renewcommand{\arraystretch}{1.08}
				\begin{tabular}{@{}p{0.27\textwidth}p{0.68\textwidth}@{}}
					\toprule
					Column & Description \\
					\midrule
					\texttt{Jname} & System name in JHHMM$\pm$DDMM format. \\
					\texttt{system\_RA\_deg} & Catalogue reference right ascension in degrees (J2000). \\
					\texttt{system\_DEC\_deg} & Catalogue reference declination in degrees (J2000). \\
					\texttt{z\_d}, \texttt{z\_s} & Adopted component-attributed deflector and source redshifts from the completed reviews; blank if not adopted. \\
					\texttt{redshift\_status} & Review status, including supported, tentative, or unresolved component attribution. \\
					\texttt{novelty} & \texttt{first\_complete\_pair} (four systems), \texttt{first\_zd} (16), or \texttt{not\_claimed}. \\
					\texttt{n\_spectra}, \texttt{n\_positions} & Number of spectral records and independent fibre positions, respectively; positions are grouped within $0\farcs5$. \\
					\texttt{TARGETIDs} & Semicolon-separated DESI identifiers; import as text to preserve all digits. \\
					\texttt{component\_assignment} & Attribution of the measured redshift to a deflector, source, or unresolved component. \\
					\texttt{adopted\_redshift\_reference} & Reference or analysis supporting the adopted component redshifts. \\
					\texttt{discovery\_catalogue} & Original input/discovery catalogue code. \\
					\texttt{reference\_ids} & Keys linking to \texttt{LGAL124\_references.csv}. \\
					\texttt{references} & Associated discovery and redshift references, with available links. \\
					\texttt{prior\_catalogue\_redshifts} & Historical compiled values with their original PHOTO, SPECTRO or SINGLE measurement labels; distinct from adopted values. \\
					\texttt{notes} & Spectral evidence, component ambiguity, rejected solutions and other object-specific qualifications. \\
					\bottomrule
				\end{tabular}
				
			\end{threeparttable}
		\end{table*}
		
		\section{DQ catalogue compiled in this work}
		\label{app:dq}
		
		\begin{figure*}[t]
			\centering
			\includegraphics[width=0.98\linewidth]{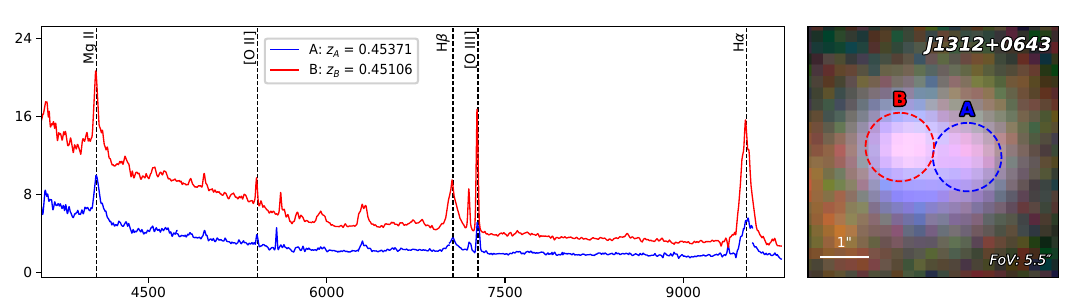}
			\caption{The left panel shows the DESI spectra of components A (blue) and B (red) of the dual quasar J1312+0643, with dashed lines marking emission-line wavelengths at the mean redshift. The right panel shows the DESI-LS $grz$ image, with colour-matched dashed circles marking the $1\farcs5$-diameter DESI fibres.}
			\label{fig:dqJ1312}
		\end{figure*}
		
		We compile a catalogue of 61 DQ systems and candidates based on their spectroscopic coverage and velocity differences. 
		Of these, 48 spectroscopically confirmed DQs and eleven DQ candidates are drawn from the main sample in Table~\ref{tab:classification_comparison}. 
		The remaining two systems come from the multiple-position sample and consist of confirmed DQs accompanied by an additional nearby galaxy at a different redshift. 
		The PQs observed with P200/DBSP are discussed separately in Appendix~\ref{app:P200_other} and are not included in this DQ catalogue. 
		
		The machine-readable catalogue, \texttt{DQ\_systems\_61.csv}, includes all associated \texttt{TARGETID}s, system coordinates, spectroscopic classifications, component and system redshifts, angular and projected separations, and velocity differences, along with relevant flags.\textsuperscript{\ref{fn:appendix_catalogues}} 
		For systems with one or two spectroscopic positions, we adopt the DESI \texttt{MEAN\_FIBER\_RA} and \texttt{MEAN\_FIBER\_DEC} of the representative spectrum or the mean of the two positions, respectively, while for the two multiple-position systems, we average only the quasar positions. 
		The system redshift is the mean of two reliable quasar redshifts, or the single available quasar value; it is left blank when no secure quasar redshift is available. 
		The line-of-sight velocity difference is calculated as $|\Delta v|=c|z_2-z_1|/[1+(z_1+z_2)/2]$ when both quasar redshifts are reliable, and is otherwise left undefined. 
		
		For the 59 systems in the previous DQ compilation, we compare with the catalogues of \citet{Jing2025DQ,Lemon2020DQ,Tang2021DQ,Chen2021DQ,He2023}, using a matching radius of $1\arcsec$ for individual components, and with published quasar-pair identifications \citep{Inada2008,Chan2023}. Forty-eight systems match the catalogue of \citet{Jing2025DQ}; seven of these also have associations with the other catalogues, and one further candidate has a published identification. The remaining 10 systems, comprising J1312+0643 and nine candidates, are absent from the comparison sources checked. J0939-0214 and J1537+0435 are later additions to this compilation and are not included in that literature cross-match. J0939-0214 remains a DQ candidate because its individual compact-source identities are unconfirmed; J1537+0435 is classified as a DQ from the visual reassessment, with two independent DESI QSO spectra at $z_1=3.493932$ and $z_2=3.521706$ and $|\Delta v|=1847.1~\mathrm{km\,s^{-1}}$. No discovery claim is made for these two additions.
		
		Figure~\ref{fig:dqJ1312} presents the DESI-LS colour image and the selected DESI spectra of J1312+0643, retaining the higher-S/N observation of component A from two records with fibre centres separated by $0\farcs133$, together with the spectrum of component B. Both spectra are classified as \texttt{QSO}, with \texttt{ZWARN=0} and \texttt{COADD\_FIBERSTATUS=0}, and have pipeline redshifts $z_A=0.453709$ and $z_B=0.451056$, yielding $z_{\rm sys}=0.452383$ and $|\Delta v|=547.6~\mathrm{km\,s^{-1}}$, while their measured fibre separation of $1\farcs494$ corresponds to $r_{\rm p}=8.90$~kpc. The unsmoothed [O\,\textsc{iii}] profiles differ between the components, with A showing a double-peaked structure reproduced in its repeat observation and B dominated by the bluer peak, supporting the DQ interpretation.
		
	\end{appendix}
	
\end{document}